\documentclass[pra,twocolumn,aps,showpacs,amsmath,amssymb,superscriptaddress,floatfix]{revtex4-1}

\usepackage{color}
\usepackage{bm}
\usepackage{dcolumn}
\usepackage{graphicx}
\usepackage{braket}
\usepackage{lipsum}
\usepackage{multirow}

\newcommand{\AAA}{\text{AAA}}
\newcommand{\AubAnd}{Aubry-Andr\'{e}}
\newcommand{\bLam}{\bar{\Lambda}}
\newcommand{\dn}{\downarrow}
\newcommand{\eps}{\epsilon}
\newcommand{\imp}{\text{imp}}
\newcommand{\host}{\text{host}}
\newcommand{\KPM}{\text{KPM}}
\newcommand{\loc}{\text{loc}}
\newcommand{\tAA}{\text{AA}}
\newcommand{\teps}{\tilde{\eps}}
\newcommand{\tDelta}{\tilde{\Delta}}
\newcommand{\tH}{\tilde{H}}

\newcommand{\tot}{\text{tot}}
\newcommand{\trho}{\tilde{\rho}}
\newcommand{\up}{\uparrow}
\newcommand{\tV}{\tilde{V}}
\newcommand{\veps}{\varepsilon}

\usepackage{xcolor}
\usepackage{xparse,xcoffins}
\ExplSyntaxOn
\NewCoffin\imagecoffin
\NewCoffin\labelcoffin
\keys_define:nn { miguel/label }
 {
  label   .tl_set:N = \l_miguel_label_tl,
  labelbox .bool_set:N = \l_miguel_label_box_bool,
  labelbox .default:n = true,
  fontsize .tl_set:N = \l_miguel_label_size_tl,
  fontsize .initial:n = \footnotesize,
  pos .choice:,
  pos/nw .code:n = \tl_set:Nn \l_miguel_label_pos_tl { left,up },
  pos/ne .code:n = \tl_set:Nn \l_miguel_label_pos_tl { right,up },
  pos/sw .code:n = \tl_set:Nn \l_miguel_label_pos_tl { left,down },
  pos/se .code:n = \tl_set:Nn \l_miguel_label_pos_tl { right,down },
  pos/n .code:n = \tl_set:Nn \l_miguel_label_pos_tl { hc,up },
  pos/w .code:n = \tl_set:Nn \l_miguel_label_pos_tl { left,vc },
  pos/s .code:n = \tl_set:Nn \l_miguel_label_pos_tl { hc,down },
  pos/e .code:n = \tl_set:Nn \l_miguel_label_pos_tl { right,vc },
  pos .initial:n = nw,
  unknown .code:n   = \clist_put_right:Nx \l_miguel_label_clist
                       { \l_keys_key_tl = \exp_not:n { #1 } }
 }
\clist_new:N \l_miguel_label_clist
\box_new:N \l_miguel_label_box
\box_new:N \l_miguel_label_image_box
\NewDocumentCommand{\xincludegraphics}{O{}m}
 {
  \group_begin:
  \tl_clear:N \l_miguel_label_tl
  \clist_clear:N \l_miguel_label_clist
  \keys_set:nn { miguel/label } { #1 }
  \tl_if_empty:NTF \l_miguel_label_tl
   {
    \miguel_includegraphics:Vn \l_miguel_label_clist { #2 }
   }
   {
    \SetHorizontalCoffin\imagecoffin
     {
      \miguel_includegraphics:Vn \l_miguel_label_clist { #2 }
     }
    \SetHorizontalCoffin\labelcoffin
     {
      \raisebox{\depth}
       {
        \bool_if:NTF \l_miguel_label_box_bool
         { \fcolorbox{white}{white}{\l_miguel_label_size_tl\l_miguel_label_tl} }
         { \l_miguel_label_size_tl\l_miguel_label_tl }
       }
     }
    \SetVerticalPole\imagecoffin{left}{3pt+\CoffinWidth\labelcoffin/2}
    \SetVerticalPole\imagecoffin{right}{\Width-3pt-\CoffinWidth\labelcoffin/2}
    \SetHorizontalPole\imagecoffin{up}{\Height-3pt-\CoffinHeight\labelcoffin/2}
    \SetHorizontalPole\imagecoffin{down}{3pt+\CoffinHeight\labelcoffin/2}
    \use:x{\JoinCoffins\imagecoffin[\l_miguel_label_pos_tl]\labelcoffin[vc,hc]} 
    \TypesetCoffin\imagecoffin
   }
   \group_end:
 }
\NewDocumentCommand{\setlabel}{m}
 {
  \keys_set:nn { miguel/label } { #1 }
 }
\cs_new_protected:Nn \miguel_includegraphics:nn
 {
  \includegraphics[#1]{#2}
 }
\cs_generate_variant:Nn \miguel_includegraphics:nn { V }
\ExplSyntaxOff

\usepackage[pdftex]{hyperref}
\hypersetup{colorlinks = true, urlcolor = blue, linkcolor = blue, citecolor = blue}

\begin{document}
\title{The \AubAnd\ Anderson model: \\ Magnetic Impurities Coupled to a Fractal Spectrum}
\author{Ang-Kun Wu}
\affiliation{Department of Physics and Astronomy, Center for Materials Theory, Rutgers University, Piscataway, New Jersey 08854, USA}
\author{Daniel Bauernfeind}
\affiliation{Center for Computational Quantum Physics, Flatiron Institute, New York, New York 10010, USA}
\author{Xiaodong Cao}
\affiliation{Center for Computational Quantum Physics, Flatiron Institute, New York, New York 10010, USA}
\author{Sarang Gopalakrishnan}
\affiliation{Department of Physics, Pennsylvania State University, University Park, Pennsylvania 16802, USA}
\author{Kevin Ingersent}
\affiliation{Department of Physics, University of Florida, Gainesville, Florida 32611-8440, USA}
\author{J. H. Pixley}
\affiliation{Department of Physics and Astronomy, Center for Materials Theory, Rutgers University, Piscataway, New Jersey 08854, USA}
\affiliation{Center for Computational Quantum Physics, Flatiron Institute, New York, New York 10010, USA}
\date{\today}

\begin{abstract}
The interplay between incommensurability and strong correlations is a challenging open issue. It is explored here via numerical renormalization-group (NRG) study of models of a magnetic impurity in a one-dimensional quasicrystal. The principal goal is to elucidate the physics at the localization transition of the \AubAnd\ Hamiltonian, where a fractal spectrum and multifractal wave functions lead to a critical \AubAnd\ Anderson (AAA) impurity model with an energy-dependent multifractal hybridization function.
This goal is reached in three stages of increasing complexity: (1) Anderson impurity models with uniform fractal hybridization functions are solved to arbitrarily low temperatures $T$. Below a Kondo temperature, these models approach a fractal strong-coupling fixed point where impurity thermodynamic properties are oscillatory in $\log_b T$ about negative average values determined by the spectrum's fractal dimension $D_F < 1$, with $b$ set by the fractal self-similarity near the Fermi energy.
(2) An impurity hybridizing uniformly with all conduction states of the critical AAA model is shown to approach the fractal strong-coupling fixed point corresponding to $D_F = 0.5$ and $b\simeq 14$.
(3) When the multifractal wave functions of the critical AAA model are taken into account, low-$T$ impurity thermodynamic properties are again negative and oscillatory, but with a more complicated structure than in (2). Under sample-averaging, the mean and median Kondo temperatures exhibit power-law dependences on the Kondo coupling with exponents characteristic of different fractal dimensions. We attribute these signatures to the impurity probing a distribution of fractal strong-coupling fixed points with decreasing temperature. 
To treat the AAA model, the numerical renormalization group (NRG) is combined with the kernel polynomial method (KPM) to form a general, efficient treatment of hosts without translational symmetry in arbitrary dimensions down to a temperature scale set by the KPM expansion order. Implications of our results for heavy-fermion quasicrystals and other applications of the NRG+KPM approach are discussed.
\end{abstract}

\maketitle


\section{Introduction}

\subsection{Background and Aims of this Work}

Strongly correlated electronic systems host qualitatively new, emergent phenomena that are of both fundamental interest and experimental relevance.
Quantum impurity models such as the Kondo model \cite{kondo1964resistance,hewson1997kondo}, which was first used to describe iron impurities in a metal, represent a particularly simple type of correlated quantum many-body system: they consist of a strongly interacting local region (the ``impurity'') either embedded in a noninteracting metallic host or, in the context of quantum dots \cite{cronenwett1998tunable,pustilnik2004kondo}, tunnel-coupled to conducting leads.
Despite the simplicity of the noninteracting host degrees of freedom, which makes these problems more tractable than generic correlated systems and has allowed significant progress in their understanding, impurity models provide quintessential examples of asymptotic freedom and nonperturbative phenomena such as Kondo screening \cite{PhysRevLett.23.89,PhysRevB.1.4464}, while also being rich enough to host boundary quantum critical phenomena \cite{PhysRevB.57.14254,si2001locally,PhysRevLett.89.076403,PhysRevLett.109.086403,fritz2013physics,PhysRevB.88.245111,nahum2022fixed}.
Our understanding of bulk correlated materials, such as heavy-fermion systems \cite{si2014kondo} and high-temperature superconductors, draws heavily on insights from impurity models. Indeed, state-of-the-art numerical techniques such as the dynamical mean-field theory \cite{RevModPhys.68.13} and its extensions map the correlated electron problem to a self-consistent quantum impurity model.

While the most physically relevant impurity problems---the Anderson \cite{PhysRev.124.41} and Kondo models---are solvable for clean, noninteracting electronic hosts, much less is understood about their behavior in inhomogeneous systems. The interplay between disorder and strong correlations, and more generally the nature of quantum phase transitions in inhomogeneous systems, remain challenging open problems. Progress has been made for impurity systems with quenched randomness, in which one can treat the randomness as uncorrelated and average over it, yielding a non-Fermi liquid ground state \cite{PhysRevLett.69.1113,miranda1996kondo,PhysRevLett.96.117209,PhysRevLett.103.126401,PhysRevB.85.115112,PhysRevB.90.201101,chakravarty2000kondo,PhysRevLett.99.247202}. However, many experimentally relevant systems have correlated-but-aperiodic \cite{jagannathan2010open} (or nearly aperiodic) spatial inhomogeneity.
Examples include quasicrystals \cite{PhysRevLett.53.2477}, incommensurate optical lattices \cite{roati2008anderson}, and moir\'e materials \cite{balents2020superconductivity,he2021moire,fu2020magic}.
Each case provides experimental evidence for correlated phases \cite{PhysRevLett.90.177205,schreiber2015observation,kamiya2018discovery}, such as quantum criticality without tuning in the Yb-Al-Au heavy-fermion quasicrystal \cite{deguchi2012quantum,matsukawa2016pressure,PhysRevB.96.241102,ishimasa2018interpretation} and observation of insulating phases at integer fillings of the moir\'e unit cell in magic-angle twisted bilayer graphene \cite{cao2018unconventional,cao2018correlated,wu2021chern}. 
To date, however, there is no theoretical framework that handles both the aperiodic inhomogeneity and the strong interactions on the same footing.

\begin{figure}[t!]
\centering
\includegraphics[width=0.47\textwidth]{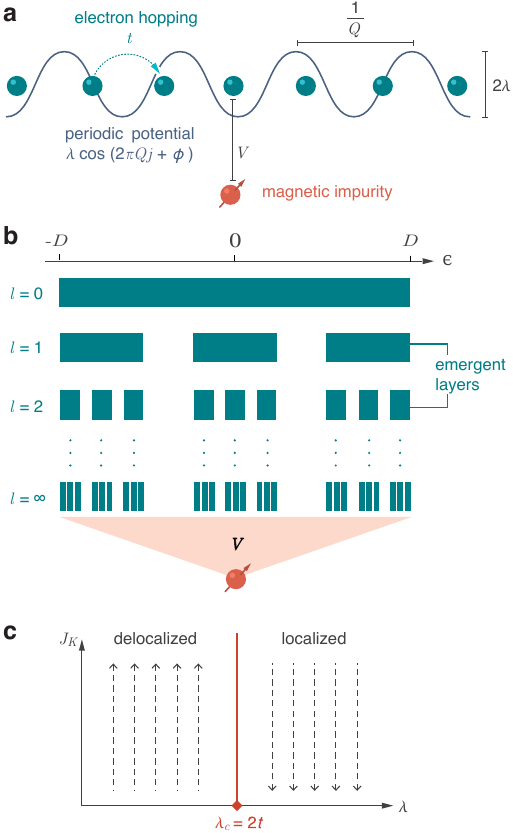}
\caption{\label{fig:model+phase-diagram}%
(a) Schematic of the AAA model describing a magnetic impurity (red circle) hybridizing with matrix element $V$ with the middle site of an \AubAnd\ chain (green circles).
(b) An impurity hybridizes with an emergent, self-similar electron band formed by dividing a uniform band of halfwidth $D$ into subbands separated by gaps, and then repeatedly dividing each subband into narrower ones. For iteration number $l\to\infty$, the band evolves into a fractal containing no interval of nonzero width.
(c) Schematic phase diagram for the AAA model in its Kondo limit [Eq.\ \eqref{eq:H_K}]: Renormalization-group flow of the Kondo coupling $J_K$ in a typical system for bare $J_K\ll D$. In the delocalized phase ($\lambda < \lambda_c$), a divergent $J_K$ signals many-body quenching of the impurity spin. In the localized phase ($\lambda > \lambda_c$), vanishing of $J_K$ is associated with an asymptotically free impurity spin. (For $\lambda > \lambda_c$, sufficiently large bare values of $J_K/D$ can cause local screening of the impurity spin.) The orange diamond marks the critical point ($\lambda = \lambda_c$) that is the main focus of this work.}
\end{figure}

In this work, we take an initial step toward developing such a framework by studying Kondo physics in an electronic host described by a spinful version of the \AubAnd\ (AA) tight-binding model for a one-dimensional quasicrystal \cite{aubry1980analyticity} (related to the Harper model for band electrons in a magnetic field \cite{harper1955single}). Increasing the strength $\lambda$ of a smooth, periodic potential that is incommensurate with the $\lambda = 0$ lattice [as depicted in Fig.\ \ref{fig:model+phase-diagram}(a)] drives a localization-delocalization quantum phase transition \cite{RevModPhys.80.1355} without a mobility edge \cite{aubry1980analyticity,avila2009ten} [shown schematically in Fig.\ \ref{fig:model+phase-diagram}(c)]. 
Precisely at the transition, the host has a fractal energy spectrum [reflected in the iterative sequence of minibands in Fig.\ref{fig:model+phase-diagram}(b)] and critical single-particle wave functions \cite{PhysRevLett.43.1954,aubry1980analyticity,harper1955single,dominguez2019aubry,deng2017many}; the global density of states (global DOS) is a nonuniform fractal \cite{PhysRevB.14.2239, hiramoto1992electronic, Azbel1964,xu1987fractal, wu2021fractal} while the local DOS (or LDOS) on any tight-binding site exhibits multifractal character \cite{PhysRevLett.51.1198,hiramoto1992electronic,PhysRevLett.120.207604}. To solve our \AubAnd\ Anderson (AAA) impurity problem, we introduce a ``KPM+NRG'' computational approach that combines the numerical renormalization group (NRG) for nonperturbative solution of quantum impurity models \cite{RevModPhys.47.773,RevModPhys.80.395} with the kernel polynomial method (KPM) \cite{RevModPhys.78.275} for efficiently evaluating the global or local DOS of inhomogenous hosts in arbitrary dimensions \cite{RevModPhys.78.275}.

To identify the characteristics that distinguish the Kondo problem considered here from more conventional versions, it is useful to review how a magnetic impurity interacts with its electronic environment.
The Anderson impurity model \cite{PhysRev.124.41} fully captures tunneling of electrons between the impurity level and the host in an energy-dependent hybridization function that is proportional to the host's local density of state (local DOS or LDOS) at the impurity site.
One can expect criticality of a quasicrystalline host to have two effects on the impurity. First, the energy eigenstates near the Fermi energy are highly nonuniform in space \cite{hiramoto1992electronic,PhysRevB.96.045138}; depending on its location, the impurity may be either very weakly or very strongly coupled to any given host state, resulting in an LDOS very different from the global DOS $\rho(\eps)$.
Thus, the Kondo temperature $T_K$---the characteristic scale for the many-body screening of the impurity's magnetic moment---should become broadly distributed, as also seen in random systems \cite{PhysRevLett.115.036403,PhysRevLett.69.1113,miranda1996kondo,PhysRevLett.96.117209,PhysRevLett.103.126401,PhysRevB.85.115112,PhysRevB.90.201101}.
Another effect, specific to quasicrystals and the focus of the current manuscript, is that the DOS itself becomes fractal at the critical point \cite{hiramoto1992electronic}: the eigenenergies cluster in flat minibands, separated from one another by a self-similar hierarchy of gaps [see Fig.\ \ref{fig:model+phase-diagram}(b)].
At a fixed band filling, the DOS is almost always infinite at the Fermi energy $\eps_F$, and one cannot expect the Kondo temperature to exhibit the dependence $\log T_K\sim -1/\rho(\eps_F)$ that holds in conventional metallic hosts \cite{hewson1997kondo}.

\begin{figure*}[t!]
\setlabel{pos=nw,fontsize=\normalsize,labelbox=false}
\xincludegraphics[scale=0.325,label=a]{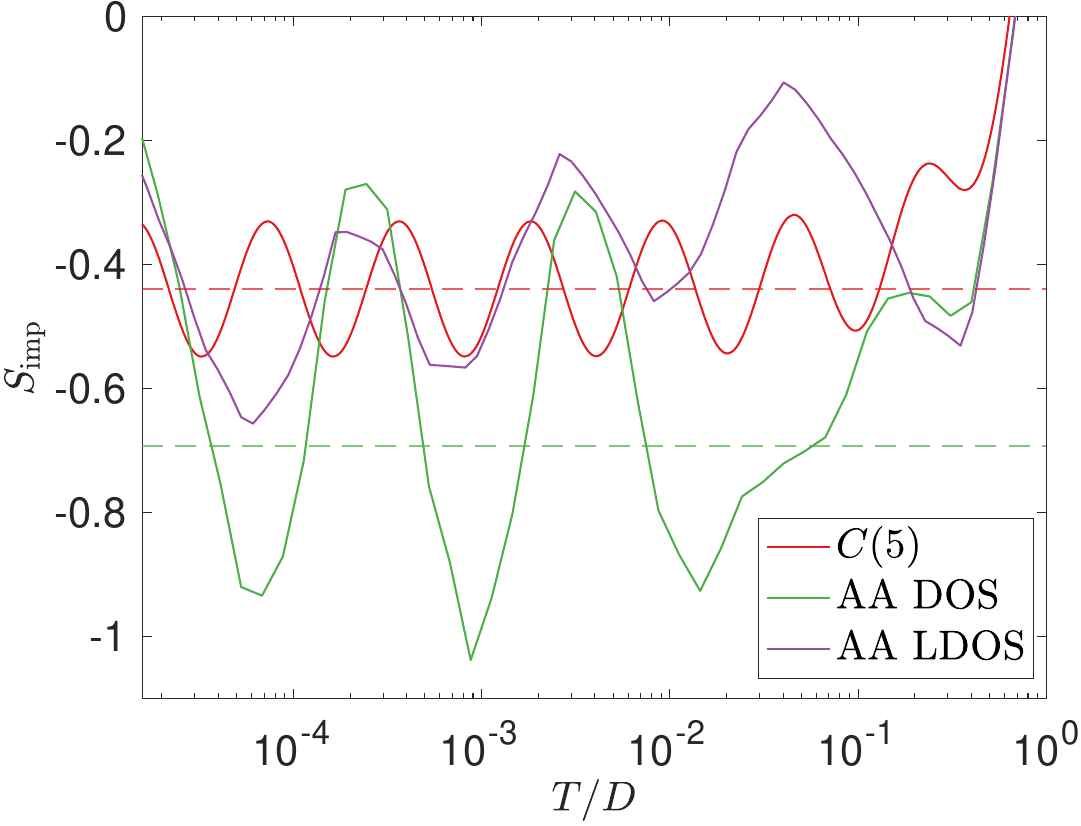}
\xincludegraphics[scale=0.325,label=b]{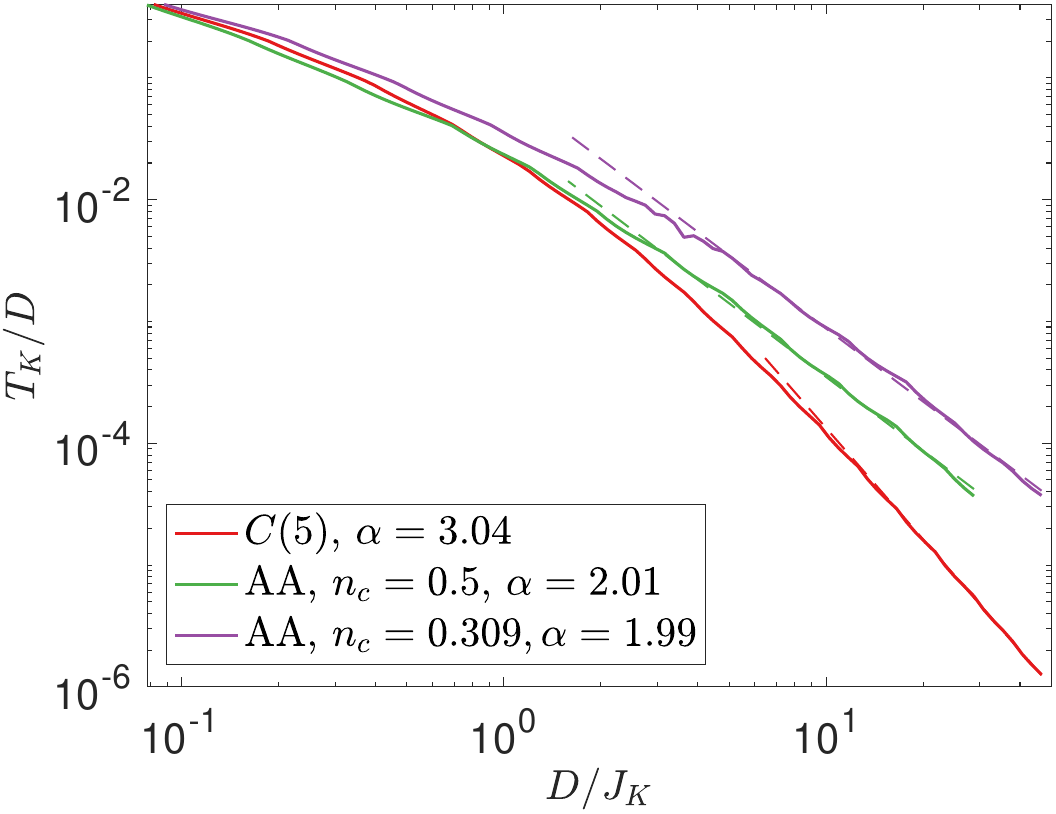}
\xincludegraphics[scale=0.325,label=c]{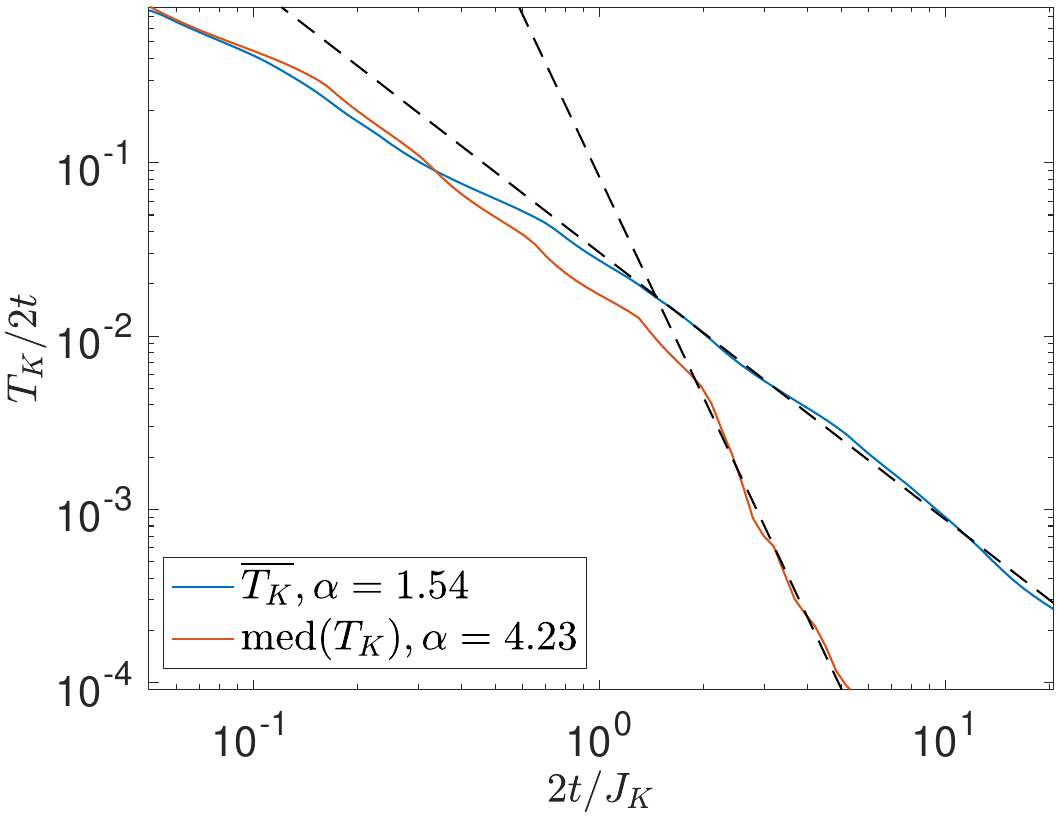}
\caption{\label{fig:key-results}%
Summary of main results: (a) Impurity entropy $S_{\imp}$ vs temperature $T$ for an Anderson impurity strongly hybridizing with one of three hosts: a uniform Cantor-set fractal $C(5)$ (see Sec.\ \ref{subsec:fractal-host}) at band filling $n_c = 0.5$, and the global DOS and LDOS of the critical \AubAnd\ (AA) model [Eq.\ \protect\eqref{eq:H_AA} with $\lambda = 2t$] at $n_c = 0.309$. LDOS data are averaged over $100$ random impurity positions. The upper (lower) dashed line marks the average value expected for the fractal dimension $D_F=\log_5 3$ of $C(5)$ ($D_F= 0.5$ of the AA model). (b) Kondo temperature $T_K$ vs reciprocal Kondo coupling $1/J_K$ for $C(5)$ at $n_c = 0.5$ and for the AA DOS at $n_c = 0.5, \, 0.309$. Dashed lines are small-$J_K$ fits to $T_K \sim J_K^\alpha$ with $\alpha$ values shown in the legends. (c) Mean $\overline{T_K}$ and median $\text{med}(T_K)$ over $500$ random impurity positions vs $1/J_K$ for the AA LDOS at $n_c=0.309$. Technical details (see Secs. \protect\ref{sec:models} and \protect\ref{sec:numerics}): impurity parameters $U=-2\eps_d = D$, treating the AA model for $L = 10^6$ sites to KPM expansion order $N_C=10^5$. Also (a) hybridization $V=1.6D$ and NRG discretization $\Lambda=5^{1/4}$ ($\Lambda=3$) for $C(5)$ (AA model); (b) $\Lambda = 5$; (c) $\Lambda=8$.}
\end{figure*}

To explore the preceding general expectations and achieve a deeper understanding of how critical wave functions and a fractal spectrum impact the physics of a strongly interacting impurity, we progress in three stages, gradually incorporating more features of the full problem of interest. (1) We isolate the effect of a fractal spectrum (neglecting the wave function contribution) by treating a local magnetic level that hybridizes [see Fig.\ \ref{fig:model+phase-diagram}(a)] with an idealized energy spectrum following a well-understood fractal pattern, namely, a uniform Cantor set. This simplification allows robust NRG solution for thermodynamic properties down to arbitrarily low temperatures and the identification of characteristic signatures of fractality. (2) We apply the KPM+NRG approach to an impurity mixing with the global DOS of the critical host, a problem that also neglects the wave-function contribution but takes account of the specific form of the fractal spectrum for the model quasicrystal. Although we cannot access such low temperatures as in (1), we are able to establish with confidence (for large, finite systems at two different band fillings) that the infrared limit exhibits the same signatures of fractality. (3) We perform a KPM+NRG study of the full model of interest, using the first two stages to guide the interpretation of results.

\subsection{Overview of Principal Results}

This section provides a summary of the main findings from the three stages of our study, illustrated in Fig.\ \ref{fig:key-results} in terms of two physical properties whose precise definitions appear in Sec.\ \ref{subsec:observables}. The first is $S_{\imp}$, the impurity contribution to the thermodynamic entropy, representing the difference between the total entropy of the combined host-impurity system and the total entropy of the host alone. The second property shown in Fig.\ \ref{fig:key-results} is the Kondo temperature $T_K$, already introduced above as the characteristic scale for screening of the impurity magnetic moment by the host. For a magnetic impurity in a conventional metal \cite{hewson1997kondo}, (a) $S_{\imp}$ remains near $\ln 2$ \cite{Note1} over a range of intermediate temperatures where the impurity acts as a spin one-half degree of freedom before crossing over below $T_K$ to approach a low-temperature ``strong-coupling'' limit of zero, and (b) $T_K$ is exponentially sensitive to the effective impurity-host exchange coupling $J_K$. 

As noted above, we have explored three classes of models: (1)~an impurity coupled to a host with a Cantor-set spectrum, (2)~an impurity nonlocally coupled to the AA model, i.e., replacing the LDOS with the global DOS, and (3)~an impurity locally coupled to the AA model. While case~(3) is the most physically relevant, it is also the least tractable.

\textbf{Case (1)} might arise if the DOS were fractal but the wave functions remained delocalized. 
This could occur, for example, if the ``impurity spin'' were a nonlocal two-level system such as a nonlinear oscillator \cite{PhysRevLett.107.063601}. 
In this case, we have found a \textit{fractal strong-coupling fixed point\/} with the highly unusual feature [see data labeled ``$C(5)$'' in Fig.\ \ref{fig:key-results}(a)] that the impurity's thermodynamic properties, such as its entropy $S_\imp$, exhibit oscillations that are periodic in $\log_b T$ about a negative value that is determined by the fractal dimension $D_F<1$ of the spectrum [defined in Eq.\ \eqref{eq:D-box}]. The period of these oscillations is set by the self-similarity of the fractal DOS under multiplicative rescaling $\eps-\eps_F \to (\eps-\eps_F)/b$ of energies about the Fermi energy $\eps_F$, whereas the oscillation phase depends on the band filling. Kondo screening sets in around a temperature $T_K\sim J_K^{\alpha}$, where $\alpha=1/(1-D_F)$ for small $J_K$ [see Fig.\ \ref{fig:key-results}(b)]. Both this power-law dependence of $T_K$ and the negative temperature-averaged values of impurity thermodynamic properties reproduce the behaviors of a system with a smooth (nonfractal) DOS exhibiting a singularity $\rho(\epsilon)\propto |\eps - \eps_F|^r$ at the Fermi energy, where $r = D_F - 1 < 0$. We have verified these results numerically to arbitrarily low temperatures.

\textbf{Case (2)} considers the global DOS of the critical AA model, which has a fractal dimension $D_F = 0.5$ and an energy self-similarity factor $b\simeq 14$. A magnetic impurity hybridizing with this DOS (amounting to a uniform coupling to all conduction electrons of the one-dimensional host) exhibits thermodynamic properties that are both qualitatively and quantitatively consistent with the fractal strong-coupling scenario  over the temperature range that we can probe; see $S_\imp$ data labeled ``AA DOS'' in Fig.\ \ref{fig:key-results}(a) and $T_K$ curves labeled ``AA'' in Fig.\ \ref{fig:key-results}(b).

\textbf{Case (3)} builds on intuition and insight from the fractal strong-coupling fixed point to interpret data for the full AAA impurity model. For individual samples representing specific impurity locations within the host, the system appears to probe several different fractal strong-coupling fixed points as it flows to strong coupling, as exemplified by often-large fluctuations in impurity thermodynamic quantities about temperature-dependent average values. Sample-averaging the impurity thermodynamic quantities brings out log-temperature oscillations about a background value that drifts slowly with temperature; see data labeled ``AA LDOS'' in Fig.\ \ref{fig:key-results}(a). Importantly, the oscillations qualitatively resemble those from case (2), implying that sample averaging is similar to working with the global DOS. The multifractal wave functions at the delocalization-localization transition lead to a broad distribution of Kondo temperatures, with a clear tail in its cumulative distribution function towards vanishing Kondo coupling $J_K$. As a result, the mean and median Kondo temperatures differ in the small-$J_K$ limit, although they both follow power-law forms as shown in Fig.\ \ref{fig:key-results}(c), indicative of a singular hybridization function. However, in contrast to case (2), the power laws $T_K\sim J_K^{\alpha}$ do not have $\alpha$ values that are simply related to a single fractal dimension.

\subsection{Outline of the Rest of the Paper}

The remainder of the paper is organized as follows: Section \ref{sec:models} defines the models studied, while Section \ref{sec:numerics} describes the numerical methods used and the observable properties that we compute. Sections \ref{sec:Cantor}, \ref{sec:fractal-AA}, and \ref{sec:full-AAA} present in turn more detailed results from the three stages of our investigation. Discussion and conclusions appear in Sec.\ \ref{sec:discussion}. Appendix \ref{app:KPM+NRG} lays outs the KPM+NRG approach, shows that it reproduces the pure-NRG treatment of two specific model hosts, and yields excellent agreement with the density-matrix renormalization-group \cite{RevModPhys.77.259} for the AAA impurity model at smaller system sizes. Appendices \ref{app:self-similar} and \ref{app:AAA-NRG} address other technical details.

\section{Models}
\label{sec:models}

We are interested in describing a magnetic impurity with an on-site repulsion embedded in a quasicrystalline host. One of the simplest possible descriptions of such a host is the AA model of a one-dimensional tight-binding chain of spin-1/2 electrons subjected to an incommensurate potential. We will find it advantageous to make further simplifications to separately understand the effects of a fractal energy spectrum and multifractal wave functions, both of which occur at the critical point of the AA model.

\subsection{\AubAnd\ Anderson impurity model}
\label{subsec:AAA-model}

The Anderson impurity Hamiltonian for an interacting impurity level coupled to one site (hereafter called ``the impurity site'') of an otherwise non-interacting host lattice can be written as
\begin{equation}
\label{eq:H_A}
    H_\text{A} = H_\host + H_\imp + H_\text{hyb}.
\end{equation}
In the AAA model, illustrated schematically in Fig.\ \ref{fig:model+phase-diagram}(a), the host is represented by the spinful AA Hamiltonian,
\begin{multline}
\label{eq:H_AA}
    H_\host = \sum_{j=1}^L \sum_\sigma \Bigl[ t( c_{j\sigma}^\dag c_{j+1,\sigma} + \text{H.c.}) \\[-2ex]
    + \lambda \cos(2\pi Q j+\phi) \, c_{j\sigma}^\dag c_{j\sigma} \Bigr] ,
\end{multline}
where $c_{j\sigma}$ annihilates a band electron with spin $z$ component $\sigma=\:\up$ or $\dn$ at site $j$ in a one-dimensional chain of $L$ sites. This AA chain [see Fig.\ \ref{fig:model+phase-diagram}(a)] has a nearest-neighbor tight-binding hopping $t$ and a potential with a strength $\lambda$, an incommensurate wave number $Q$, and a phase $\phi$ that will be treated as a random variable to be averaged over; see Fig.\ \ref{fig:model+phase-diagram}(a)].

The second term on the right-hand side of Eq.\ \eqref{eq:H_A} is
\begin{equation}
\label{eq:H_imp}
    H_\imp = (\eps_d + \eps_F) (\hat{n}_{d\up} + \hat{n}_{d\dn}) + U \hat{n}_{d\up} \hat{n}_{d\dn} + \frac{h}{2} (\hat{n}_{d\up} - \hat{n}_{d\dn}),
\end{equation}
describing a nondegenerate impurity orbital occupied by $\hat{n}_{d\sigma} = d_\sigma^\dag d_\sigma$ electrons having spin $z$ component $\sigma$, energy $\eps_d$ measured from the host Fermi energy $\eps_F$, and an on-site repulsion $U$. The impurity is subjected to a local magnetic field $h$ that is set to zero except when calculating the local magnetic susceptibility (Sec.\ \ref{subsec:observables}) and certain results shown in Appendix \ref{subsec:KPM+NRG-compare-DMRG}. Finally,
\begin{equation}
\label{eq:H_hyb}
    H_\text{hyb} = V \sum_{\sigma} (d_\sigma^\dag c_{R\sigma} + \text{H.c.})
\end{equation}
introduces mixing between the impurity level and host lattice site $R$ with a hybridization matrix element $V$ that can be taken to be real and non-negative.

It is convenient to transform $H_\text{A}$ to the single-particle eigenbasis $\{ |\eps_k,\sigma\rangle \}$ of $H_\host$, where $\eps_k$ is the energy eigenvalue of a state annihilated by an operator $\bar{c}_{k,\sigma}$ that has wave function $\phi_{k,\sigma}(j) = \braket{j,\sigma | \eps_k}$ at lattice site $j$. $H_\imp$ is unaffected by the basis change, while the remaining parts of $H_\text{A}$ become
\begin{align}
\label{eq:H_host-bar}
    H_\host &= \sum_{k,\sigma} \eps_k \bar{c}_{k\sigma}^\dag \bar{c}_{k\sigma} , \\
    H_\text{hyb} &= V \sum_{k,\sigma} \bigl[ \phi_k(R) d_{\sigma}^\dag \bar{c}_{k\sigma} + \mathrm{H.c.} \bigr].
\end{align}
The influence of the host on the impurity is completely determined by the so-called hybridization function:
\begin{equation}
\label{eq:hyb}
    \Delta(\eps)
    = \pi V^2 \sum_k |\phi_k(R)|^2 \delta(\eps-\eps_k)
    \equiv \pi V^2 \rho_R(\eps),
\end{equation}
where $\rho_R(\eps)$ is the host LDOS per spin orientation at the impurity site $R$, to be distinguished from the global DOS (per spin orientation, per lattice site) $\rho(\eps)=L^{-1}\sum_k \delta(\eps - \eps_k)$.

For $-\eps_d,\: U\!+\!\eps_d \gg V, \: T$ \footnote{We work in units where the reduced Planck constant $\hbar$, Boltzmann's constant $k_B$, and the electron magnetic moment $g\mu_B$ take values $\hbar = k_B = g\mu_B = 1$.}, occupancy $n_d = 1$ overwhelmingly predominates, localizing a spin-1/2 degree of freedom in the impurity level. In this limit, the Schrieffer-Wolff transformation \cite{PhysRev.149.491} can be used to map $H_\text{A}$ to an effective Kondo Hamiltonian
\begin{equation}
\label{eq:H_K}
    H_\text{K} = H_\host + J_K \mathbf{S}_\imp \cdot \mathbf{s}_R + V_K \sum_\sigma c_{R\sigma}^\dag c_{R\sigma},
\end{equation}
where $V_K$ is the strength of local potential scattering from the impurity and $J_K$ is the local Kondo exchange coupling between the impurity spin $\mathbf{S}_\imp = \sum_{\alpha,\beta} d_\alpha^\dag {\textstyle\frac{1}{2}} \boldsymbol{\sigma}_{\alpha\beta} d_{\beta}$ and the host spin $\mathbf{s}_R =\sum_{\alpha,\beta}c_{R\alpha}^\dag {\textstyle\frac{1}{2}}\boldsymbol{\sigma}_{\alpha\beta}c_{R\beta}$ at the impurity site.
For simplicity, in this paper we focus on particle-hole-symmetric impurities, i.e., $\eps_d = -U/2$, for which cases the Schrieffer-Wolff transformation gives
\begin{equation}
\label{eq:SW-couplings}
    J_K / D = 8 V^2 / U, \quad V_K = 0.
\end{equation}
We note that a non-zero potential scattering $V_K$ can also be generated due to asymmetry of the hybridization function about the Fermi energy.

In the case of the AA host, sample averaging can be performed by varying the phase $\phi$, so without loss of generality we can couple the impurity to the middle lattice site $j=R=L/2$. We apply open boundary conditions to the AA chain and set $Q=(\sqrt{5}-1)/2$, the reciprocal of the golden ratio.
The properties of the AA band are strongly dependent on the filling. Following Ref.\ \onlinecite{PhysRevB.100.165116}, we focus on a fixed band filling (rather than a fixed chemical potential) to try to avoid the Fermi energy falling in a large energy gap in our finite-size simulations. Guided by this earlier work, which identified nongapped fillings over system sizes on the order of $L=10^4$, the filling of the conduction band
\begin{equation}
\label{eq:n_c}
    n_c = \frac{1}{2L} \sum_{j=1}^L \sum_\sigma \braket{ c_{j\sigma}^\dag c_{j\sigma}}
\end{equation}
is taken to be $0.309$ per spin per site, while the half-filled case $n_c=1/2$ is also studied for comparison.

A great advantage of the AA model is that its phase diagram is known exactly from duality transformations \cite{aubry1980analyticity,sokoloff1985unusual} as well as from the Bethe-ansatz for commensurate approximants \cite{PhysRevLett.72.1890,PhysRevLett.73.1134,PhysRevLett.81.2112}. 
The model has a localization-delocalization transition at $\lambda_c=2t$ for all eigenenergies (i.e., without a mobility edge) as sketched for $J_K=0$ in Fig.\ \ref{fig:model+phase-diagram}(c). The LDOS reflects this transition---e.g., through its geometric mean value $\exp\overline{\ln\rho_R}$~\cite{PhysRevLett.114.146601}, with the averaging taking place over both $\phi$ and $\eps$---allowing us to predict the low-temperature behavior of a strongly correlated impurity on either side of the transition. We note that averaging over impurity location or $\phi$ yield equivalent results. Throughout the delocalized phase ($\lambda < \lambda_c$), the host states are spatially extended, so a typical impurity will hybridize with an LDOS that, in the thermodynamic limit $L\to\infty$, is featureless around $\eps = \eps_F$. The impurity will therefore exhibit conventional Kondo physics, with even very weak Kondo couplings such that $\rho_R(\eps_F) J_K \ll 1$ resulting in local-moment screening at temperatures $T$ much below a local Kondo temperature $T_K \propto \exp[-1/\rho_R(\eps_F)J_K]$. By contrast, in the localized phase $\lambda>\lambda_c$, the band wave functions are exponentially localized. Hence, an impurity coupled to a typical site $R$ will hybridize only with a discrete subset of band states $|\eps_k\rangle$. The smallest value of $|\eps_k - \eps_F|$ over this subset defines a gap scale $\eps_\text{gap}(R)$ such that $\Delta(\eps) = 0$ for $|\eps - \eps_F| < \eps_\text{gap}(R)$. As a result, the physics will be similar to that of a magnetic impurity in a band insulator, where Kondo screening occurs only if $J_K$ exceeds a threshold value, while for weaker Kondo couplings the impurity moment becomes asymptotically free as $T\to 0$. This phenomenon is summarized by the schematic RG flows in Fig.\ \ref{fig:model+phase-diagram}(c).

A complete solution of the AAA Hamiltonian at the critical point of the AA model remains a nontrivial and challenging task. In the following, we develop a novel numerical approach to solve this problem by integrating the KPM for computing the LDOS into the NRG method. The NRG and KPM methods are both formulated for a dimensionless spectrum contained within the interval $[-1, \, 1]$. With this in mind, we identify the greatest particle or hole excitation energy above the Fermi energy $\eps_F$ as
\begin{equation}
\label{eq:D}
    D = D_\host + |\eps_F|,
\end{equation}
where $D_\host = \sup_k |\eps_k|$ is the half-bandwidth of $H_\host$. In the case of the AA model, both $D_\host$ and (for $n_c \ne \frac{1}{2}$) $\eps_F$ depend on the incommensurate potential strength $\lambda$ entering Eq.\ \eqref{eq:H_AA}. We then define a reduced band energy
\begin{equation}
\label{eq:eps-tilde}
    \teps = (\eps-\eps_F) / D,
\end{equation}
as well as a reduced DOS, LDOS, and hybridization function
\begin{align}
\label{eq:rho-tilde}
    \trho(\teps)& = D\rho(D\teps), \quad
    \trho_R(\teps) = D\rho_R(D\teps), \\[0.5ex]
\label{eq:Delta-tilde}
    \tDelta(\teps)& = D \Delta(D\teps)/(\pi V^2),
\end{align}
all of which are unit-normalized and necessarily vanish for $|\teps|>1$. Finally, we define reduced Hamiltonians
\begin{equation}
\label{eq:H-tilde}
\tH_\text{A} = H_\text{A}/D, \quad \tH_\text{K} = H_\text{K}/D.
\end{equation}
containing reduced parameters $U/D$, $\eps_d/D$, $V/D$, $J_K/D$, and $V_K/D$.

The results presented in this paper were all computed for fixed $U = -2\eps_d = D$, with $V$ being varied to control the Kondo coupling $J_K$.
The KPM+NRG technique is restricted to temperatures exceeding a scale set by the finite energy resolution of the KPM. For this reason, before turning to results for the AAA model, we first consider a simpler model that can be studied to arbitrarily low temperatures. 

\begin{figure}[t!]
\centering
\includegraphics[width = 0.49\textwidth]{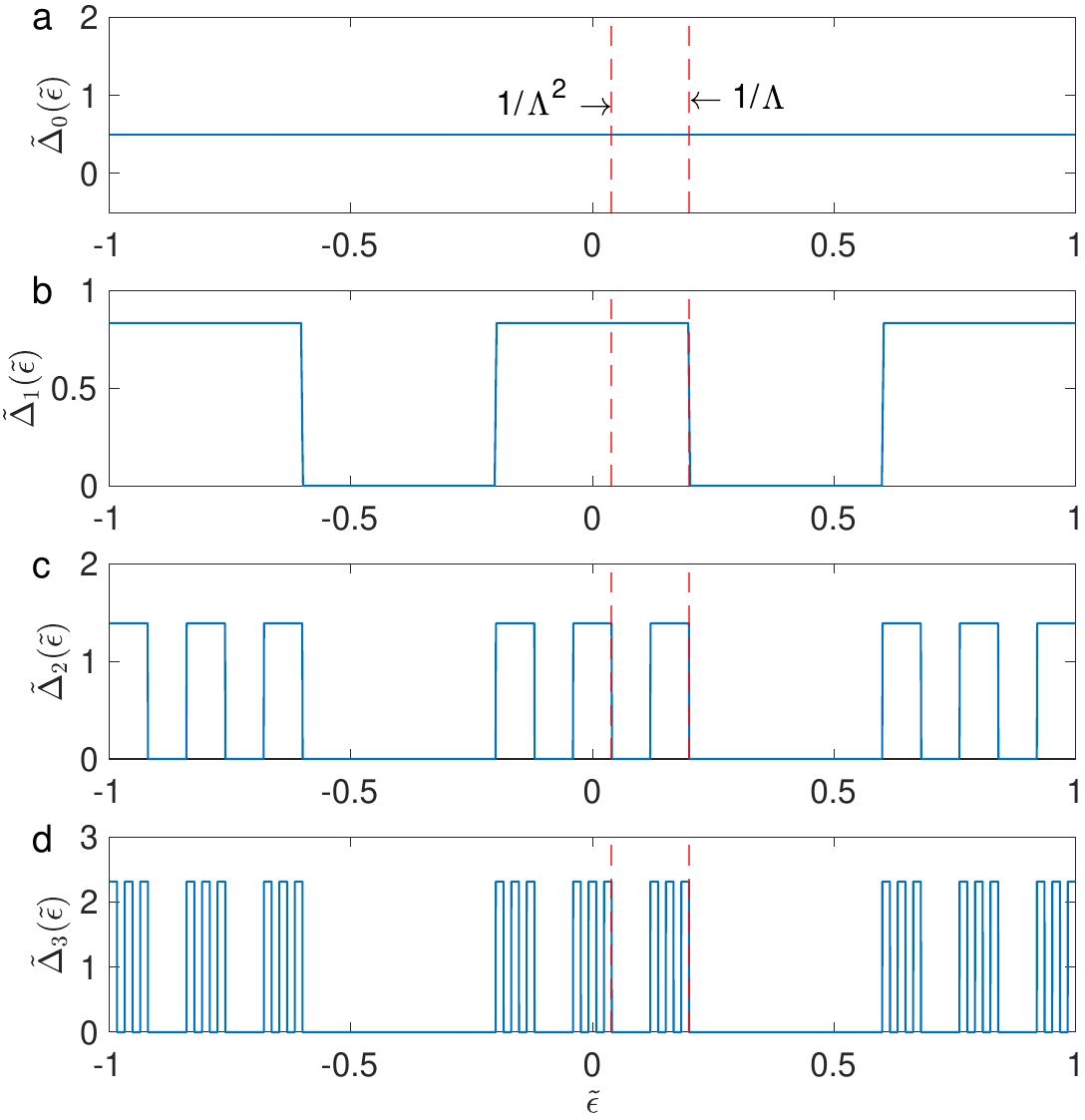}
\caption{\label{fig:5-3-Cantor-DOS}%
Reduced hybridization functions $\tDelta_l(\teps)$ approximating a fractal $1/5$ Cantor set: (a) Uniform initial hybridization function $\tDelta_0(\teps)$. (b)-(d) First three approximants $\Delta_l(\teps)$ formed by iteratively dividing each interval into five equal parts labeled 1 through 5 and removing the two even-numbered parts. The vertical red dashed lines mark the lower bounds $\Lambda^{-m}, \, m = 1, \, 2$ of the first two logarithmic bins in the NRG discretization of the hybridization function for discretization $\Lambda=5$ and offset $z=1$. For $l\ge m$, $\teps=\Lambda^{-m}$ lies at the upper edge of an energy range in the support of $\tDelta_l(\teps)$.}
\end{figure}

\subsection{Anderson impurity model in a fractal host}
\label{subsec:fractal-host}

As outlined above, the effect of the host in an Anderson impurity model is fully captured via a hybridization function [Eq.\ \eqref{eq:hyb}] that can be interpreted as the convolution of two parts: an energy spectrum $\eps_k$ that determines the global DOS and the probability weight $|\phi_k(R)|^2$ of each single-particle eigenstate, both of which contribute to the LDOS. At the localization transition point of the one-dimensional quasicrystal, the DOS is expected to assume a fractal form while the LDOS (and hence the hybridization function) should be multifractal. The full multifractal AAA model will be addressed in Sec.\ \ref{sec:full-AAA}. However, we first seek insight from two examples from a simpler class of Anderson impurity models having uniform fractal hybridization functions. Such a hybridization function has a unique value $0<D_F<1$ of the box-counting dimension
\begin{equation}
\label{eq:D-box}
   D_F = \lim_{\veps\to 0} \frac{\log N(\veps)}{\log\veps^{-1}}
\end{equation}
where $N(\veps)$ is the number of non-overlapping boxes of width $\Delta\teps=\veps$ required to cover the support of the reduced hybridization function $\tDelta(\teps)$.

One way to generate a fractal hybridization function is through a finite subdivision rule, i.e., $\Delta(\eps)=\lim_{l\rightarrow\infty} \Delta_l(\eps)$ with $\Delta_{l+1}(\eps)=\mathcal{R}\Delta_l(\eps)$. Here, $\mathcal{R}$ is a discrete transformation that reduces the support of a fractal approximant function, yielding a new approximant that exhibits fractal scaling down to a finer energy resolution.

Section \ref{sec:Cantor} focuses on a hybridization function $\Delta_{C(4M+1)}(\eps)$ described by a uniform $1/(4M+1)$ Cantor set where $M$ is a positive integer. Starting with a flat-top $\Delta_0(\eps)= \frac{1}{2} \pi V^2 \, \Theta(1-|\eps/D|)$, where $\Theta(x)$ is the Heaviside function, one forms $\Delta_l(\eps)$ for $l=1, \, 2,\, 3\, \ldots$ by taking each contiguous energy range over which $\Delta_{l-1}(\eps)>0$ and performing three steps: (1) Divide the range into $4M+1$ equal-width intervals labeled $1$ to $4M+1$ in order of ascending central energy. (2) Set $\Delta_l(\eps)=0$ throughout each of the $2M$ even-numbered intervals. (3) Set $\Delta_l(\eps) = (4M+1)(2M+1)^{-1} \Delta_{l-1}(\eps)$ throughout the $2M+1$ odd-numbered intervals so that $\int_{-D}^D \Delta_l(\eps) \, d\eps = \pi V^2$ for all $l$. This finite subdivision rule has been designed so that $\Delta_{C(4M+1)}(\eps)$ has nonvanishing integrated weight over energy ranges arbitrarily close to $\eps=0$ (in contrast to the situation in a band insulator.) Figure \ref{fig:5-3-Cantor-DOS} illustrates the first three iterations of the rule for $M=1$.

Since the support of $\Delta_l(\eps)$ consists of $N_l = (2M+1)^l$ subbands, each of width $W_l = 2D/(4M+1)^l$, Eq.\ \eqref{eq:D-box} gives the fractal dimension of $\Delta_{C(4M+1)}$ as
\begin{equation}
\label{eq:D_C}
D_{C(4M+1)} = \lim_{l\to\infty} \frac{\log N_l}{\log(D/W_l)} = \frac{\log(2M+1)}{\log(4M+1)}.
\end{equation}
Another fractal characteristic of $\Delta_{C(4M+1)}(\eps)$ is self-similarity under energy rescaling about infinitely many different reference energies. For example, if $\eps_0$ lies at the center of a retained interval beginning with approximant $\Delta_l(\eps)$---and is thus also at the center of a retained interval for all higher-order approximants---then $\Delta(\eps_0+\Delta\eps)=\Delta(\eps_0+\Delta\eps/(4M+1))$ for $|\Delta\eps| < 3W_l/2$. As we shall see, self-similarity about the Fermi energy will be of particular consequence for the fractal Anderson impurity problem.

\begin{figure}[t!]
\centering
\includegraphics[width=0.48\textwidth]{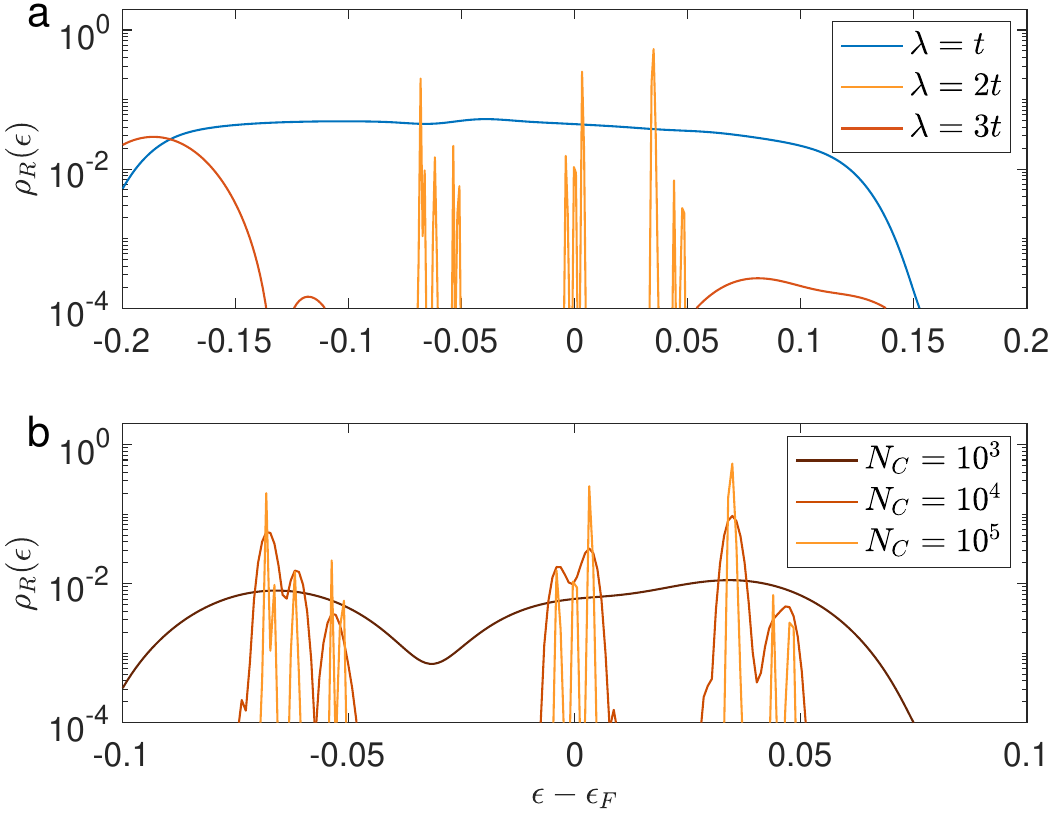}
\includegraphics[width=0.48\textwidth]{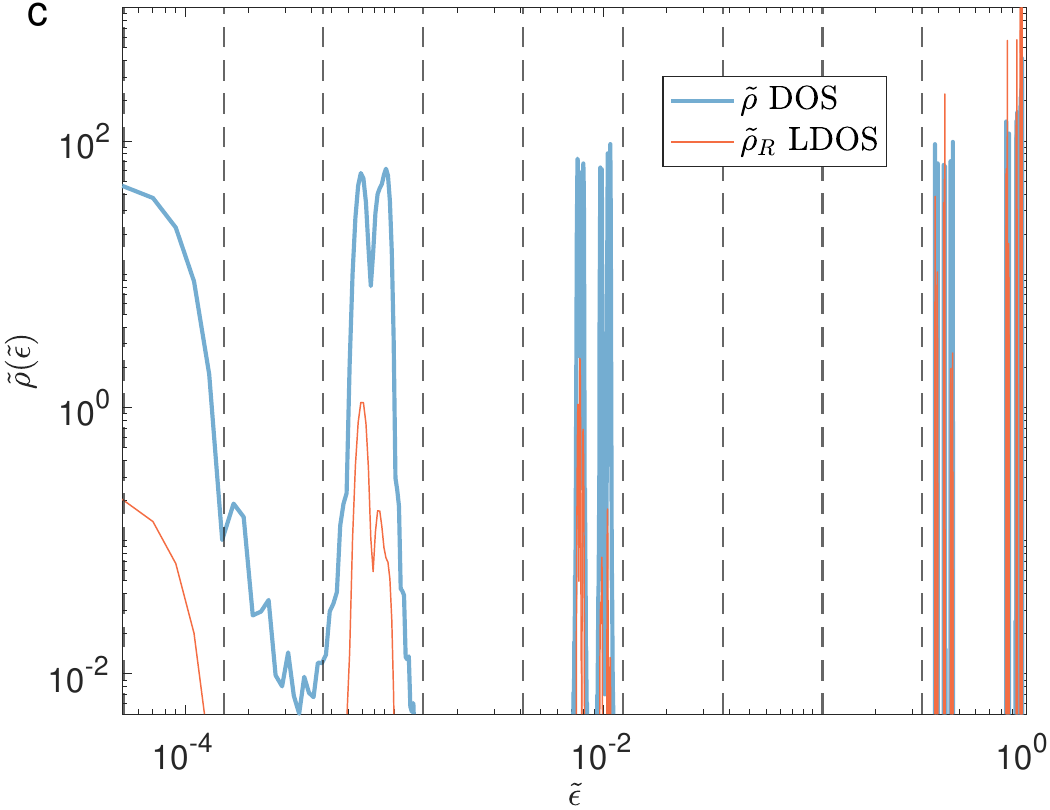}
\caption{\label{fig:AA-DOS+LDOS}%
DOS and LDOS at the middle site $R=L/2$ for a single realization $\phi=0$ of the \AubAnd\ model with lattice size $L = 10^6$, filling $n_c = 0.309$, and various values of the KPM expansion parameter $N_C$.
(a) LDOS for $N_C=10^5$ in the delocalized ($\lambda=t$) and localized ($\lambda=3t$) phases and at the critical point ($\lambda=\lambda_c=2t$). (b) Critical LDOS for different $N_C$ values. The self-similarity of the LDOS emerges with increasing $N_C$. (c) Log-log plots of the critical DOS and LDOS over positive values of $\teps$, with tildes denoting reduced quantities as defined in Sec.\ \ref{subsec:AAA-model}. The dashed lines mark the boundaries of NRG energy bins at $\teps=\pm\Lambda^{-m}$ ($m=1, \, 2, \, 3, \ldots$) for discretization parameter $\Lambda=3$. Plots for $\teps<0$ (not shown) show very similar behavior.}
\end{figure}

Appendix \ref{app:self-similar} briefly treats a related class of hybridization functions $\Delta_{C(4M+3)}(\eps)$ for positive integer $M$ that can be constructed by a variant of the above finite subdivision rule in which each nonzero energy range of $\Delta_{l-1}(\eps)>0$ is divided into $4M+3$ equal-width windows, and one sets $\Delta_l(\eps)=0$ throughout each \textit{odd}-numbered interval. Such a hybridization function has fractal dimension
\begin{equation}
\label{eq:D_Cbar}
D_{C(4M+3)} = \frac{\log(2M+1)}{\log(4M+3)}
\end{equation}
and is self-similar under rescalings $\eps \to \eps_0 + (\eps-\eps_0)/(4M+3)$ about $\eps_0 = 0$ and a countable infinity of other points.
Also considered in Appendix \ref{app:self-similar} is a hybridization function
\begin{equation}
\label{eq:Delta-log-bins}
   \Delta_{S(b)}(\eps) =
   \begin{cases} \displaystyle
       \frac{\pi V^2}{2}( 1\! + \! b^{-1/2} ) & b^{-(m+1/2)} < |\eps/D| \le b^{-m} \!, \\[1.5ex]
       0 & \text{otherwise} ,
   \end{cases}
\end{equation}
for $b>1$ and $m = 0,\, 1, \, 2, \, \ldots$. This function is not fractal: it has a box-counting dimension equal to its topological dimension of $1$ and exhibits self-similarity under rescalings $\eps \to \eps_0 + (\eps-\eps_0)/b$ about a single reference energy $\eps_0=0$. Comparison between properties in the strong-coupling (Kondo) limit of the Anderson model with hybridization functions $\Delta_{C(4M+1)}$, $\Delta_{C(4M+3)}$, and $\Delta_{S(b)}$ allows us to separate signatures of fractality from ones that arise merely from self-similarity about the Fermi energy.

The Cantor-set hybridization function has the advantage of lending itself rather naturally to treatment using the NRG method, allowing nonperturbative solution of the corresponding Anderson impurity model down to asymptotically low temperatures. We do not expect such hybridization functions to occur in physical settings where the impurity is a spatially local degree of freedom. However, there are experimentally relevant settings in which the ``impurity spin'' is a spatially nonlocal object, such as a strongly anharmonic eigenmode of an optical resonator \cite{PhysRevLett.107.063601}. If we consider a degenerate Fermi gas coupled to a nonlocal two-level system of this type, it is plausible that the hybridization function will be roughly proportional to the total density of states. We leave a more detailed discussion of this potential experimental realization to future work.

A second route to obtaining a fractal $\Delta(\eps)$ is for a magnetic impurity to hybridize with the global DOS of a critical quasicrystalline host, rather than the LDOS that also contains information about site-specific wave functions. Section \ref{sec:fractal-AA} addresses the Anderson model resulting from coupling an impurity to the global DOS of the AA model at its critical point $\lambda = \lambda_c = 2t$. Figure \ref{fig:AA-DOS+LDOS} illustrates the DOS and one particular LDOS for this critical host. The DOS exhibits a fractal structure \cite{wu2021fractal} that can be discerned in the self-similar arrangement of peaks that all have the same height, similar to those that emerge from the uniform Cantor set construction. The LDOS shares the self-similar energy structure of the DOS, but the peaks have different heights from one another, reflecting the inhomogeneity of the eigenfunctions. Coupling to this LDOS yields the multifractal Anderson impurity problem studied in Sec.\ \ref{sec:full-AAA}.

\section{Numerical Approach}
\label{sec:numerics}

\subsection{Methods}
\label{subsec:methods}

To solve each impurity model of interest we use the NRG approach \cite{RevModPhys.47.773,RevModPhys.80.395}. The reduced electronic band energy range $-1 < \teps < 1$ is divided into bins $\teps_{m+1} < \pm\teps < \teps_m$, where
\begin{equation}
\label{eq:teps_m}
  \teps_0 = 1, \qquad
  \teps_m = \Lambda^{1-z-m} \quad \text{for } m = 1, \, 2, \, \ldots.
\end{equation}
Here, $\Lambda>1$ is a dimensionless discretization parameter and $z>0$ is an offset parameter that can be averaged over to remove certain artifacts of the energy binning \cite{RevModPhys.80.395, PhysRevB.57.14254,PhysRevB.41.9403,PhysRevB.49.11986}; throughout this paper, $z=1$ unless explicitly stated otherwise. The continuum of band states within each bin is replaced by a single state: the particular linear combination of bin states that couples to the impurity degrees of freedom. Following this logarithmic discretization step, the Lanczos procedure \cite{lanczos1950iteration} is used to map the reduced Anderson Hamiltonian to the limit $N\to\infty$ of
\begin{gather}
\label{eq:H_N}
    \tH_N = \tH_\imp + (V/D) \sum_\sigma(d_\sigma^\dag f_{0\sigma} + f_{0\sigma}^\dag d_\sigma) + \sum_{\sigma} \tH_{0,N,\sigma} \\
\label{eq:H_N,sigma}
    \begin{split}
    \tH_{n_0,N,\sigma} =
    & \sum_{n=n_0}^N \veps_n f_{n\sigma}^\dag f_{n\sigma} \\
    & + \sum_{n=n_0}^{N-1} t_n (f_{n\sigma}^\dag f_{n+1,\sigma} + f_{n+1,\sigma}^\dag f_{n\sigma})\bigr] ,
    \end{split}
\end{gather}
describing a nearest-neighbor tight-binding chain (the ``Wilson chain'') with sites $n_0=0, \, 1, \, 2, \ldots, N$ coupled to the impurity only at its end site $0$. The annihilation operator $f_{0\sigma}$ is identical to $c_{R\sigma}$ entering Eq.\ \eqref{eq:H_hyb}. (Our notation departs slightly from, but is entirely equivalent to, that of Refs.\ \onlinecite{bulla1997anderson}, \onlinecite{PhysRevB.57.14254} and \onlinecite{RevModPhys.80.395}.)

As reviewed in Appendix \ref{subsec:chain-mapping}, the NRG tight-binding parameters $\veps_n$ and $t_n$ are defined entirely in terms of zeroth and first moments of the hybridization function over each energy bin:
\begin{equation}
\label{eq:alpha,beta}
    \alpha_m^{\pm} = \pm \! \int_{\pm\teps_{m+1}}^{\pm\teps_{m}} \!\! \tDelta(\teps) \, d\teps , \quad
    \beta_m^{\pm} = \pm \! \int_{\pm\teps_{m+1}}^{\pm\teps_{m}} \!\! \teps \, \tDelta(\teps) \, d\teps.
\end{equation}
Due to the separation of energy scales introduced by the discretization parameter $\Lambda>1$, the hopping coefficients decay exponentially with increasing $n$ as $t_n\propto \bLam^{-n/2}$. It has previously been found that $\bLam=\Lambda$ if $\tDelta(\teps)$ is nonvanishing as $\teps$ approaches zero from both sides, whereas $\bLam=\Lambda^2$ if $\tDelta(\teps)=0$ on one side of $\eps=0$ (as is the case at the top or bottom of an electronic band, or in the treatment of a dispersive bosonic bath) \cite{PhysRevLett.91.170601}. We will find that other values of $\bLam$ can be realized for a fractal hybridization function. Whatever the specific value of $\bLam$, it is useful to define a scaled hopping coefficient
\begin{equation}
\label{eq:xi_n}
    \xi_n = \bLam^{n/2} \, t_n .
\end{equation}

If $\tDelta(\teps)=\tDelta(-\teps)$ for every $\teps$, then (a) $\alpha_m^\pm = \alpha_m$ and $\beta_m^\pm = \pm\beta_m$ for every $m$, and (b) $\veps_n=0$ for all $n$. Absent this strict particle-hole symmetry, $\veps_n$ also decays at least as fast as $\bLam^{-n/2}$.

The exponential decay of tight-binding parameters along the Wilson chain allows a systematic, iterative solution of a series of finite-chain problems $\tH_N, \, N=0, \, 1, \, 2, \ldots$. Constraints of computer memory and processing time require that only a subset of many-body the eigenstates of $\tH_N$---typically, the $N_s$ states $\ket{E_{N,r}}$ of lowest energy $E_{N,r}$---be retained to construct the basis for $\tH_{N+1}$. Although it is impractical to extend calculations to the continuum limit $\Lambda\to 1$ and $N_s\to\infty$, solutions of $\tH_N$ turn out to provide a good account of thermodynamic properties at reduced temperatures $T/D$ of order $\bLam^{-N/2}$. Advantage can be taken of conserved quantum numbers---such as the total charge (electron number measured from half filling) and the total spin $z$ component---to reduce the Hamiltonian matrix into block diagonal form and thereby reduce the computational burden of finding the eigensolution.

For uniform Cantor-set hybridization functions, the integrals in Eqs.\ \eqref{eq:alpha,beta} can be computed for approximants $\tDelta_l(\teps)$ of increasing $l$. Within a fairly small number of iterations, one reaches converged values for the tight-binding coefficients for low-numbered Wilson-chain sites and can infer the $l\to\infty$ asymptotes of $t_n$ and $\veps_n$ for larger $n$.
 
For an AA host, especially at criticality, finding the energy eigenstates of a sufficiently large system, then constructing the hybridization function and obtaining its moments over logarithmic bins in order to compute the tight-binding parameters of the NRG Wilson chain, becomes a computationally prohibitive task.
 An alternative to exact diagonalization of the host Hamiltonian is the KPM \cite{RevModPhys.78.275}, an efficient and stable numerical technique that can be used to represent the spectral density of large matrices as an expansion in Chebyshev polynomials. The representation requires the spectrum to be rescaled to lie within $[-1,1]$, which can be accomplished as described at the end of Sec.\ \ref{subsec:AAA-model}. After this rescaling, the KPM representation of the reduced hybridization function is
\begin{equation}
\label{eq:Delta-expansion}
    \tDelta(\teps) = \frac{1}{\pi\sqrt{1-\teps^2}} \bigg( g_0 \mu_0 + 2 \sum_{n=1}^{N_C-1} g_n \mu_n T_n(\teps) \bigg),
\end{equation}
where $T_n(x)$ is the $n$th Chebyshev polynomial of the first kind and
\begin{gather}
    g_n = \frac{1}{N_C + 1} \biggl[ (N_C + 1 - n) \cos\frac{\pi n}{N_C + 1} \\
    + \sin\frac{\pi n}{N_C + 1} \cot\frac{\pi}{N_C + 1} \biggr] 
\end{gather}
is a coefficient of the Jackson kernel that is used to remove the Gibbs phenomenon created by truncating the series after $N_C$ terms  \cite{RevModPhys.78.275}. With this kernel, the KPM expansion has an energy resolution near $\teps = 0$ of
\begin{equation}
\label{eq:KPM-res}
    \delta\teps = \pi/N_C .
\end{equation}

When the hybridization function is computed using the LDOS at the impurity site $R$, the moments of the expansion that must be computed are
\begin{equation}
\label{eq:mu_n-R}
    \mu_n=\bra{R,\sigma}T_n(\tH_\host)\ket{R,\sigma},
\end{equation}
where $\ket{R,\sigma}$ (for $\sigma=\:\uparrow$ or $\downarrow$) is a single-particle host state at the impurity site, and for any single-particle state $\ket{\alpha}$, $\ket{\alpha_n} \equiv T_n(\tH_\host)\ket{\alpha}$ can be computed via a set of recursion relations
$\ket{\alpha_0} = \ket{\alpha}$,
$\ket{\alpha_1} = \tH_\host \ket{\alpha_0}$, and
$\ket{\alpha_n} = 2\tH_\host \ket{\alpha_{n-1}} - \ket{\alpha_{n-2}}$ for $n\ge 2$.

If the hybridization is instead calculated in terms of the global DOS, one replaces Eq.\ \eqref{eq:mu_n-R} by
\begin{equation}
\label{eq:mu_n-avg}
    \mu_n=\frac{1}{L} \, \text{Tr}[T_n(\tH_\host)],
\end{equation}
where $L$ is the number of host lattice sites and the trace can be approximated by stochastic evaluation with random vectors $N_r$ \cite{gull1989maximum,PhysRevLett.70.3631,silver1994densities}. Our computations used $N_r=100$ random vectors, leading to a relative error in $\mu_n$ of order $1/\sqrt{LN_r}$, and we take $L=10^6$ throughout, unless otherwise specified.

Appendix \ref{app:KPM+NRG} shows how Eq.\ \eqref{eq:Delta-expansion} can be analytically combined with the NRG to yield $\alpha_m^\pm$ and $\beta_m^\pm$ as weighted sums over terms in the KPM expansion. This circumvents any numerical integration of the hybridization function and provides a convergent and controlled evaluation of the Wilson-chain coefficients. The accuracy of this approach is tested in Appendix \ref{subsec:KPM+NRG-compare-coeffs} against analytic calculation of the NRG tight-binding parameters for one particularly tractable hybridization functions and in Appendix \ref{subsec:KPM+NRG-compare-DMRG} against density-matrix RG results for the AAA model.

\subsection{Observables}
\label{subsec:observables}

Our results focus on a pair of impurity thermodynamic properties, each expressed as the difference $X_\imp = X_\tot - X_\tot^{(0)}$ between $X_\tot$, the total value of a quantity $X$ in the coupled impurity-host system, and $X_\tot^{(0)}$, its counterpart for the same host in the absence of the impurity.
The first property of interest is the impurity spin susceptibility defined through \cite{Note1} $\chi_{\tot}(T) = \beta (\braket{S_{\tot,z}^2} - \braket{S_{\tot,z}}^2)$, where $\beta = 1/T$ and $S_{\tot,z}$ is the total spin-$z$ component: $S_{\tot,z} = S_{\imp,z} + \frac{1}{2}\sum_n (f_{n\up}^\dag f_{n\up} - f_{n\dn}^\dag f_{n\dn})$ with $S_{\imp,z}=\frac{1}{2} (n_{f\up} - n_{f\dn})$ being the $z$ component of the impurity spin operator defined after Eq.\ \eqref{eq:H_K} \footnote{In Sec.\ \protect\ref{subsec:observables}, we define $S_{\tot,z}$ and $S_{\imp,z}$ to be the $z$ components of spins, while $S_\tot$ and $S_\imp$ (without a subscript $z$) denote entropies. Appendix \protect\ref{subsec:KPM+NRG-compare-DMRG} makes use of the same $S_{\imp,z}$ as well as a local entanglement entropy $S_\text{loc}$. Other sections of the paper reference only the impurity entropy $S_\imp$ and the impurity spin vector $\textbf{S}_\imp$.}. We also consider the impurity entropy defined via $S_\tot = \beta \braket{H} + \ln Z$, where $H$ is the Hamiltonian and $Z=\text{Tr}\exp(-\beta H)$ is the grand canonical partition function for zero chemical potential [after the rescaling in Eq.\ \eqref{eq:eps-tilde}]. Although $X_\tot$ and $X_\tot^{(0)}$ are both expected to be non-negative, nothing prevents their difference $\chi_{\imp}$ from assuming negative values.

In the NRG treatment, $Z=\sum_{r} \exp(-\beta E_{N,r})$ and $X_\tot= Z^{-1} \sum_r \exp(-\beta E_{N,r}) \braket{ N,r | X | N,r}$ for the coupled host-impurity system are evaluated as traces over many-body eigenstates $|N,r\rangle$ having energies $E_{N,r}$.
The NRG spectrum at iteration $N$ is used to compute $X_\tot$ at temperatures $T_N(\bar{\beta})\simeq \bLam^{-(N-1)/2}/\bar{\beta}$ \cite{Note1} where $\bar{\beta}$ is of order 1 \cite{RevModPhys.47.773,RevModPhys.80.395}; the results presented in this paper were computed for $\bar{\beta}=0.9$ and $0.9\bLam^{-1/2}$.
The corresponding quantity $X_\tot^{(0)}$ can be calculated in terms of single-particle eigenvalues of the Wilson chain, as described in more detail in Sec. \ref{subsec:Cantor-SC}.

In conventional metallic hosts, the many-body screening of an Anderson impurity degree of freedom reveals itself in a monotonic reduction of the impurity entropy from a value $S_\imp\simeq \ln 2$ at intermediate temperatures (where the impurity occupancies $n_d = 0$ and $2$ initially become frozen out) toward $\lim_{T\to 0} S_\imp=0$. There is a parallel, monotonic reduction of $T\chi_\imp$ (which can be interpreted as being proportional to the square of the effective impurity moment) from $1/4$ toward $0$. In such canonical settings---and in the limit of temperature $T$ and non-thermal parameters such as frequency $\omega$ and magnetic field $B$ that are all small compared with the half-bandwidth $D$---each physical property is solely a function of $T/T_K$, $\omega/T_K$, $B/T_K$, etc. Here, the Kondo temperature $T_K$ serves as the sole energy scale describing the approach of impurity properties toward their values in the Kondo strong-coupling ground state. Moreover $T_K$ can be defined as the temperature at which a chosen property crosses through a threshold value en route from local-moment to strong-coupling behavior, with one common convention being \cite{RevModPhys.47.773}
\begin{equation}
\label{eq:T_K-std}
    T_K\chi_\imp(T_K) = 0.0701.
\end{equation}

For the present work, we find it preferable to adopt in place of Eq.\ \eqref{eq:T_K-std} the alternative definition
\begin{equation}
\label{eq:T_K-loc}
    T_K\chi_\loc(T_K) = 0.0701,
\end{equation}
where
\begin{equation}
    \chi_\loc(T)
    = -\frac{\partial \braket{S_{\imp,z}(T,h)}}{\partial h}\biggr|_{h=0}
    =\lim_{h\to 0} -\frac{\braket{S_{\imp,z}(T,h)}}{h} ,
\end{equation}
is the static local spin susceptibility describing the response to the local magnetic field $h$ entering Eq.\ \eqref{eq:H_imp}. In the Kondo regime of conventional metallic hosts \cite{PhysRevB.45.5368}, 
\begin{equation}
    \chi_\imp(T) = [1 + \rho(\eps_F) J_K + \ldots] \, \chi_\loc(T),
\end{equation}
making Eqs.\ \eqref{eq:T_K-std} and \eqref{eq:T_K-loc} essentially equivalent. However, in hosts where the hybridization function vanishes \cite{PhysRevB.52.14436,PhysRevB.57.14254} or diverges \cite{PhysRevB.88.195119} continuously on approach to the Fermi energy, it is the approach of $T\chi_\loc$ to zero from above that signals Kondo screening of the impurity local moment, while $T\chi_\imp$ can exhibit non-monotonic temperature variation and/or approach a non-vanishing $T=0$ limit. As will be seen in Secs.\ \ref{sec:Cantor}--\ref{sec:full-AAA}, impurities in fractal and multifractal hosts exhibit rather similar behaviors, leading us to define the Kondo temperature through the local susceptibility.

\section{Uniform Cantor Set Spectra}
\label{sec:Cantor}

This section presents results for the Anderson impurity model with a uniform Cantor set $\Delta(\eps)$. A hybridization function of this type is made up of an uncountably infinite number of points, contains no interval of nonzero length, and has zero measure over its entire range $|\eps|\le D$. The relative simplicity of the finite subdivision rules for creating Cantor sets allows an NRG treatment of the Anderson impurity model down to asymptotically low temperatures. The considered hybridization functions satisfy $\Delta(\eps) = \Delta(-\eps)$. Except where explicitly stated to the contrary, we assume that the Fermi energy is located at $\eps_F=0$ so the reduced hybridization function obeys $\tDelta(\teps) = \tDelta(-\teps)$.

Section \ref{subsec:fractal-host} specifies a finite subdivision rule for creating the level-$l$ approximant $\Delta_l(\eps)$ to $\Delta_{C(4M+1)}(\eps)$ with $M$ a positive integer. Since $\Delta_l(\eps)>0$ for all $|\eps|< D / (4M+1)^l$, a host described by this hybridization function behaves like a conventional metal on temperature and energy scales much smaller than $D/ (4M+1)^l$. For energies $\eps$ such that $D / (4M+1)^l \ll |\eps| \ll D$, by contrast, $\Delta_l(\eps)$ has a hierarchy of gaps of widths ranging from $2 D / (4M+1)^l$ to $2D / (4M+1)$. In the limit $l\to\infty$, this gap structure extends all the way down to $\eps=0$.

We show in this section that on a coarse-grained level defined by a specific choice of NRG discretization parameter, namely $\Lambda = 4M+1$, $\Delta_{C(4M+1)}(\eps)$ is equivalent to a continuous hybridization function that diverges on approach to $\eps = 0$ according to a power law that reflects the fractal dimension of the $1/(4M+1)$ Cantor set. However, when $\Lambda$ is reduced toward 1 to explore the continuum (nondiscretized) limit of the Anderson impurity model, one finds---as detailed in Appendix \ref{subsec:5-3-chain-mapping}---that the hierarchical gap structure of $\Delta_{C(4M+1)}(\eps)$ creates additional structure in the $n$ dependence of the Wilson-chain coefficients $t_n$ and $\veps_n$ entering Eq.\ \eqref{eq:H_N,sigma}. By calculating the single-particle eigenvalues of the Wilson-chain Hamiltonian, we identify a \textit{fractal strong-coupling} limit of the Anderson/Kondo model with a Cantor-set hybridization function. This regime exhibits thermodynamic signatures that distinguish it from those obtained for a divergent continuous $\Delta(\eps)$. The section ends with full NRG many-body results showing how thermodynamic properties evolve with decreasing temperature toward the fractal strong-coupling limit. The focus throughout will be on the uniform $1/5$ Cantor set, with brief mention of results for $C(4M+1)$ with $M>1$ and two other families of self-similar hybridization functions discussed in Appendix \ref{app:self-similar}.

\subsection{Wilson-chain description of Cantor-set hybridization functions}
\label{subsec:Cantor-chains}

The tight-binding coefficients $\veps_n$ and $t_n$ entering Eq.\ \eqref{eq:H_N,sigma}, the Wilson-chain description of $H_\host$, are fully determined by the set of moments $\alpha_m^\pm$ and $\beta_m^\pm$ defined in Eqs.\ \eqref{eq:alpha,beta}. Since $\Delta_l(\eps)=\Delta_l(-\eps)$ for every approximant to $\Delta_{C(4M+1)}(\eps)$, we need only compute $\alpha_m=\alpha_m^\pm$ and $\beta_m=\pm\beta_m^\pm$, with symmetry dictating that $\veps_n = 0$ for all $n$.

The NRG mapping of a hybdridization $\Delta(\eps)$ can be performed using any value $\Lambda>1$ of the Wilson discretization parameter. However, the self-similarity of $\Delta_{C(4M+1)}$ most clearly reveals itself by considering
\begin{equation}
\label{eq:Lambda-seq}
    \Lambda_k = (4M+1)^{1/2^k} \quad \text{for } k = 0, \, 1, \, 2, \, \ldots
\end{equation}
In practice, NRG calculations will be performed for small values of $k$, but allowing for $k\to\infty$ provides a route for approaching the continuum limit $\Lambda = 1$.

\subsubsection{$\Lambda=4M+1$ Wilson chain: Equivalence to a power-law divergent hybridization function}
\label{subsubsec:Cantor-Lambda=4M+1}

For an offset parameter $z=1$ entering Eq.\ \eqref{eq:teps_m}, the choice $\Lambda=4M+1$ places the NRG bin boundaries $\pm\teps_m$ at the upper/lower edges of the central nonvanishing range of $\tDelta_m(\teps)$, as illustrated in Fig.\ \ref{fig:5-3-Cantor-DOS} for $M=1$ and $m=1, \, 2$. A consequence of this alignment is that $\alpha_m$ and $\beta_m$ cease to change with increasing $l$ once $l>m$. Due to the self-similarity of $\tDelta_{C(4M+1)}(\teps)$ under $\teps \to \teps/(4M+1)$, it is straightforward to see that for $l\to\infty$ and for $m\ge 0$,
\begin{equation}
\label{eq:alpha,beta-Lambda=4M+1}
\begin{aligned}
    \alpha_m[C(4M+1)]&
      = \frac{M}{(2M+1)^{m+1}} , \\
    \beta_m[C(4M+1)]&
      = \frac{2M(M+1)}{[(2M+1)(4M+1)]^{m+1}}.
\end{aligned}
\end{equation}

It is instructive to compare Eqs.\ \eqref{eq:alpha,beta-Lambda=4M+1} with the corresponding moments for a continuous, power-law-divergent hybridization function $\Delta_{P(r)}$ that has the reduced form
\begin{equation}
\label{eq:Delta-power}
    \tDelta_{P(r)}(\teps) = \frac{1}{2}(1+r)|\teps|^r
\end{equation}
with $-1 < r < 0$ \cite{PhysRevB.88.195119}:
\begin{equation}
\label{eq:alpha,beta-power-law}
\begin{aligned}
    \alpha_m[P(r)]& = \frac{\Lambda^{1+r}-1}{2 \Lambda^{(m+1)(1+r)}}, \\
    \beta_m[P(r)]& = \frac{(1+r)(\Lambda^{2+r}-1)}{2 (2+r) \Lambda^{(m+1)(2+r)}}.
\end{aligned}
\end{equation}
For $\Lambda=4M+1$, $\alpha_m[P(r)]$ becomes identical to $\alpha_m[C(4M+1)]$ provided that
\begin{equation}
\label{eq:r-C}
    r=\frac{\log(2M+1)}{\log(4M+1)} - 1 = D_{C(4M+1)} - 1 ,
\end{equation}
where $D_{C(4M+1)}$ is the fractal dimension of the $1/(4M+1)$ Cantor set given in Eq.\ \eqref{eq:D_C}.
This choice also yields
\begin{equation}
\label{eq:a_4M+1}
\frac{\beta_m[C(4M+1)]}{\beta_m[P(r)]} = \frac{2(M+1)}{4M+3} \biggl[ 1 + \frac{\log(4M+1)}{\log(2M+1)} \biggr] \equiv a_{4M+1}.
\end{equation}
Examination of Eqs.\ \eqref{eq:recursion}--\eqref{eq:A} shows that the Wilson-chain representations of the two hybridization functions must satisfy $t_n[C(4M+1)]/t_n[P(r)] = a_{4M+1}$, an overall multiplicative factor that can be absorbed into rescaling of the half-bandwidth $D$ and the impurity parameters $U$, $\eps_d$, and $V$. In both cases, the hopping parameters satisfy
\begin{equation}
\label{eq:tstar}
    \lim_{n\to\infty} \Lambda^{n/2} t_n =
    \begin{cases}
    t^* & \text{for $n$ even}, \\    
    t^*\Lambda^{-r/2} & \text{for $n$ odd}.
    \end{cases}
\end{equation}
This is precisely the relation reported in Eq.\ (3.3) of Ref.\ \onlinecite{PhysRevB.57.14254} for positive values of $r$ describing a power-law vanishing of the hybridization function at the Fermi energy---a case to which Eqs.\ \eqref{eq:alpha,beta-power-law} also apply.

The preceding analysis of Wilson-chain coefficients leads to the conclusion that a $\Lambda = 4M+1$, $z=1$ NRG treatment of the $\Delta_{C(4M+1)}$ hybridization function will yield properties equivalent to a $\Lambda = 4M+1$, $z=1$ NRG treatment of a continuous hybridization function $\Delta_{P(D_{C(4M+1)}-1)}$. As a result, the integrals $\alpha_m^\pm$ and $\beta_m^\pm$ over bin $m$ [see Eqs. \eqref{eq:alpha,beta}] acquire a simple power-law dependence on the index $m$.
Appendix \ref{app:self-similar} shows that the same equivalence exists between the $\Lambda=4M+3$, $z=1$ NRG treatments of $\Delta_{C(4M+3)}$ and $\Delta_{P(D_{C(4M+3)}-1)}$, as well as between the $\Lambda=b$, $z=1$ NRG treatments of $\Delta_{S(b)}$ and $\Delta_{P(0)}$ (i.e., a flat-top hybridization function).

\subsubsection{Approaching the continuum limit $\Lambda = 1$}
\label{subsubsec:Cantor-Lambda-to-1}

The equivalence between the $\Lambda=4M+1$, $z=1$ NRG treatments of hybridization functions $\Delta_{C(4M+1)}$ and $\Delta_{P(D_{C(4M+1)}-1)}$ arises because this particular combination of $\Lambda$ and $z$ perfectly aligns the logarithmic energy bins with the self-similarity of the fractal hybridization function about the Fermi energy. Each bin boundary $\teps_m$ in Eq.\ \eqref{eq:teps_m} coincides exactly with the top of a subband (see Fig.\ \ref{fig:5-3-Cantor-DOS}). 
Alignment of the NRG bin boundaries with subband edges is disrupted by a change in $\Lambda$ and/or $z$. Thus, we expect such a change to cause the NRG description of $\Delta_{C(4M+1)}$ to deviate from that of $\Delta_{P(D_{C(4M+1)}-1)}$.

Appendix \ref{app:self-similar} discusses the evolution of the Wilson-chain hopping coefficients $t_n$ for the uniform $1/5$ Cantor set as one progresses through the sequence of discretizations specified in Eq.\ \eqref{eq:Lambda-seq}. The appendix also summarizes observations concerning the $t_n$ coefficients for two other families of hybridization functions. This analysis leads to the following conclusions concerning the NRG discretization of any hybridization function that (a) is particle-hole symmetric, i.e., $\tDelta(\teps) = \tDelta(-\teps)$, and (b) satisfies the discrete self-similarity property $\tDelta(\teps) = \tDelta(\teps/b) = \tDelta(\teps/b^2) = \ldots$ for all $|\teps|$ below some upper cutoff and for $b$ taking some smallest value greater than 1 (to exclude a constant hybridization):

\noindent
(1) If $\Lambda=b^{1/p}$ with $p$ being a positive integer, then $\tDelta(\teps)$ is nonzero for at least some energies within $q>0$ of the $p$ NRG energy bins that cover each energy range $b^{-z-m'} < \teps < b^{1-z-m'}$, with $q$ taking the same value for all positive integers $m'$. The scaled hopping coefficients $\xi_n$ defined in Eq.\ \eqref{eq:xi_n} with $\bLam=b^{1/q}\equiv \Lambda^{p/q}$ satisfy $\lim_{n\to\infty}\xi_{n+2q} = \xi_n$, or equivalently
\begin{equation}
\label{eq:t_n-exact}
    \lim_{n\to\infty} t_{n+2q} / t_n = 1 / b.
\end{equation}

\noindent
(2) For generic values of $\Lambda$ that are not roots of $b$, the scaled hopping coefficients $\xi_n$ do not exhibit exact periodicity. We conjecture that there exists a $\bLam = b^{\,q_1/q_2}$, where $q_1$ and $q_2$ are positive integers, such that the scaled hopping coefficients $\xi_n$ defined in Eq.\ \eqref{eq:xi_n} remain within a bounded range, neither diverging nor vanishing as $n\to\infty$.

One can regard $2q$ as a measure of the complexity of the hybridization function: the number of hopping coefficients required to faithfully describe $\tDelta(\teps)$ over a factor of $b$ change in energy when coarse-graining with a discretization parameter $\Lambda=b^{1/p}$. As $\Lambda\to 1^+$ (i.e., $p\to\infty$), one expects $2q$ to diverge, reflecting the increasing structure of the Cantor-set hybridization function when viewed with an ever-finer energy resolution $\Delta(\log\eps) = \log\Lambda$. In this way, the fractal nature of the hybridization function is encoded in the Wilson chain and thereby makes its way into physical observables. By contrast, the Wilson-chain hopping coefficients for a power-law hybridization function obey Eq.\ \eqref{eq:tstar}, or equivalently, $\lim_{n\to\infty}t_{n+2}/t_n=1/\Lambda$, where the complexity remains constant at $2q=2$ but the right-hand side approaches $1$ in the continuum limit due to the absence of any intrinsic self-similarity scale.

The remainder of Sec. \ref{sec:Cantor} explores manifestations of self-similarity and fractality in thermodynamic properties. We begin in Sec.\ \ref{subsec:Cantor-SC} by analyzing the low-temperature limit, while higher-temperature crossover phenomena will be the focus of Sec. \ref{subsec:Cantor-NRG}.

\subsection{Strong-coupling limit}
\label{subsec:Cantor-SC}

\begin{table*}[t!]
\centering
\setlength{\tabcolsep}{6pt}
\renewcommand{\arraystretch}{1.2}
\begin{tabular}{l|ccD{.}{.}{4}D{.}{.}{5}D{.}{.}{3}@{\hspace{4ex}}|D{.}{.}{7}D{.}{.}{6}D{.}{.}{5}D{.}{.}{5}D{.}{.}{4}}
    & \multicolumn{5}{c|}{Power law $r = D_{C(5)}-1$} & \multicolumn{5}{c}{Cantor set $C(5)$} \\[.5ex]
    \cline{2-6} \cline {7-11}
    $\Lambda$ & \multicolumn{1}{c}{\rule[-1.2ex]{0pt}{4ex} $A_{T\chi}$} & \multicolumn{1}{c}{$A_S$} & \multicolumn{1}{c}{$\phi_{T\chi}(1)$} & \multicolumn{1}{c}{$\phi_{T\chi}(0.5)$} &
    \multicolumn{1}{c|}{$\Delta\phi$} &
    \multicolumn{1}{c}{$A_{T\chi}$} & \multicolumn{1}{c}{$A_S$} & \multicolumn{1}{c}{$\phi_{T\chi}(1)$} & \multicolumn{1}{c}{$\phi_{T\chi}(0.5)$} & \multicolumn{1}{c}{$\Delta\phi$} \\
    \hline
    \rule{0pt}{3ex} 5 & 0.012002 & 0.12786 & 1.6338 & 4.7754 & 0.7104 & 0.012002 & 0.12786 & 0.2965 & 0.2965 & 0.7104 \\
    $5^{1/2}$ & $9.87\times 10^{-5}$ & $5.87\times 10^{-4}$ & 0.3122 & 3.4538 & 1.9724 & 0.012002 & 0.12786 & 0.2965 & 0.3548 & 0.7104 \\
    $5^{1/4}$ & $2.09\times 10^{-9}$ & $1.24\times 10^{-8}$ & 2.935 & 6.077 & 0.852 & 0.010227 & 0.10894 & 0.3548 & 0.4017 & 0.7104 \\
    $5^{1/8}$ & & & & & & 0.009741 & 0.10377 & 0.4078 & 0.4252 & 0.7104 \\
    $5^{1/16}$ & & & & & & 0.009540 & 0.10163 & 0.4268 & 0.4283 & 0.7104
\end{tabular}
\caption{\label{tab:SC-oscillations}%
Comparison between oscillatory components of strong-coupling impurity thermodynamic properties [Eqs. \eqref{eq:X-Cantor-SC}--\eqref{eq:f_X}] for an $r = D_{C(5)}-1 \simeq -0.3174$ power-law hybridization function and a uniform $1/5$ Cantor set [or $C(5)$] hybridization function, both at half filling (i.e., for Fermi energy $\eps_F=0$): Variation with NRG discretization $\Lambda$ of the amplitudes $A_X$ (for NRG offset $z = 1$) and phases $\phi_X(z)$ (for $z = 1, \, 0.5$) entering Eq.\ \eqref{eq:f_X} for the magnetic susceptibility ($X=T\chi$) and the entropy ($X=S$). The oscillations have base $b=\Lambda$ for the power law and $b=5$ for the Cantor set. The phase difference $\Delta\phi(z) = \phi_S(z) - \phi_{T\chi}(z)$ takes the same value for $z=1$ and $0.5$ to within the uncertainty of estimates (which is $\pm 1$ or better in the last digit).}
\end{table*}

The strong-coupling limit of the Anderson impurity model is reached when $V\to\infty$ for finite values of $U$ and $\eps_d$. In a metallic host, the strong-coupling RG fixed point describes the asymptotic low-temperature physics for any nonzero bare value of the hybridization $V$ \cite{PhysRevB.21.1003,PhysRevB.21.1044}. In a gapped host \cite{PhysRevB.57.5225} or a semimetal \cite{PhysRevLett.64.1835,bulla1997anderson,PhysRevB.57.14254,PhysRevB.70.214427,PhysRevLett.107.076404,PhysRevB.88.245111,PhysRevLett.89.076403}, strong coupling is reached only if the bare value of $V$ exceeds a critical value; otherwise, the zero-temperature limit is described by a free-local-moment RG fixed point at which the impurity retains an unquenched spin-1/2 degree of freedom. The central goal of the present work is to understand the fate of an impurity spin in a fractal or multifractal host.

We begin by focusing on situations exhibiting strict particle-hole symmetry, where $U = -2\eps_d$ and $\tDelta(\teps) = \tDelta(-\teps)$. At strong coupling, the degrees of freedom in the impurity level and on site 0 of the Wilson chain become frozen out through some superposition of spin singlet formation (i) between two electrons in the impurity level with site $0$ unoccupied, (ii) between two electrons on site $0$ with an empty impurity level, and (iii) between one electron each in the impurity and on chain site $0$ \cite{PhysRevB.21.1003}. The remainder of the Wilson chain is effectively free, so the reduced NRG Hamiltonian describing the strong-coupling limit is
$\tH_\text{N}^\text{(SC)} = \sum_{\sigma} \tH_{1,N,\sigma}$ with $\tH_{1,N,\sigma}$ defined in Eq.\ \eqref{eq:H_N,sigma}.
Since $\tH_N^\text{(SC)}$ is quadratic, it is numerically straightforward (at least for $N$ up to a few hundred) to find its single-particle eigenvalues $\eta_n^{(1,N)}$, $n = 1, \, 2, \, \ldots, N$. The host by itself is described by another quadratic NRG Hamiltonian $\tH_N^\text{(0)} = \sum_{\sigma} \tH_{0,N,\sigma}$ with single-particle eigenvalues $\eta_n^{(0,N)}$, $n = 0, \, 1, \, 2, \, \ldots, N$. One can therefore compute the strong-coupling impurity contribution to a thermodynamic property $X$ for temperatures $T\simeq D\bLam^{-N/2}$ as
\begin{align}
\label{eq:X_imp,SC}
    X_\imp^\text{(SC)}(T)
    & \equiv X_\tot^\text{(SC)}(T) - X_\tot^{(0)}(T) \notag\\[0.5ex]
    & = X(1,N,1/T) - X(0,N,1/T),
\end{align}
with the magnetic susceptibility of a Wilson chain consisting of sites $n_0$ through $N$ being given by
\begin{equation}
    T\chi(n_0,N,\beta) = \frac{1}{8} \sum_{n=n_0}^N \text{sech}^2 \Bigl( \beta D \eta_n^{(n_0,N)} / 2 \Bigr)
\end{equation}
and the corresponding entropy by
\begin{multline}
    S(n_0,N,\beta) = 2 \sum_{n=n_0}^N \biggl\{ \ln \Bigl[ 1 + \exp \Bigl(-\beta D \eta_n^{(n_0,N)} \Bigr) \Bigr] \\
    + \beta D \eta_n^{(n_0,N)} \Bigl[ \exp \Bigl( \beta D \eta_n^{(n_0,N)} \Bigr) + 1 \Bigr]^{-1} \biggr\} .
\end{multline}

We have evaluated these strong-coupling properties for the first five members of the sequence $\Lambda=5^{1/2^k}$ in the NRG treatment of the $\tDelta_{C(5)}(\teps)$ hybridization function as well as the continuous, divergent $\tDelta_{P(D_{C(5)}-1)}(\teps)$. For $\Lambda=5$ and $z=1$, as discussed in Sec.\ \ref{subsubsec:Cantor-Lambda=4M+1}, the Wilson chains describing $\tDelta_{C(5)}$ and $\tDelta_{P(D_{C(5)}-1)}$ are related by $t_n[C(5)]=a_5 t_n[P(r)]$, where $a_5\simeq 1.409$ is defined in Eq.\ \eqref{eq:a_4M+1}. Since the Wilson chain encodes all relevant information about the host, for the $\Lambda=5$, $z=1$ discretization the strong-coupling thermodynamic properties for the uniform $1/5$ Cantor set at temperature $T$ must be identical to those of the $r = D_{C(5)}-1$ power-law problem at temperature $a_5 T$. In both cases, the properties have an oscillatory temperature dependence
\begin{align}
\label{eq:X-Cantor-SC}
    X_\imp(T) & = X_\imp^\text{(SC,$r$)} + f_X(\Lambda, T),
\end{align}
where
\begin{equation}
\label{eq:Tchi,S-SC}
    T\chi_\imp^\text{(SC,$r$)} = r/8, \qquad S_\imp^\text{(SC,$r$)} = 2r\ln 2 
\end{equation}
are the continuum-limit strong-coupling values for the power-law hybridization function \cite{PhysRevB.88.195119}, while
\begin{equation}
\label{eq:f_X}
   f_X(T) \simeq A_X \sin[2\pi\log_b(T/D) + \phi_X] ,
\end{equation}
with $b$ to be defined shortly. For $r<0$, Eq.\ \eqref{eq:Tchi,S-SC} yields a negative impurity entropy. The occurrence of $S_\imp(T) < 0$ violates no fundamental thermodynamic principle; it just indicates that at temperature $T$, the total entropy of the coupled host-impurity system is less positive than the total entropy of the host by itself.

For hybridization functions that are featureless near the Fermi energy, $\log T$ oscillations are known \eqref{eq:teps_m} to be artifacts of the NRG discretization \cite{PhysRevB.49.11986} that have (a) base $b=\Lambda$, (b) an amplitude $A_X\propto\exp(-\pi^2/\Lambda)$, and (c) a phase $\phi_X(z) = \phi_X(0) + 2\pi z$ that allows removal of the oscillations by averaging over the offset parameter $z$ entering Eq.\ \eqref{eq:teps_m}. Similar characteristics hold for power-law hybridization functions from the class defined in Eq.\ \eqref{eq:Delta-power}. Table \ref{tab:SC-oscillations} lists parameters of the oscillatory term in the magnetic susceptibility and the entropy for the $r = D_{C(5)}-1 \simeq -0.3174$ power-law. The amplitudes $A_{T\chi}$ and $A_S$ entering Eq.\ \eqref{eq:f_X} fall off rapidly as $\Lambda$ is reduced, with the oscillations becoming almost undetectable for $\Lambda \le 5^{1/4}$. The table also shows that the phase $\phi_{T\chi}$ differs by $\pi$ for offset parameters $z = 1$ and $z = 0.5$, allowing the oscillations to be largely removed, even for $\Lambda = 5$, by averaging each property over just these two $z$ values.

Table \ref{tab:SC-oscillations} demonstrates that the thermodynamics for $\tDelta_{C(5)}(\teps)$ evolve very differently along the sequence $\Lambda_k$ defined in Eq.\ \eqref{eq:Lambda-seq}. With increasing $k$, (a) the oscillation period remains pinned at base $b = 5$, (b) the amplitudes $A_{T\chi}$ and $A_S$ appear to approach nonzero limiting values, and (c) the phases $\phi_{T\chi}$ and $\phi_S$ approach the same values for $z=1$ and $0.5$, precluding elimination of the oscillations by averaging over $z$. (The equivalence between the $\Lambda=5$ Wilson chains for $\tDelta_{C(5)}$ and $\tDelta_{P(D_{C(5)-1})}$ holds only for $z$ equal to an integer. For any non-integer value of $z$, the two hybridization functions have very different Wilson-chain coefficients.) Even though there is some change of $\phi_{T\chi}$ with $\Lambda$ and $z$, $\phi_S-\phi_{T\chi}$ varies very little. These observations indicate that the $\log T$ oscillations are not merely artifacts of the NRG technique, but intrinsic features of the fractal strong-coupling fixed point that survive in the continuum limit $\Lambda\to 1$.

The uniform Cantor-set hybridization functions $\tDelta_{C(4M+3)}(\teps)$ discussed in Appendix \ref{subsec:Cantor-4M+3} are self-similar under an energy rescaling $\teps\to\teps/(4M+3)$. We have verified that the case $M=1$ leads to sinusoidal oscillations of strong-coupling impurity thermodynamic properties as functions of $\log_b T$ with base $4M+3=7$ about average values corresponding to a power-law hybridization function with $r = D_{C(7)} - 1 \simeq -0.4354$. The oscillations appear to approach a nonzero amplitude in the continuum limit $\Lambda\to 1$. The amplitude of the $C(7)$ oscillations for $\Lambda=7^{1/16}$ is approximately twice the amplitude of the $C(5)$ oscillations for $\Lambda=5^{1/16}$.

We have also determined numerically that the nonfractal self-similar hybridization function $\tDelta_{S(b)}$ defined in Eq.\ \eqref{eq:Delta-log-bins} has strong-coupling impurity thermodynamic properties that oscillate as functions of $\log_b T$ about average values of zero. For $b=5$ and $7$ with $\Lambda = b^{1/16}$, the $S(b)$ oscillation amplitudes are approximately 90\% of those for $C(b)$. Strikingly, the phase difference $\phi_S - \phi_{T\chi}$ is the same for $S(b)$ and $C(b)$. Analysis of $S(b)$ over the range $2\le b\le 7$ suggests that the amplitudes go as $A_X \propto \exp(-\text{const.}/b^2)$.

So far, this section has focused on strong-coupling properties under conditions of strict particle-hole symmetry. In a metallic host---which can be thought of as corresponding to a power-law hybridization function with exponent $r=0$---there is not a single strong-coupling fixed point, but rather a line of them described by a family of effective Hamiltonians
\begin{equation}
    \tH_N^\text{SC}(V) = \sum_\sigma \Bigl( \tH_{0,N,\sigma} + \tV_\text{K,eff} f_{0\sigma}^{\dag} f_{0\sigma} \Bigr)
\end{equation}
parameterized by an effective potential scattering at the impurity site that can take any value $-\infty \le \tV_\text{K,eff} \le \infty$  \cite{PhysRevB.21.1003}. Different degrees of particle-hole symmetry in the bare problem---tuned, for instance, by the impurity level asymmetry $2\eps_d+U$ and/or the position of the Fermi energy $\eps_F$---result in flow to different strong-coupling fixed points. The particle-hole-symmetric fixed point $\tH_N^\text{SC}$ introduced earlier in the section corresponds to $\tV_\text{K,eff}=\pm\infty$ plus a shift of $\mp 1$ in the total charge quantum number. By contrast, in a host that has a power-law divergent hybridization function $\tDelta_{r<0}$, particle-hole asymmetry is irrelevant in the strong-coupling regime (so long as the hybridization divergence remains pinned to the Fermi energy) \cite{PhysRevB.88.195119}.

For Cantor-set hybridization functions $\tDelta_{C(b)}$ with $b=4M+1$ or $4M+3$, we find that particle-hole asymmetry, particularly as controlled by the location of the Fermi energy, plays a role different from that for $r=0$ and $r<0$. Most importantly, for all cases studied where $\eps_F$ lies at a point in the Cantor set, we find $S_\imp$ and $T\chi_\imp$ to exhibit $\log_b T$ oscillations about the values expected for an $r = D_F - 1$ power-law hybridization function. The amplitude of the oscillations is greatest when $\eps_F$ lies at a high-symmetry point corresponding to the center of one of the retained intervals in all approximant hybridization functions $\tDelta_{l'}(\teps)$ for $l'\ge l$, in which case $\tDelta(\teps)$ is particle-hole symmetric for $|\teps| < 3 b^{-l}$. The oscillation amplitude is smallest when $\eps_F$ lies at the upper/lower edge of an interval in some $\tDelta_l(\teps)$, where $\tDelta(\teps)$ exhibits a gap spanning $0<\pm\teps< 2b^{-l}$ but has an integrated weight of $(2M+1)^{-l}$ over $0\le \mp\teps\le 2b^{-l}$. Cases where $\eps_F$ lies at a more generic point in the Cantor set lead to oscillations of intermediate amplitude. Both the amplitude $A_X$ and phase $\phi_X$ entering Eq.\ \eqref{eq:f_X} seem to take the same values for all locations of $\eps_F$ corresponding to a given type (interval center, interval edge, or other location) but to differ between types. At this stage, we cannot rule out further subdivision of one or more of these three types of location. However, we have found no sign of any variation in either the oscillation period or the average values about which the oscillations occur.

The results reported in this section point to the existence of a \textit{fractal strong-coupling fixed point} for fractal hybridization functions having an exact self-similarity about the Fermi energy: $\tDelta(\teps) = \tDelta(\teps/b)$ for all $|\teps|$ smaller than some upper cutoff and for $b$ having some smallest value greater than $1$. At this fixed point, the impurity contributions to the magnetic susceptibility and entropy vary periodically in $\log_b T$ around negative average values. These oscillations, whose amplitude grows with increasing $b$, can be attributed to the self-similarity of the hybridization function. The negative average values result from a coarse-grained equivalence between a hybridization with fractal dimension $D_F <1 $ and a power-law hybridization function with a negative exponent
\begin{equation}
\label{eq:r-vs-D_F}
  r = D_F - 1 .   
\end{equation}
These features of the strong-coupling thermodynamic properties serve as a signature of host fractality in Anderson and Kondo problems.

\subsection{NRG results}
\label{subsec:Cantor-NRG}

\begin{figure}[t!]
\begin{center}
\includegraphics[width=0.44\textwidth]{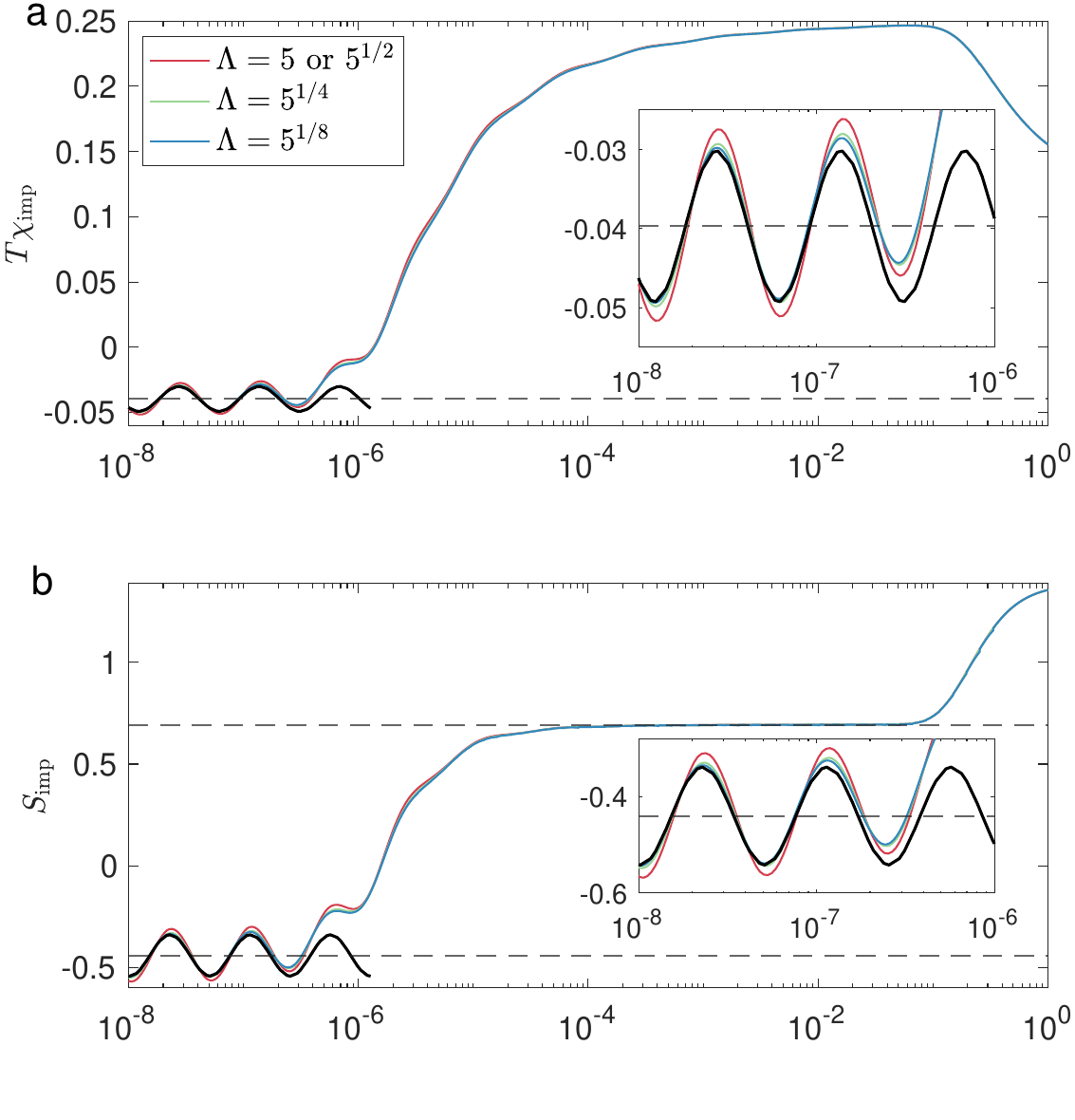}
\includegraphics[width=0.44\textwidth]{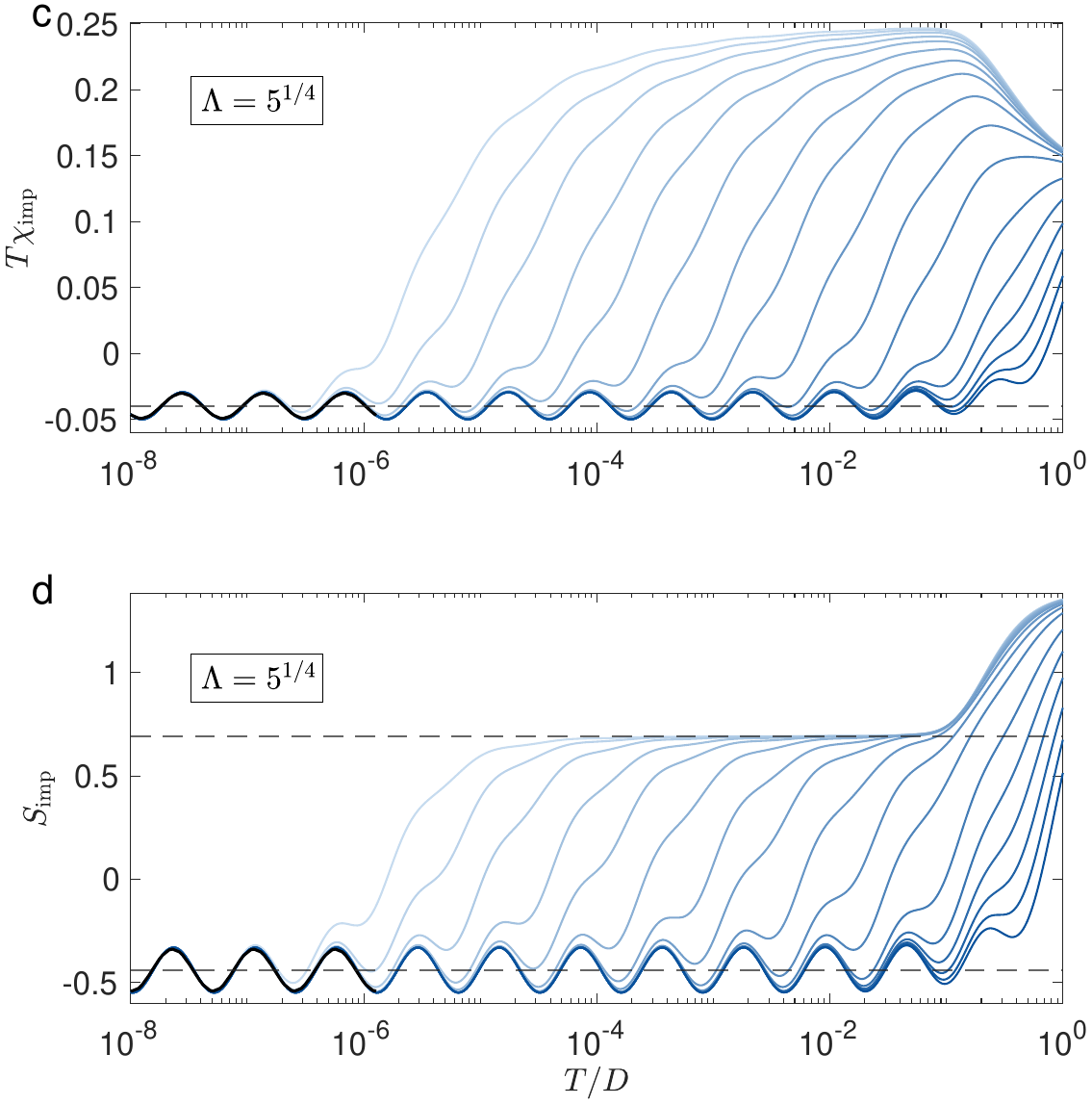}
\caption{\label{fig:Tchi,S-5-3-Cantor}%
Temperature dependence of impurity thermodynamic properties for an Anderson impurity with $U= -2\eps_d = D$ and a uniform 1/5 Cantor-set hybridization function: (a) Magnetic susceptibility $T\chi_\imp$ and (b) entropy $S_\imp$ for hybridization $V=0.05D$ and different NRG discretizations $\Lambda = 5$ (equivalent to $5^{1/2}$), $5^{1/4}$, and $5^{1/8}$, retaining up to $1000$, $15\,500$, and $31\,000$ many-body eigenstates, respectively. Solid black curves plot strong-coupling fixed point properties computed via Eq.\ \eqref{eq:X_imp,SC} with $\Lambda=5^{1/16}$, while dashed black lines represent the local-moment value $S_\imp = \ln 2$ and the strong-coupling values [Eqs.\ \eqref{eq:Tchi,S-SC}] for a power-law hybridization [Eq.\ \eqref{eq:Delta-power}] with exponent $r$ given in Eq.\ \eqref{eq:r-C}. Insets show the low-temperature properties on a magnified scale. (c) $T\chi_\imp$ and (d) $S_\imp$ for $\Lambda=5^{1/4}$ and $V/D$ spanning 0.05 (top curve) to 1.6 (bottom). Black lines are as in (a), (b).}
\end{center}
\end{figure}

Having resolved the strong-coupling thermodynamic properties of an Anderson impurity in a host with a uniform Cantor-set hybridization function via analysis of quadratic fixed-point Hamiltonians, we now turn to the full temperature dependence obtained via NRG many-body solutions of Eqs.\ \eqref{eq:H_N} and \eqref{eq:H_N,sigma}.

Figures \ref{fig:Tchi,S-5-3-Cantor}(a) and \ref{fig:Tchi,S-5-3-Cantor}(b) respectively plot $T \chi_{\imp}$ and $S_{\imp}$ as functions of temperature for the uniform $1/5$ Cantor-set hybridization $\Delta_{C(5)}$ with fixed impurity parameters $U=-2\eps_d = D$, $V=0.05D$, and the band discretizations $\Lambda = 5^{1/2^k}, \, k = 0,\, 1,\, 2\,, 3$ discussed in Sec.\ \ref{subsubsec:Cantor-Lambda-to-1}. With decreasing temperature, the system crosses over around $T\sim U$ from a free-impurity regime characterized by $T\chi_\imp \simeq 1/8$, $S_\imp \simeq \ln 4$ to a local-moment regime in which $T\chi_\imp \simeq 1/4$ and $S_\imp \simeq \ln 2$. Upon further decrease in the temperature, there is a second crossover to the strong-coupling regime analyzed in Sec.\ \ref{subsec:Cantor-SC}, in which the properties oscillate about the values $T\chi_\imp \simeq -0.040$ and $S_\imp \simeq -0.44$ corresponding to Eqs.\ \eqref{eq:Tchi,S-SC} for a power-law hybridization function with $r = D_{C(5)} - 1 \simeq -0.3174$. The insets to Figs.\ \ref{fig:Tchi,S-5-3-Cantor}(a) and \ref{fig:Tchi,S-5-3-Cantor}(b) show the data over the lowest temperature range on a magnified scale. With increasing $k$, one observes a convergence of the full NRG results at low temperatures toward the strong-coupling properties (solid black lines) calculated within the single-particle analysis of Sec.\ \ref{subsec:Cantor-SC}. 

Figures \ref{fig:Tchi,S-5-3-Cantor}(c) and \ref{fig:Tchi,S-5-3-Cantor}(d) show impurity thermodynamic properties for the same case $U=-2\eps_d = D$, but with fixed $\Lambda = 5^{1/4}$ and a range of different hybridizations $V$. With increasing $V$, the high-temperature crossover from free-impurity to local-moment behavior at first becomes less pronounced, with $T \chi_\imp$ not rising as close to $1/4$ and $S_\imp$ showing a less pronounced plateau near $\ln 2$; there remain clear signs of a second crossover, representing Kondo screening of an impurity moment, with a Kondo scale that can be defined through Eq.\ \eqref{eq:T_K-std}. For larger hybridizations, by contrast, signatures of a local-moment regime disappear, to be replaced by a direct crossover from the free-impurity regime to strong coupling. In all cases, however, the asymptotic low-temperature behavior is the strong-coupling regime analyzed in Sec.\ \ref{subsec:Cantor-SC}.

To summarize Section \ref{sec:Cantor}, the low-temperature behavior of a magnetic impurity coupled to a uniform Cantor-set hybridization function is governed by a fractal strong-coupling fixed point with properties that reflect both the self-similarity and the fractal dimension of the host spectrum. Self-similarity of the spectrum under multiplicative rescaling  $\epsilon-\epsilon_F \rightarrow (\epsilon-\epsilon_F)/b$ manifests in periodic oscillations of impurity thermodynamic quantities with $\log_b T$, while the fractal dimension $D_F <1$ causes these oscillations to occur about negative mean values identical to those for a power-law hybridization function [Eq.\ \eqref{eq:Delta-power}] with an exponent given by Eq.\ \eqref{eq:r-vs-D_F}.

Self-similarity under multiplicative rescaling is a general feature of fractals, suggesting that the results of this section extend, at least qualitatively, to other fractal hosts. We next consider an Anderson impurity coupled to DOS of the critical AA model, which has a fractal form that can be described by a nonuniform subdivision rule, and show that in this case too the low-temperature physics is described by fractal strong-coupling (both at and away from particle-hole symmetry).

\section{Fractal Spectrum of the \AubAnd\ Model}
\label{sec:fractal-AA}

\begin{figure}[t!]
\centering
\includegraphics[width=0.49\textwidth]{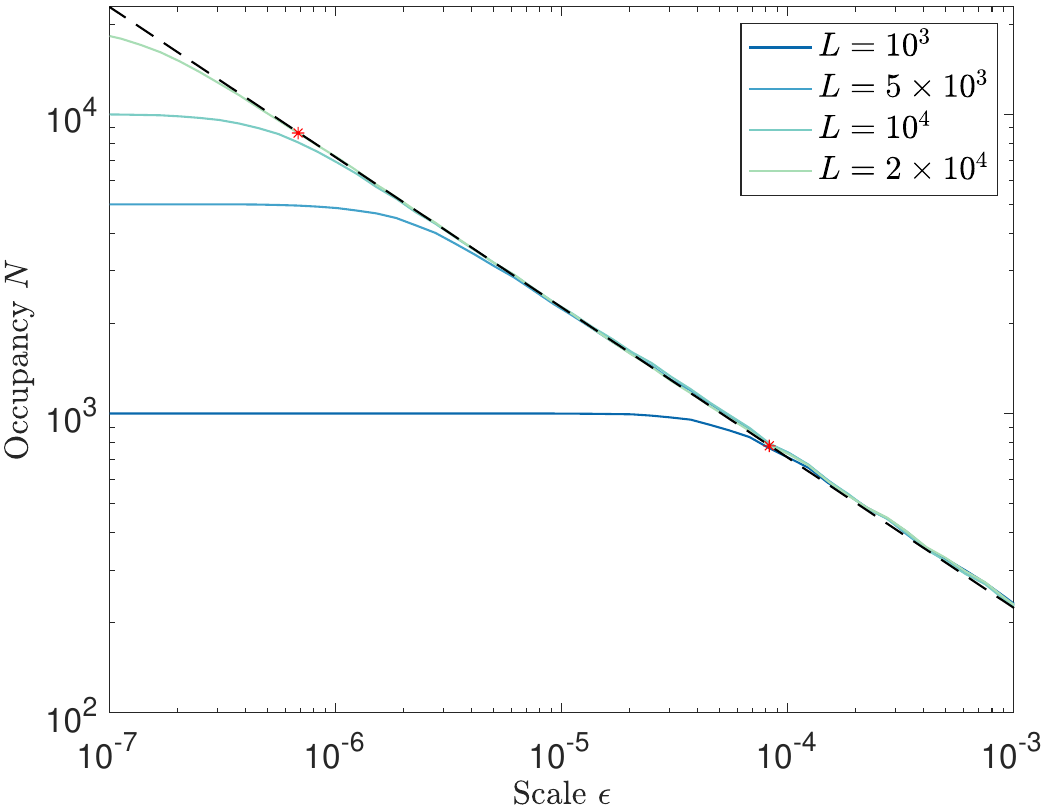}
\caption{Box-counting for the \AubAnd\ spectrum at the critical point $\lambda=2t$: Number $N$ of non-overlapping boxes required to cover the spectrum vs box width $\epsilon$ for fixed $\phi=0$, based on exact diagonalization for various lattice sizes $L$ labeled in the legend. The dashed line marks linear regression of the $L=2\times 10^4$ data over the range between the red stars, with a slope $-0.5000\pm 0.0015$.}
\label{fig:AAFractalDim}
\end{figure}

This section presents results for the Anderson impurity model with a hybridization function $\Delta_\tAA(\eps)$ set by the global DOS of a critical AA model defined in Eq.\ \eqref{eq:H_AA} with $Q=(\sqrt{5}-1)/2$ and $\lambda = 2t$.
The spectrum for this critical AA model can be reproduced by iterated non-uniform subdivision of the bandwidth according to rules \cite{wu2021fractal} that (i) are considerably more complicated than those that generate $\Delta_{C(4M+1)}$ and $\Delta_{C(4M+3)}$ treated in Sec. \ref{sec:Cantor} and (ii) reveal self-similarity of the DOS under rescaling of energies by a factor $b = 13.74$.

Figure \ref{fig:AAFractalDim} show numerical results based on exact diagonalization of $Q=(\sqrt{5}-1)/2$, $\lambda = 2t$ AA chains up to length $L=2\times 10^4$. These box-counting data lead to the conclusion, via Eq.\ \eqref{eq:D-box}, that the spectrum has fractal dimension $D_\tAA = 0.5000\pm 0.0015$. Therefore, study of $\Delta_\tAA(\eps)$ provides a natural bridge between the fractal ``toy'' models investigated in Sec.\ \ref{sec:Cantor} and  the full AAA model (to be treated in Sec.\ \ref{sec:full-AAA}) that has a distribution of fractal dimensions due to sampling of multifractal wave functions by the LDOS.

\begin{figure}[t!]
\centering
\includegraphics[width=0.48\textwidth]{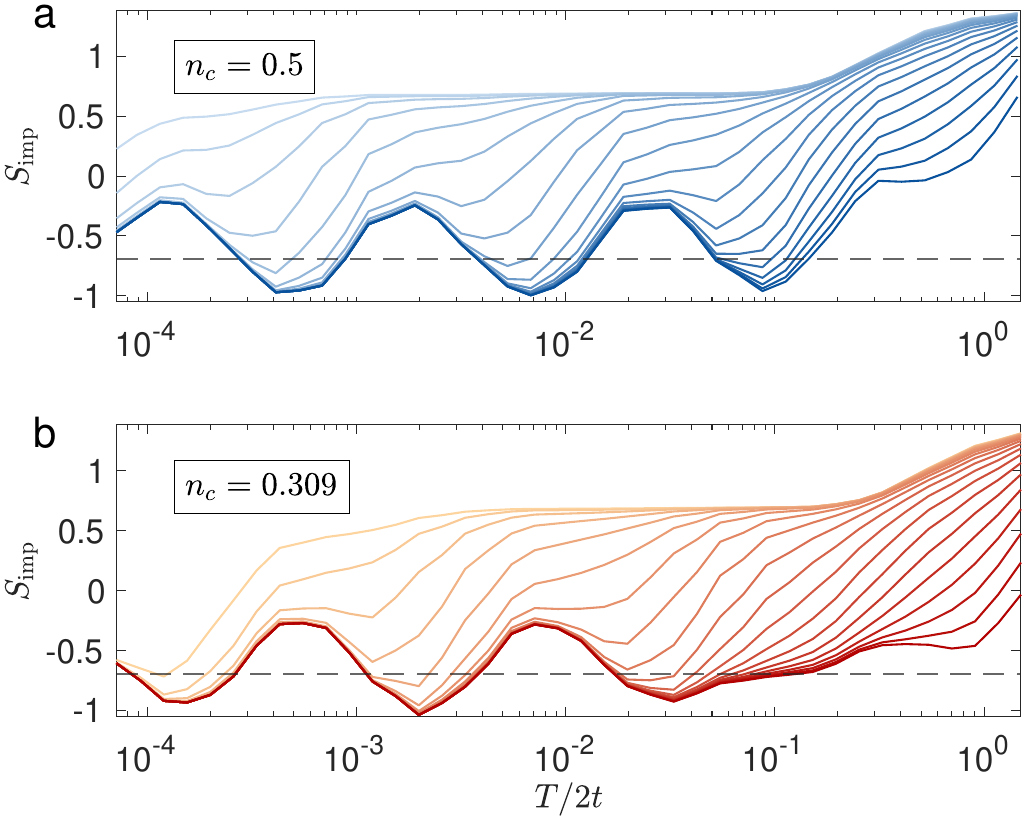}
\caption{\label{fig:S-AA}%
Impurity entropy $S_\imp$ vs $T$ for an Anderson impurity having $U = -2\eps_d = D$ hybridizing with the global DOS of a critical \AubAnd\ chain consisting of $L=10^6$ sites at filling (a) $n_c = 0.5$ with half-bandwidth $D=2.60t$, (b) $n_c=0.309$ with $D=4.53t$. KPM+NRG results for $V/D$ spanning 0.06 (top curve) to 1.58 (bottom) with $N_C = 10^5$, $\Lambda = 3$, and $N_s = 4^7$.
Horizontal dashed lines mark the strong-coupling value in Eqs.\ \eqref{eq:Tchi,S-SC} for a power-law hybridization given by Eq.\ \eqref{eq:Delta-power} with exponent $r = D_\tAA - 1 = -0.5$.}
\end{figure}

Whereas in Sec.\ \ref{sec:Cantor} it was possible to obtain the Wilson-chain coefficients analytically or via relatively straight-forward computation, for $\Delta_\tAA$ we must rely on numerically intensive methods. We employ the KPM+NRG approach described in Sec.\ \ref{subsec:methods} and Appendix \ref{app:KPM+NRG} to compute the hybridization function in Eq.\ \eqref{eq:hyb} for a system size of $L=10^6$, sufficiently large that the lowest temperature that can be reached is set not by the level spacing $\simeq 4t/L$  but rather by the KPM energy resolution [Eq.\ \eqref{eq:KPM-res}] associated with the finite expansion order $N_C=10^5$.
Since the global density of states is unaffected by the random phase of the potential it suffices to consider a single phase choice $\phi = 0$. We consider both a particle-hole-symmetric band corresponding to filling [Eq.\ \eqref{eq:n_c}] $n_c=1/2$ as well as an asymmetric case $n_c=0.309$. It should be noted that the half-bandwidth $D$ defined in Eq.\ \eqref{eq:D} depends on $\lambda$ and also on $n_c$. Throughout this section and Sec.\ \ref{sec:full-AAA} we omit plots of $T\chi_\imp$ vs $T$ because the magnetic susceptibility data do not add materially to the physical understanding that can be drawn just from $S_\imp$.

\begin{figure}[h!]
\centering
\includegraphics[width=0.49\textwidth]{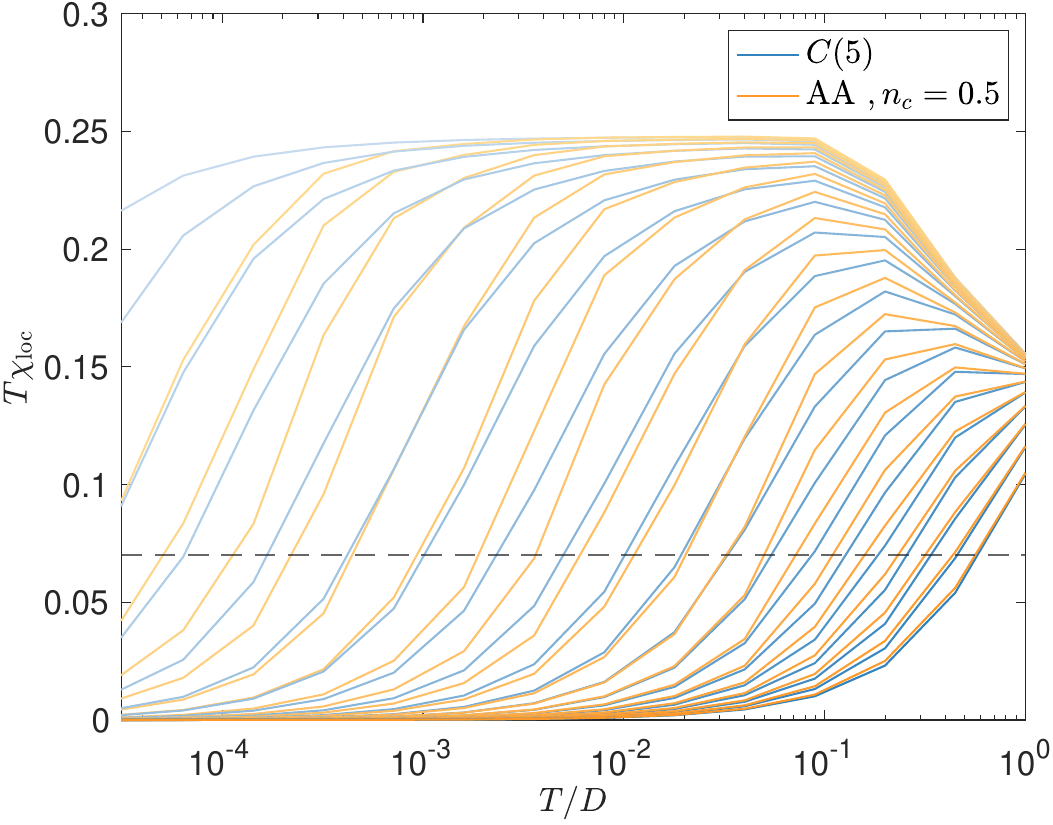}
\caption{\label{fig:Tchi_loc-fractals}%
Local susceptibility $T\chi_\loc$ vs $T/D$ for an Anderson impurity interacting with a uniform $1/5$ Cantor set [$C(5)$] hybridization function or with the global DOS of a critical \AubAnd\ chain at $n_c=0.5$, calculated for $U=-2\eps_d=D=2.60t$, $L=10^6$, $N_C=10^5$, $\Lambda=5$, $N_s=6000$, and matrix elements $V/D$ spanning 0.06 (top curves) to 1.58 (bottom). The horizontal dashed line marks the value $T\chi_\loc = 0.0701$ used to define $T_K$.}
\end{figure}

Figure \ref{fig:S-AA}(a) plots the temperature dependence of $S_\imp$ for $U=-2\eps_d = D$, $n_c = 0.5$, and a range of different hybridizations $V$, while Fig.\ \ref{fig:S-AA}(b) shows its $n_c = 0.309$ counterpart. In each case, $S_\imp$ approaches its value at the fractal strong-coupling fixed point, oscillating about the negative value given by Eq.\ \eqref{eq:Tchi,S-SC} with $r = D_\tAA - 1 = -0.5$. The oscillations are approximately sinusoidal in $\log T$ with a period $\log b$ that reflects the self-similarity of the AA spectrum under rescaling of energies by a multiplicative factor $b = 13.72$ \cite{wu2021fractal}. The oscillation amplitude is greater than seen for the $1/5$ Cantor set in Fig.\ \ref{fig:Tchi,S-5-3-Cantor}, for which the self-similarity factor is $b = 5$. This is consistent with our finding for the models studied in Sec.\ \ref{sec:Cantor} that the amplitude grows with increasing self-similarity factor.
While the amplitude and period of the strong-coupling oscillations is the same to within our numerical resolution for $n_c = 0.5$ and $n_c = 0.309$ (respectively at and away from particle-hole symmetry), the phase differs between the two cases, reminiscent of the sensitivity to the location of the Fermi energy discussed in Sec. \ref{subsec:Cantor-SC}.

Figure \ref{fig:S-AA} shows that with decreasing hybridization strength, the fractal strong-coupling fixed point is approached at ever lower temperatures. We estimate the effective Kondo temperature for this crossover from the local spin susceptibility $\chi_\loc$, which (as mentioned in Sec.\ \ref{subsec:observables}) tends to have a simpler temperature variation in fractal hosts than $\chi_\imp$. Figure \ref{fig:Tchi_loc-fractals} shows $T\chi_\loc$ vs $T$ for different values of $V$, both for hybridization function $\Delta_{C(5)}$ from Sec.\ \ref{sec:Cantor} and for the critical AA hybridization function at half-filling.
Figure \ref{fig:key-results}(b) plots values of $T_K$ determined via Eq.\ \eqref{eq:T_K-loc} for the uniform $1/5$ Cantor-set hybridization function and for the critical AA DOS at fillings $n_c = 0.5$ and $0.309$. In each case, the Kondo temperature for small $J_K$ has a power-law dependence
\begin{equation}
\label{eq:T_K-vs-J}
    T_K \sim J_K^{\alpha(D_F)}, \quad \alpha(D_F)=\frac{1}{1-D_F}.
\end{equation}
This is precisely the behavior that should be expected based on the coarse-grained equivalence between a fractal hybridization function and a power-law hybridization described by Eqs.\ \eqref{eq:Delta-power} and \eqref{eq:r-vs-D_F}, given that the latter obeys $T_K \sim J_K^{-1/r}$ \cite{PhysRevB.88.195119}.

Our results have been obtained for a specific choice $Q = (\sqrt{5}-1)/2$ of the wave number entering Eq.\ \eqref{eq:H_AA}. The AA model has a delocalization-localization transition with a fractal spectrum at the same $\lambda_c=2t$ for all irrational values of $Q$ \cite{aubry1980analyticity,sokoloff1985unusual}. We believe, therefore, that for any such $Q$, the low-temperature physics of an Anderson impurity hybridizing with the global DOS of a critical AA model will be described by a fractal strong-coupling fixed point. However, it is quite possible that the quantitative details of the fixed point will depend on the specific value of $Q$. For example, there are open conjectures that for any irrational value of $Q$, the fractal dimension satisfies $D_\tAA = 1/2$ \cite{bell1989hierarchical}, $D_\tAA < 1/2$ \cite{rudinger1997hofstadter} or $D_\tAA \le 1/2$ \cite{PhysRevB.58.9881}. Variation of $D_\tAA$ with $Q$ will lead to differences in the low-temperature-averaged value of thermodynamic properties, while variation of the self-similarity factor will change the periodicity of $\log T$ oscillations in those properties about the averages.
We leave the detailed exploration of the effect of varying $Q$ on the fractal strong-coupling fixed point for future work.

\section{\AubAnd\ Anderson Impurity Model}
\label{sec:full-AAA}

Sections \ref{sec:Cantor} and \ref{sec:fractal-AA} treated Anderson impurity models in which the impurity hybridization function is determined by the global DOS of a fractal host. The current section addresses hybridization functions $\Delta_\AAA$ determined by the LDOS at the impurity site in an \AubAnd\ host. As was the case for $\Delta_\tAA$ considered in Sec.\ \ref{sec:fractal-AA}, $\Delta_\AAA$ requires a fully numerical treatment using the KPM+NRG method. We first present results exploring the Kondo physics in the host's delocalized ($\lambda<2t$) and Anderson-localized ($\lambda > 2t$) phases. We then turn to the impurity problem at the critical point $\lambda_c = 2t$ of the AA model, where the hybridization function reflects not only the fractal spectrum but also the multifractal nature of the wave functions.

\subsection{Delocalized phase}

In the delocalized phase of the AA model (accessed for $0 < \lambda < \lambda_c = 2t$), the spectrum is broken into minibands separated by hard gaps. We are interested in situations where the Fermi energy lies within a miniband, guaranteeing that the AAA model ultimately flows to its strong-coupling RG fixed point. In an RG picture, the flow to strong coupling begins at high temperatures of order the half-bandwidth $D$ as one integrates out electronic excitations having energies much greater than $T$. As the temperature decreases through a gap, however, one expects a temporary reversal of the RG flow to instead head toward the local-moment fixed point. Flow toward strong coupling resumes once the thermal scale further decreases into a miniband.

\begin{figure}[t!]
\centering
\setlabel{pos=nw,fontsize=\large,labelbox=false}
\xincludegraphics[scale=0.49,label=a]{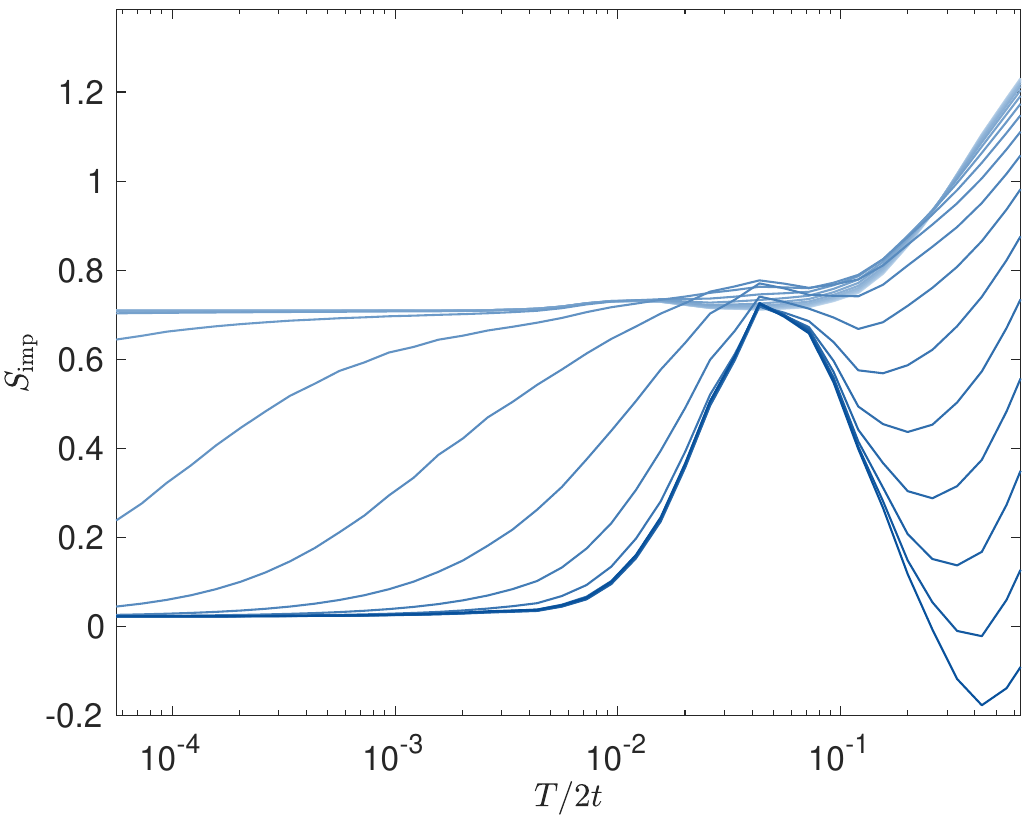}
\xincludegraphics[scale=0.49,label=b]{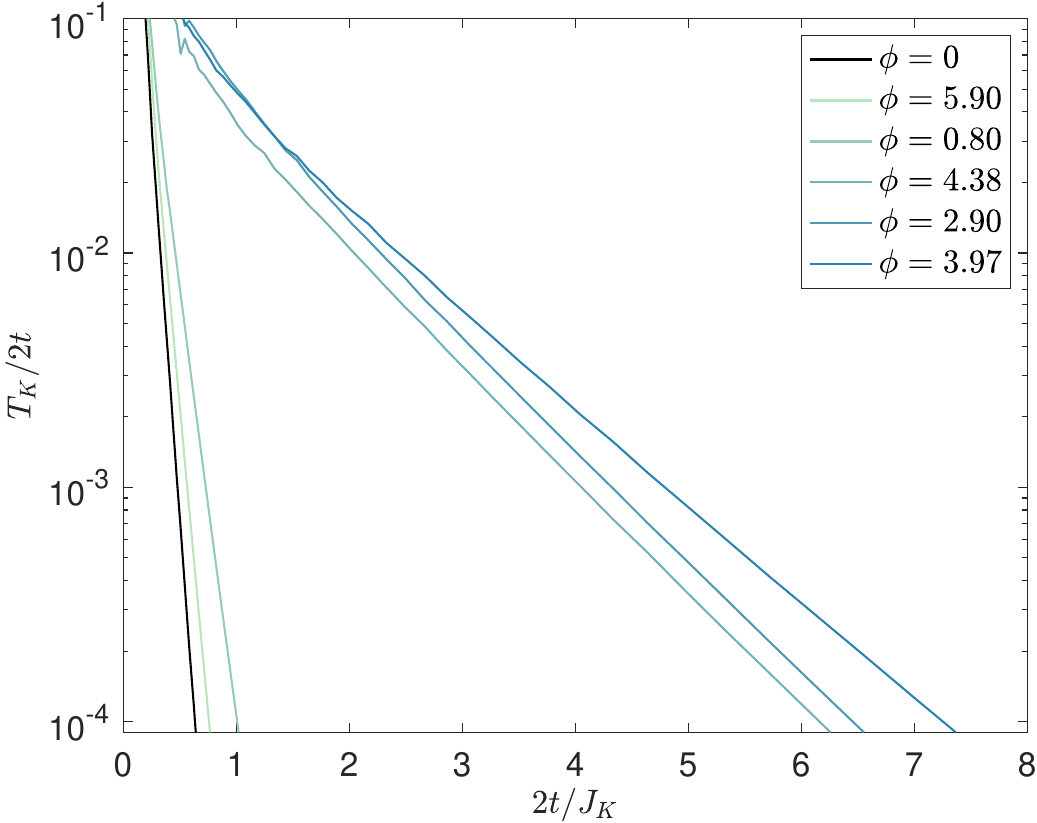}
\caption{\label{fig:AAA-lambda=t}%
Impurity properties in the AAA model at $\lambda = t$ (in the delocalized phase). An impurity having $U = -2\eps_d = D = 3.57t$ hybridizes with the local DOS at the middle of an \AubAnd\ chain consisting of $L = 10^6$ sites at filling $n_c=0.309$. Data obtained using the KPM+NRG method with $N_C=10^3$, $\Lambda = 3$, and $N_s = 5000$.  (a) Impurity entropy $S_\imp$ vs $T/2t$ for a single realization with $\phi = 0$ and matrix elements $V/D$ spanning 0.11 (top curve) to 2.83 (bottom). (b) Kondo temperature $T_K/2t$ extracted from the local magnetic susceptibility via Eq.\ \eqref{eq:T_K-loc} plotted vs $2t/J_K$ for the $\phi = 0$ sample as well as five randomly chosen values of $\phi$. For $T_K/2t\lesssim 10^{-2}$, each curve exhibits the relation $\log T_K \sim -1/J_K$ characteristic of the Kondo effect in metals.}
\end{figure}

The qualitative expectations laid out in the preceding paragraph are tested in Fig.\ \ref{fig:AAA-lambda=t}(a), which plots the temperature dependence of the impurity entropy for $\lambda=t$, deep within the delocalized phase, and for a range of hybridization matrix elements $V$. Since the LDOS is nonvanishing and featureless near the Fermi energy, it suffices to work at a relatively low KPM expansion order $N_C = 10^3$. Values $V\ll D$ cause the system to fully enter the local-moment regime, with the impurity entropy decreasing from $S_\imp = \ln 4$ in the high-temperature free-orbital regime to plateau at $S_\imp\simeq\ln 2$ over an intermediate temperature window before falling smoothly at lower temperatures toward metallic strong coupling ($S_\imp = 0$). Larger $V$ values initially set the system on course for a direct crossover from the free-orbital regime to strong coupling, while extremely large hybridizations even create a window of negative impurity contributions (meaning that the combined host-impurity system has lower total susceptibility and total entropy than the host by itself). However, what would in a simple metallic host (e.g., the AAA model with $\lambda = 0$) be a rapid approach to strong coupling is interrupted by the presence for $\lambda = t$ of several hard spectral gaps around Fermi energy. We note in particular that all curves show $S_\imp \simeq \ln 2$ at $T/2t\simeq 0.04$, suggesting that at this thermal scale, any impurity screening that took place at higher temperatures has been completely reversed. Nonetheless, with further decrease in temperature, all curves eventually approach the strong-coupling limit.

Figure \ref{fig:AAA-lambda=t}(b) plots Kondo temperatures for the AAA model at $\lambda = t$. $T_K$ is extracted from $T\chi_\loc$ vs $T$ via Eq.\ \eqref{eq:T_K-loc} for the $\phi = 0$ realizations shown in Fig.\ \ref{fig:AAA-lambda=t}(a) as well as for five randomly chosen values of $\phi$. All samples exhibit the small-$J_K$ dependence $\log T_K \sim -1/[\rho_{R}(\eps_F)J_K]$ expected in a metal. Each impurity location has a different LDOS $\rho_{R}(\eps_F)$, which changes the slope of the log-linear plot of $T_K$ vs $1/J_K$.

\subsection{AA localized phase}

In the localized phase of the AA model (reached for $\lambda > \lambda_c = 2t$), all eigenstates are spatially localized. An impurity coupled to a typical site $R$ hybridizes with only a discrete subset of band states $\ket{\eps_k}$ such that $|\eps_k - \eps_F|$ has a minimum value $\eps_\text{gap}(R) > 0$. Since the hybridization function vanishes for $|\eps - \eps_F| < \eps_\text{gap}(R)$, one expects there to be a threshold value of $V$ [or of the Kondo exchange $J_K$ given in Eq.\ \eqref{eq:SW-couplings}] for the system to reach the strong-coupling RG fixed point, while for sub-threshold couplings, the ground state instead has a decoupled impurity spin degree of freedom.

\begin{figure}[b!]
\centering
\includegraphics[width=0.49\textwidth]{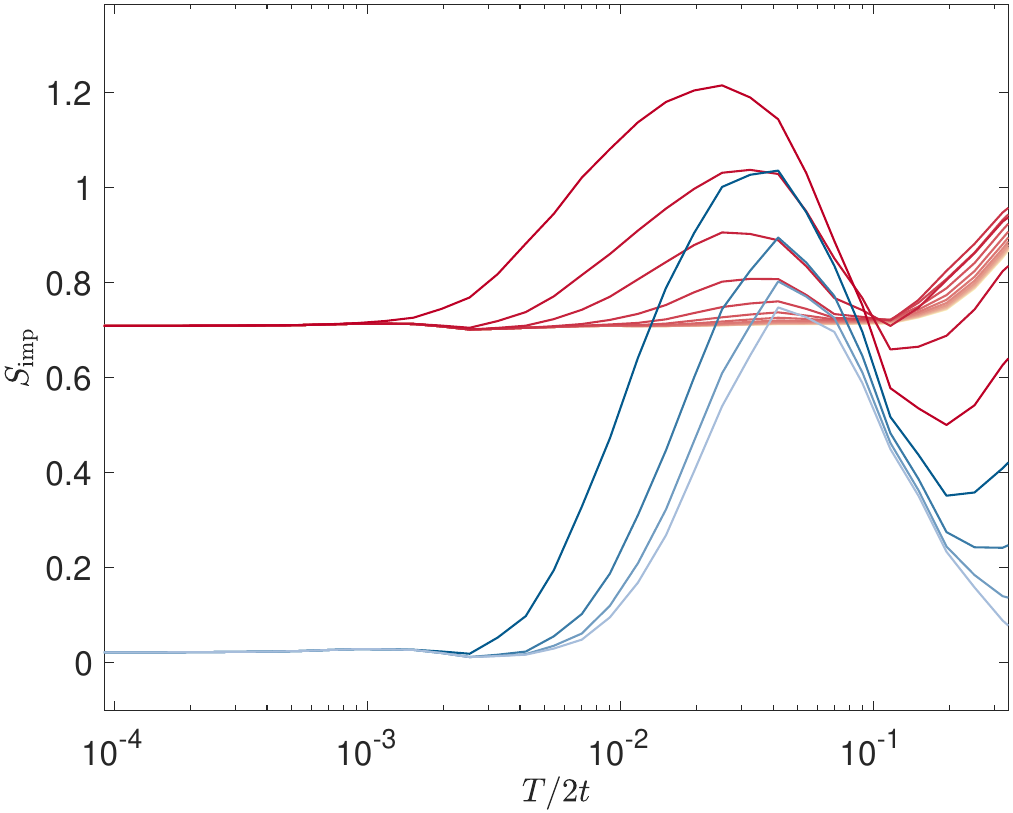}
\caption{\label{fig:S-AAA-lambda=3t}%
Impurity entropy $S_\imp$ vs $T/2t$ for the AAA model at $\lambda = 3t$ (in the localized phase). An impurity having $U = -2\eps_d = D = 5.78t$ hybridizes with the local DOS at the middle site of a single realization $\phi=0$ of an \AubAnd\ chain consisting of $L = 10^6$ sites at filling $n_c=0.309$. Data obtained using the KPM+NRG method with $N_C=10^3$, $\Lambda = 3$, and $N_s =6000$. 
Matrix elements $V/D$ from 0.06 to 0.79 (red lines) yield flow to the local-moment fixed point ($S_\imp \to \ln 2$) while higher values from 0.94 to 1.58 produce flow to strong coupling ($S_\imp \to 0$).
}
\label{localized}
\end{figure}

Figure \ref{fig:S-AAA-lambda=3t} shows the temperature variation of the impurity entropy for a number of different $V$ values at $\lambda = 3t$. The finite energy resolution of the KPM expansion [Eq.\ \eqref{eq:KPM-res}] restricts the physical validity of the results to $T \gtrsim T_\KPM = \pi D / N_C \simeq 9\times 10^{-5} (2t)$. Data for $\tilde V\in [0.06,0.79]$ (red lines in Fig.\ \ref{fig:S-AAA-lambda=3t}) show no sign of Kondo screening down to $T_\KPM$, and can be presumed to approach the local-moment fixed point. By contrast, the results for $\tilde V\in [0.94,1.58]$ (blue lines in Fig.\ \ref{fig:S-AAA-lambda=3t}) are indicative of crossover to strong coupling around a Kondo temperature much greater than $T_\KPM$. Somewhere between $V = 0.79D$ and $0.94D$ must lie a critical hybridization $V_c$ such that $T_K$ vanishes as $V$ approaches $V_c$ from above. The finite KPM resolution [Eq.\ \eqref{eq:KPM-res}] prevents evaluation of $V_c$ to high accuracy, and in any case, this quantity will be sample (i.e., $\phi$) dependent.

\subsection{AA critical point}

\begin{figure}[t!]
\centering
\includegraphics[width=0.49\textwidth]{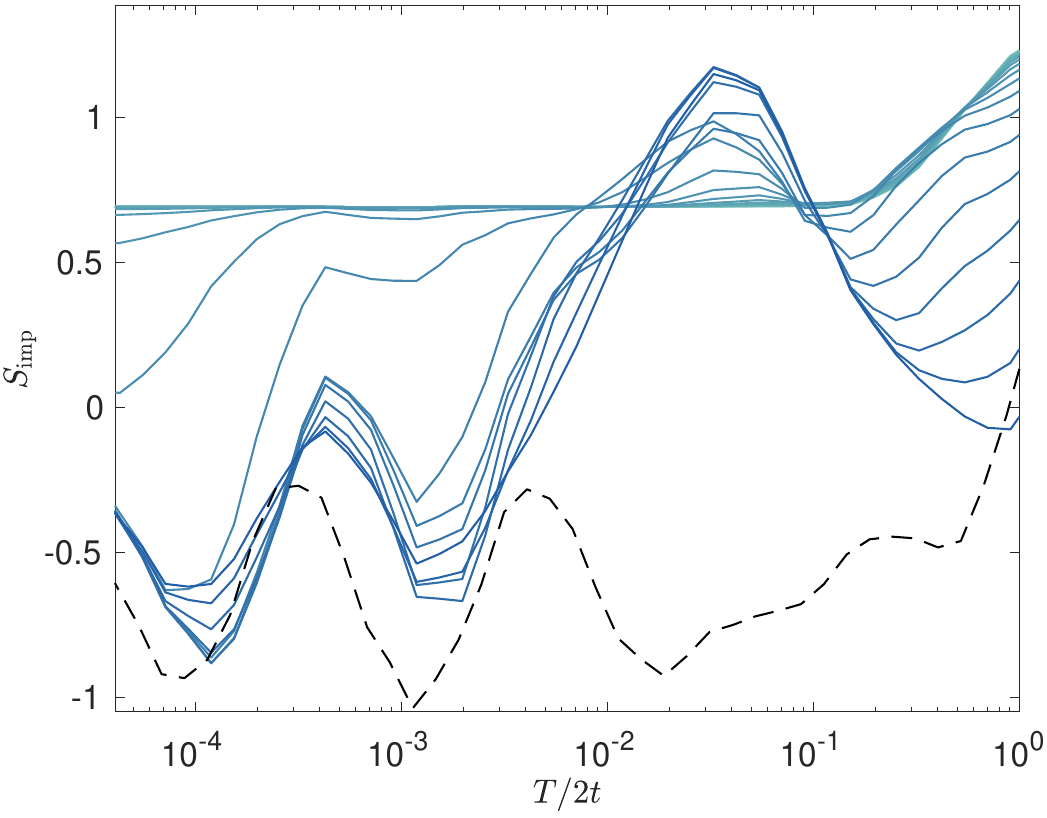}
\caption{\label{fig:S-AAA-crit-phi=0}%
Impurity entropy $S_\imp$ vs $T/2t$ for a single realization of the critical AAA model. An impurity having $U = -2\eps_d = D = 4.53t$ hybridizes with the local DOS at the middle site of an \AubAnd\ chain consisting of $L = 10^6$ sites at filling $n_c=0.309$with phase $\phi = 0$. KPM+NRG data for matrix elements $V/D$ spanning 0.06 (lightest curve) to 1.58 (darkest) with $N_C=10^5$, $\Lambda = 3$, and $N_s = 4^7$. The dashed line reproduces the largest-$V$ results for hybridization function $\Delta_\tAA$ from Fig.\ \ref{fig:S-AA}(b).}
\end{figure}

At the critical potential strength $\lambda_c = 2t$, the AA model exhibits a fractal spectrum with spatially inhomogeneous wave functions. These features combine to produce an LDOS at the impurity site whose energy variation is encoded in the NRG Wilson-chain coefficients as described in Appendix \ref{app:AAA-NRG}. As the Wilson-chain coefficients vary strongly from iteration to iteration we retain up to $N_s = 4^7 = 16\ 484$ many-body eigenstates for convergence that is demonstrated by $T\chi_\imp$ and $S_\imp$ varying only slightly on reducing $\Lambda$ from $5$ to $3$. All plots of $S_\imp$ vs $T$ present $\Lambda = 3$, $N_s = 4^7$ data, but to reduce computational time we have used $\Lambda = 8$, $N_s = 5000$ when constructing distributions of Kondo temperatures over large numbers of samples.

\begin{figure}[t!]
\centering
\includegraphics[width=0.49\textwidth]{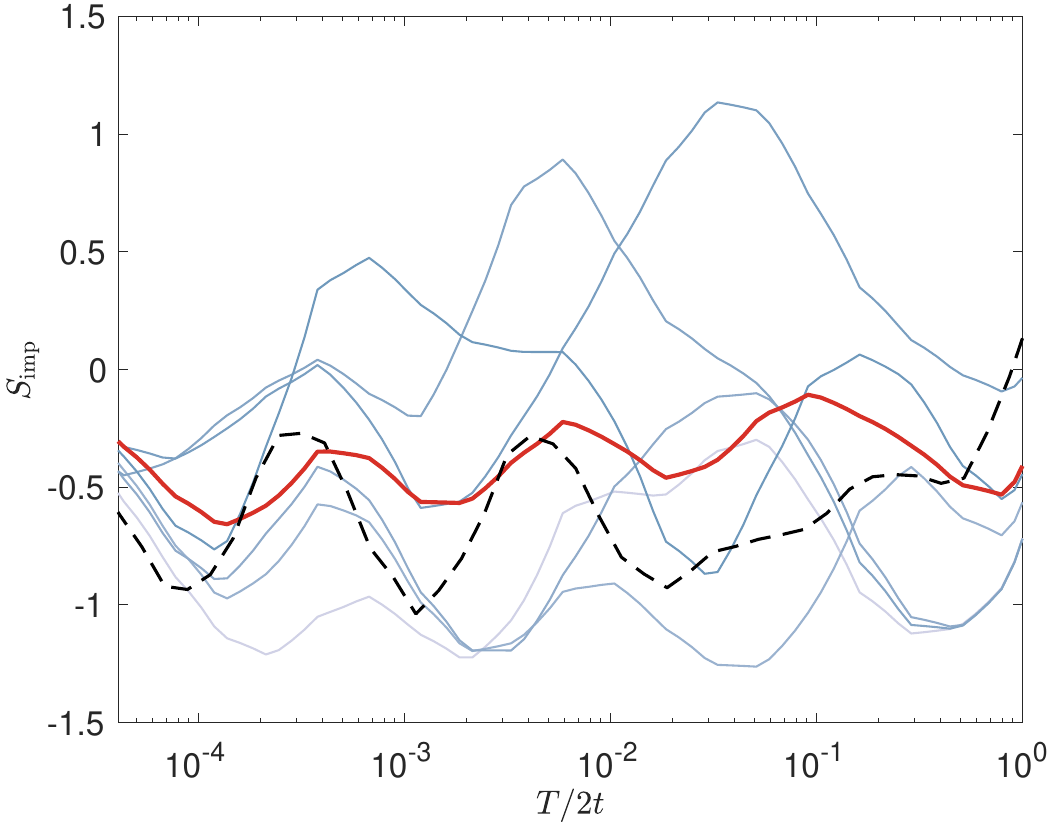}
\caption{\label{fig:S-AAA-crit-fixed-V}%
Impurity entropy $S_\imp$ vs $T/2t$ for multiple realizations of the critical AAA model. An impurity having $U = -2\eps_d = D = 4.53t$ hybridizes with matrix element $V = 1.58D$ with the local DOS at the middle site of an \AubAnd\ chain consisting of $L = 10^6$ sites at filling $n_c=0.309$. Blue curves represent single-sample results for five different randomly chosen phases $\phi$, while the red curve plots the mean $\overline{S_\imp}$ over 100 different random phases. The dashed line reproduces the largest-$V$ results for hybridization function $\Delta_\tAA$ from Fig.\ \ref{fig:S-AA}(b). All data obtained using the KPM+NRG method with $N_C=10^5$, $\Lambda = 3$, and $N_s = 4^7$.}
\end{figure}

\begin{figure}[t!]
\centering
\includegraphics[width=0.49\textwidth]{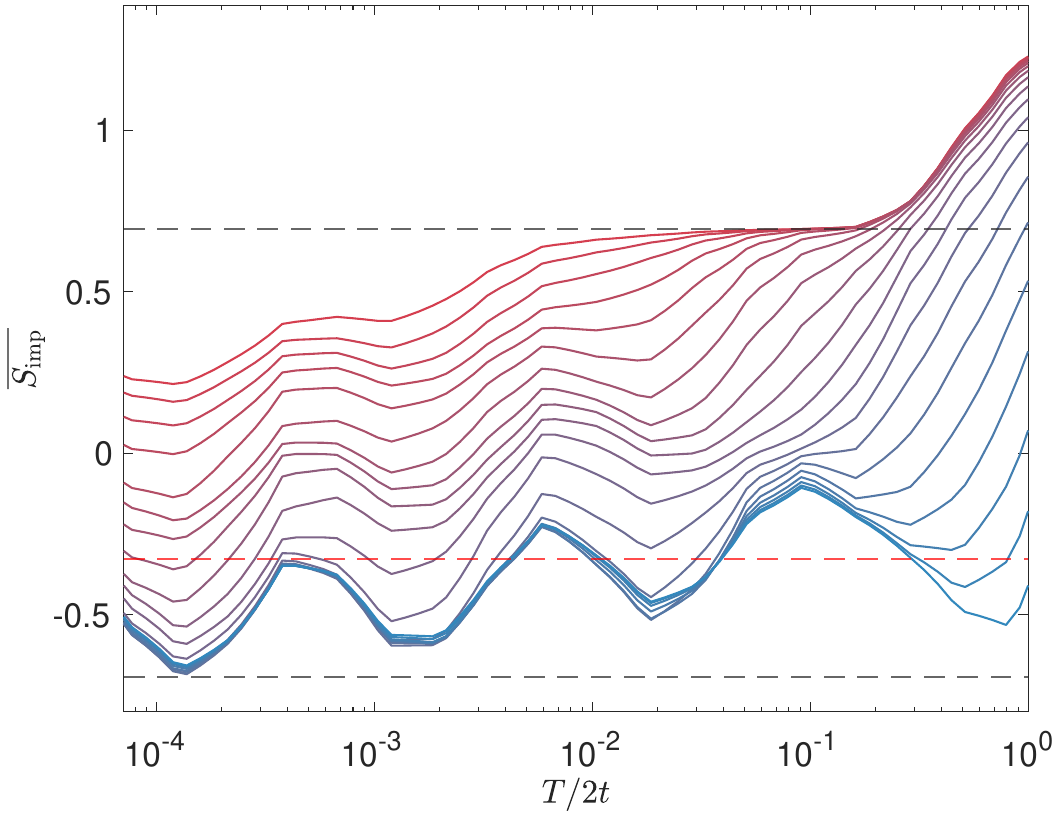}
\caption{\label{fig:S-AAA-crit-various-V}%
Sample-mean impurity entropy $\overline{S_\imp}$ vs $T/2t$ for the critical AAA model. An impurity having $U = -2\eps_d = D = 4.53t$ hybridizes with the local DOS at the middle site of an \AubAnd\ chain consisting of $L = 10^6$ sites at filling $n_c=0.309$. KPM+NRG data for matrix elements spanning 0.06 (top curve) to 1.58 (bottom) with $N_C=10^5$, $\Lambda = 3$, and $N_s = 4^7$, averaging over 100 randomly chosen values of $\phi$. Black horizontal dashed lines mark the local-moment value $S_\imp = \ln 2$ and the strong-coupling values in Eqs.\ \eqref{eq:Tchi,S-SC} for a power-law hybridization given by Eq.\ \eqref{eq:Delta-power} with exponent $r = D_\tAA - 1 = -0.5$. The red horizontal dashed line marks the strong-coupling value for $r = -1/\alpha = -0.236$ based on the median Kondo temperature data in Fig.\ \ref{fig:key-results}(c).
}
\end{figure}

We begin our discussion of KPM+NRG results for the critical AAA model by focusing on a single realization $\phi = 0$. Figure \ref{fig:S-AAA-crit-phi=0} plots the temperature dependence of the impurity entropy for a wide range of hybridization matrix elements $V$, keeping all other parameters constant. The oscillatory behavior and negative values attained at low temperatures by $S_\imp$ (and also by $T\chi_\imp$, not shown) echo the corresponding results for the hybridization function $\Delta_\tAA$ based on the global DOS of a critical AA chain (see Fig.\ \ref{fig:S-AA}). Comparison with the dashed curve in Fig.\ \ref{fig:S-AAA-crit-phi=0}, which reproduces the largest-$V$ results from Fig.\ \ref{fig:S-AA}(b), shows the spacing between turning points along the $\log T$ axis to be very similar for hybridization with the global DOS and hybridization with the LDOS at the middle site. However, the oscillations for the full AAA problem do not become truly periodic over the temperature range accessible in our KPM+NRG calculations. We attribute the more complicated temperature dependence to the LDOS sampling the fractal critical spectrum of the AA chain with different weights that depend on the amplitude of each energy eigenstate at the impurity site. This should result in the system effectively exhibiting not a single fractal dimension, but instead a distribution of fractal dimensions, each holding within its own energy window. With decreasing temperature, the system samples different fractal dimensions, each having its own strong-coupling fixed point characterized by $\log T$ oscillations of thermodynamic properties about different average values.

The scenario of LDOS multifractality suggests strong sample dependence of the physical properties. Figure \ref{fig:S-AAA-crit-fixed-V} confirms this to be the case for the impurity entropy computed at a large, fixed hybridization matrix element $V = 1.58D$ for each of five randomly chosen phases $\phi$. Both the extremal values of $S_\imp$ and the temperatures at the extrema occur show wide dispersion across samples. By contrast, the mean $\overline{S_\imp}$ over 100 randomly chosen phases (red curve) has turning points at very similar temperatures to the highest-$V$ data for hybridization function $\Delta_\tAA$ [reproduced from Fig.\ \ref{fig:S-AA}(b) as dashed curves in Fig.\ \ref{fig:S-AAA-crit-fixed-V}]. However, it is also clear that the oscillations of the sample-averaged properties are about values that are less negative than their counterparts for $\Delta_\tAA$. These observations suggest that sample averaging over the LDOS restores the self-similarity of the global DOS under energy rescaling (the feature that underlies the $\log T$ oscillations in the thermodynamic properties), while failing to reproduce the fractal dimension $D_\tAA = -0.5$ (which determines the temperature-averaged values).

\begin{figure}[t!]
\centering
\includegraphics[width=0.49\textwidth]{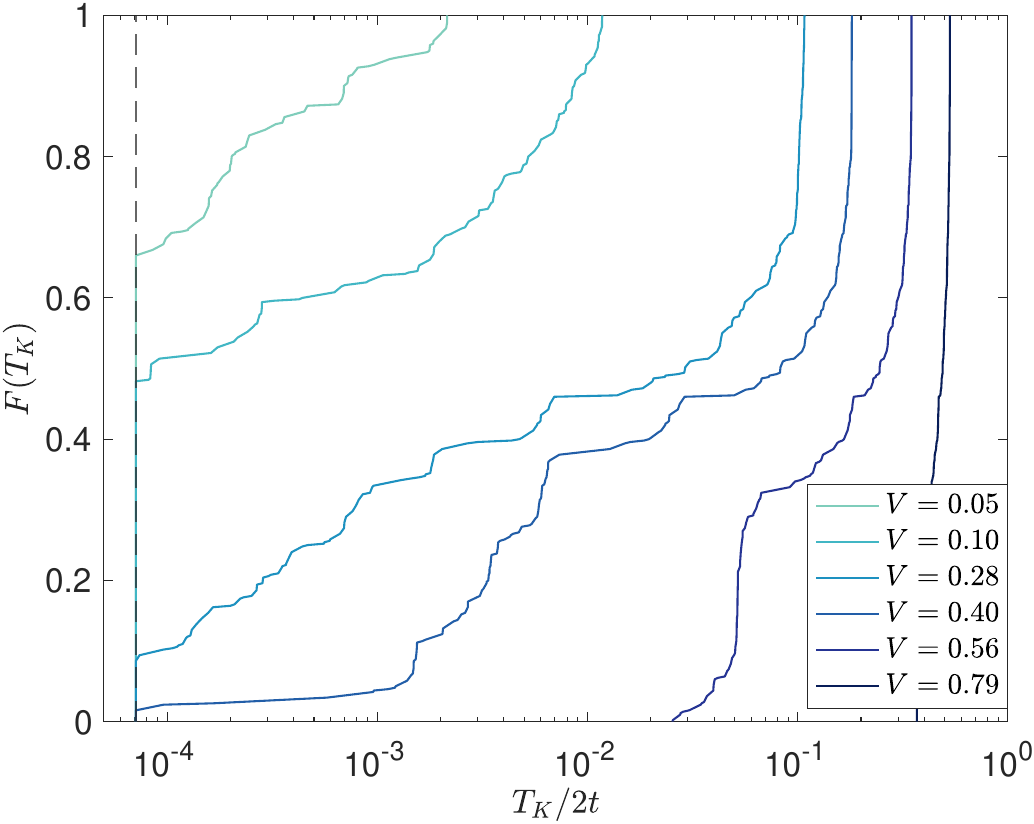}
\caption{\label{fig:T_K-dist-AAA-crit}%
Cumulative distribution $F(T_K)$ of the Kondo temperature $T_K$ in the critical AAA model for various values of the hybridization matrix $V$ (expressed in the legend as a multiple of $D$). Each distribution of $T_K$ values is extracted via Eq.\ \eqref{eq:T_K-loc} for $500$ random samples, using the KPM+NRG method with $n_c=0.309$, $U=-2\eps_d=D=4.53t$, $N_C=10^5$, $\Lambda = 8$, and $N_s = 5000$.
}
\end{figure}

Figure \ref{fig:S-AAA-crit-various-V} plots the sample-averaged impurity entropy $\overline{S_\imp}$ vs $T$ for a wide range of values of the hybridization matrix element $V$. With decreasing $T$, $\overline{S_\imp}$ appears to approach a strong-coupling limit with $\log T$ oscillations about a negative average value. However, the crossover from local-moment behavior ($\overline{S_\imp} \simeq \ln 2$) is more gradual than in the pure-fractal problems studied in Secs.\ \ref{sec:Cantor} and \ref{sec:fractal-AA}, and even the curve for the largest-$V$ appears still to be drifting downward at the lower limit $T=\pi D/N_C$ of reliability of our results. At the lowest temperature, the sample-averaged results show slow flow to the negative strong-coupling regime. This suggests that the distribution of Kondo temperatures with exchange coupling $J_K$ may be different from the behavior found for pure fractal models [see Fig.\ \ref{fig:key-results}(b)].

Figure \ref{fig:T_K-dist-AAA-crit} presents $F(T_K)$, the cumulative distribution of the Kondo temperature in the critical AAA model, calculated for a number of different values of the hybridization matrix element $V$ based on 500 random phases $\phi$. $F(T_K)$ has an initial value at $T_K = T_\KPM = \pi D / N_C \simeq 3\times 10^{-5} D$ equal to the fraction of samples that do not have a solution of Eq.\ \eqref{eq:T_K-loc} for $T_K \ge T_\KPM$, the lowest temperature that the KPM+NRG method can reliably access. For very small $V$, the probability distribution $P(T_K)= dF/dT_K$ presumably has a long tail extending to very small $T_K$ values, as a result of which $F(T_\KPM)\ge 0.5$.

Curves such as those in Fig.\ \ref{fig:T_K-dist-AAA-crit} can be used to calculate the $V$-dependence of various representative values for the $T_K$ distribution. One such is the mean $\overline{T_K}$, which for a $P(T_K)$ having significant support across many decades of $T_K$ is entirely dominated by the upper end of the range. The mean is thus little affected by our lack of knowledge of $P(T_K)$ for $T_K<T_\KPM$. This absence of information does rule out calculating the geometric mean $\exp\overline{\ln T_K}$, a quantity that is more strongly affected than the mean by the presence of very low $T_K$ values. However, for values of $V$ sufficiently large that $F(T_\KPM) < 0.5$, we can instead consider the median $\text{med}(T_K)$. 
Fig.\ \ref{fig:key-results}(c) shows the variation of $\overline{T_K}$ and $\text{med}(T_K)$ with $1/J_K$. For $J_K\ll 2t$, both measures vary as $J_K^\alpha$, similar to the behavior seen in Fig.\ \ref{fig:key-results}(b) when a impurity couples to a fractal hybridization function. However, the fitted exponents $\alpha=1.54$ for $\overline{T_K}$ and $4.23$ for $\text{med}(T_K)$ show that the latter quantity is much more sensitive to changes in the hybridization matrix element. Moreover, the presence of a tail towards vanishing $T_K$ as seen in $F(T_K)$, strongly affects the median and not the mean. Equation \eqref{eq:T_K-vs-J} can be applied to convert $\alpha$ values to effective fractal dimensions $D_F=1-1/\alpha=0.35$ and $0.76$ for $\overline{T_K}$ and $\text{med}(T_K)$,  respectively. However, the impurity entropy in Fig.\ \ref{fig:S-AAA-crit-various-V} appears to approach neither the fractal strong-coupling average value $S_\imp = -0.90$ expected for $D_F = 0.35$ nor its $D_F = 0.76$ counterpart $S_\imp = -0.33$ (red dashed line in Fig.\ \ref{fig:S-AAA-crit-various-V}). These effective $D_F$ values reflect not only geometric self-similarity, but also probability measures from critical wave functions, as well as statistics from random locations, that cannot be fully characterized by the original definition of a fractal dimension.

\section{Discussion and Conclusion}
\label{sec:discussion}

In this work we have investigated Anderson impurity problems where the host electronic degrees of freedom have a fractal energy spectrum. We have studied three classes of models. Models in the first two classes---cases (1) and (2) for which results appear in Secs.\ \ref{sec:Cantor} and \ref{sec:fractal-AA}, respectively---are simpler and ignore the effects of wave-function amplitudes on the hybridization function, but they admit an asymptotically exact solution that reveals the existence of a fractal strong-coupling fixed point. A main conclusion in this limit is that the thermodynamic response of the quantum impurity is controlled by the fractal dimension of the host spectrum, which at a coarse-grained level can be reproduced by a model with a hybridization function diverging in a power-law fashion at the Fermi energy. Thermodynamic properties exhibit $\log T$  oscillations due to contributions from minibands and gaps alternating as a function of energy.

The third class of studied models---case (3) treated in Sec.\ \ref{sec:full-AAA}---corresponds to the physically more relevant case of a quantum impurity in a quasicrystal. Here, the hybridization function acquires contributions from both the fractal spectrum and the multifractal wave functions, which can be characterized by a distribution of fractal dimensions. To solve this class of problems, we have introduced a numerical approach (dubbed KPM+NRG) that integrates the power of Wilson's NRG with the efficiency of Chebyshev expansion techniques to describe inhomogeneous host spectra in arbitrary dimensions in an efficient and accurate manner without the need to perform any diagonalization or numerical integration. This paper has focused on the case of one-dimensional quasicrystals, realized through the AA model at its critical point.

Our numerical results for the \AubAnd\ Anderson impurity model demonstrate that while the fractal nature of the density of states is divergent towards the Fermi level, wave-function-induced fluctuations produce a broad distribution of Kondo temperatures. Oscillations remain in the impurity thermodynamic properties but they are not simply set by a single fractal dimension. The strong-coupling nature of the fixed point survives, and the impurity remains Kondo-screened at the lowest energies. Exploration of the manifestations of fractality in dynamical responses will be the topic for future work. 
Going beyond the AA model, it will be interesting to incorporate other quasiperiodic models that have mobility edges \cite{PhysRevLett.114.146601,PhysRevLett.113.236403,PhysRevLett.129.103401,PhysRevLett.120.160404,PhysRevLett.126.040603} and critical phases \cite{PhysRevLett.126.080602,PhysRevLett.125.073204,lin2022general}.

In the low-energy limit, we have found that local moments are Kondo-screened both in fractal and quasicrystalline hosts. In the former setting, the Kondo temperature $T_K$ has a power-law dependence on the Kondo coupling $J_K$, consistent with hosts at a Van Hove singularity and in stark contrast to the exponential dependence in conventional metals. In the studied quasicrystals, the sample-mean and median Kondo temperatures also vary with powers of $J_K$. The mean $T_K$ has an exponent $\alpha = 1.54$ that is quantitatively quite close to the value $\alpha = 2$ found at the fractal strong-coupling fixed point for a model using the global DOS of the AA model. The higher exponent $\alpha = 4.23$ that governs the median $T_K$ reflects a broad distribution of Kondo temperatures with long tails towards vanishing Kondo coupling.

It will be fascinating to explore similar effects in quasicrystals in higher dimensions. In particular, we expect that the KPM+NRG approach can be combined with dynamical mean-field theory to describe the YbAlAu quasicrystal. 
Even before implementing such an approach, we can apply insights from the present study of impurity models to suggest why the YbAlAu quasicrystal is critical without tuning.
We have shown that the hybridization function of a fractal host is equivalent, on a coarse-grained level, to that of host at a Van Hove singularity, a situation that has been shown in the context of the Kondo lattice to produce a critical thermodynamic response \cite{PhysRevLett.109.176404,PhysRevB.84.245111}.
We therefore speculate that the critical properties without tuning in the YbAlAu quasicrystal are due to the singular hybridization function at fractal strong-coupling fixed points. This also raises an interesting connection between YbAlAu and $\beta$-YbAlB$_4$, another material that is critical without tuning \cite{matsumoto2011quantum}. It will be interesting in future work to fully understand the role of the fractal strong-coupling fixed point in the context of a Kondo lattice.

It will also be interesting to see if physical realizations can be identified of the idealized models of a magnetic impurity coupled to a hybridization function corresponding to the global DOS of a fractal host. Possible avenues for investigation include (i) systems possessing symmetries that constrain amplitudes of the host eigenstates at the impurity site, (ii) impurities coupled to tight-binding models on tree-like strucutures that could appear in electrical systems with dendritic growth \cite{dominkovics2008fractal,yang2015fractal}, and (iii) nonlocal ``impurities'' of the type mentioned in Sec.\ \ref{subsec:fractal-host}.
Finally, we foresee impactful applications of the KPM+NRG approach to treat impurities in other systems that lack translational symmetry, such as thin films, moir\'e materials, and magnetic alloys.

\section*{Acknowledgements}

We thank Piers Coleman, Vladimir Dobrosavljevi\'{c}, Gabriel Kotliar, Johann Kroha, Andrew Millis, Qimiao Si, and Romain Vasseur for useful discussions. 
We are grateful to Lucy Reading-Ikkanda for creating the schematic diagrams in Fig. 1.
A.W.\ and J.H.P.\ are partially supported  by National Science Foundation (NSF) CAREER Grant No. DMR-1941569 and the Alfred P. Sloan Foundation through a Sloan Research Fellowship. S.G.\ acknowledges support from NSF DMR-2103938. K.I.\ acknowledges support from DMR-1508122. S.G., K.I., and J.H.P.\ acknowledge hospitality by the Aspen Center for Physics, where part of this work was completed and which is supported by NSF grant PHY-1607611. The authors acknowledge the following research computing resources: the Beowulf cluster at the Department of Physics and Astronomy of Rutgers University, and the Amarel cluster from the Office of Advanced Research Computing (OARC) at Rutgers, The State University of New Jersey.

\appendix

\section{The KPM+NRG Approach}
\label{app:KPM+NRG}

This appendix presents a ``KPM+NRG'' approach for solving models of quantum impurities coupled to hosts without translational symmetry. The appendix begins with a brief review of the mapping of the host term in the Anderson impurity Hamiltonian [i.e., $H_\host$ entering Eq.\ \eqref{eq:H_A}] to an approximate NRG description in terms of a tight-binding Wilson chain whose on-site energies and nearest-neighbor hoppings depend solely on moments of the hybridization function $\Delta(\eps)$ over an (in principle) infinite set of energy bins spanning ranges of equal width in $\log|\eps - \eps_F|$.

Previous numerical investigations of quantum impurities embedded in electronic systems that lack translational symmetry have relied upon obtaining the eigenenergies and eigenstates using exact diagonalization (ED), an approach that has restricted the studies to small system sizes (e.g. Refs.\ \onlinecite{PhysRevLett.115.036403,PhysRevB.85.115112}).
An NRG treatment of quantum impurities in such a host typically requires the ED energy levels to be artificially broadened so that $\Delta(\eps)$ has nonvanishing moments in energy bins arbitrarily close to the Fermi energy. This broadening washes out any singular structure in the energy spectrum (such as that expected, for example, in a fractal host) and ensures that the NRG treatment reveals the low-energy Kondo physics expected in a metallic host.

The KPM+NRG approach described in the remainder of this appendix avoids ED by writing the hybridization function in terms of a KPM expansion that allows one to reach large system sizes of order $10^6$ sites, regardless of the spatial dimensionality. The KPM representation combines well with the NRG because it allows the parameters of the Wilson tight-binding chain to be computed efficiently without performing any numerical integration to find moments of $\Delta(\eps)$. We validate this new technique through comparisons of (1) KPM+NRG Wilson-chain coefficients with those obtained analytically or through other numerical means for simple algebraic forms of the $\Delta(\eps)$, and (2) observables involving impurity degrees of freedom with density-matrix renormalization-group (DMRG) results for one-dimensional host systems up to size $L=500$.

\subsection{Wilson-chain mapping}
 \label{subsec:chain-mapping}
 
This Appendix briefly reviews the NRG mapping of a discretized version of a host band Hamiltonian to a semi-infinite tight-binding Wilson chain. As described, for instance, in Ref.\ \onlinecite{RevModPhys.80.395}, this mapping transforms Eq.\ \eqref{eq:H_host-bar} to
\begin{equation}
    \tH_\host = \sum_{n=0}^{\infty} \sum_\sigma \Bigl[ \veps_n f_{n\sigma}^{\dag} f_{n\sigma}
    + t_n \bigl( f_{n\sigma}^{\dag} f_{n+1,\sigma} + \text{H.c.} \bigr) \Bigr] .
\end{equation}
The tight-binding coefficients are defined via a set of recursion relations
\begin{subequations}
\label{eq:recursion}
\begin{align}
\label{eq:veps_n}
    \veps_n& = \sum_{m=0}^{\infty} \bigl( \eps_m^+ \, u_{nm}^2 + \eps_m^- \, v_{nm}^2 \bigr) \\
\label{eq:t_n}
    t_n u_{n+1,m}& = (\eps_m^+ - \veps_n) u_{nm} + t_{n-1} u_{n-1,m}, \\
    t_n v_{n+1,m}& = (\eps_m^- - \veps_n) v_{nm} + t_{n-1} v_{n-1,m}, \\
\label{eq:u,v-normalize}
    1& = \sum_{m=0}^{\infty} \bigl( u_{n+1,m}^2 + v_{n+1,m}^2 \bigr) ,
\end{align}
\end{subequations}
where $t_{-1} = 0$ and 
\begin{equation}
    \eps_m^\pm = \beta_m^\pm / \alpha_m^\pm
\end{equation}
with $\alpha_m^\pm$ and $\beta_m^\pm$ as defined in Eqs.\ \eqref{eq:alpha,beta}. Equations \eqref{eq:recursion} are iterated starting from $n = 0$ with
\begin{equation}
\label{eq:u,v_0m}
    u_{0m} = \sqrt[\leftroot{-4}\uproot{6}+]{\alpha_m^+ / A}, \quad
    v_{0m} = \sqrt[\leftroot{-4}\uproot{6}+]{\alpha_m^- / A},
\end{equation}
where
\begin{equation}
\label{eq:A}
    A = \sum_m (\alpha_m^+ + \alpha_m^-) = \int_{-1}^1 \tDelta(\teps) \, d\teps ,
\end{equation}
a quantity that equals $1$ if $\tDelta(\teps)$ is unit-normalized as we have assumed.
Substituting Eqs. \eqref{eq:u,v_0m} into Eq.\ \eqref{eq:veps_n} yields
\begin{equation}
\label{eq:veps_0}
    \veps_0 = A^{-1} \sum_m (\beta_m^+ + \beta_m^-) = \frac{D^2}{\pi V^2} \int_{-1}^1 \teps \, \tDelta(\teps) \, d\teps .
\end{equation}
The values of all other tight-binding coefficients $\veps_n$ and $t_n$ depend on the band discretization parameter $\Lambda$. 
Truncating the Wilson chain at $N+1$ sites labeled $0 \le n \le N$ yields the reduced Hamiltonian $\sum_{\sigma} H_{0,N,\sigma}$ with $H_{0,N,\sigma}$ defined in Eq.\ \eqref{eq:H_N,sigma}.

\subsection{KPM+NRG Formulation}
\label{subsec:KPM+NRG-formulation}

In order to apply the NRG method, one needs to calculate $\alpha_m^\pm$ and $\beta_m^\pm$ defined in Eqs.\ \eqref{eq:alpha,beta}, i.e., the zeroth and first moments of $\tDelta(\teps)$ over reduced energy bins $\teps_{m+1} < \pm \teps < \teps_m$ with $\teps_m$ given in Eq.\ \eqref{eq:teps_m}. Using $T_n(x) = \cos(n\arccos x)$, and defining $\theta_m = \arccos\teps_m$ with $0\le \theta_m \le \pi/2$ for $m = 0,\ , 1, \, 2, \, \ldots$, one can show that
\begin{widetext}
\begin{equation}
    \pm\int_{\pm\teps_{m+1}}^{\pm\teps_m} \frac{T_n(\teps)}{\sqrt{1 - \teps^2}} \, d\teps =
    \begin{cases}
        \theta_{m+1} - \theta_m & \text{for } n = 0, \\[1ex]
        {\displaystyle\frac{(\pm 1)^n}{n}} ( \sin n\theta_{m+1} - \sin n\theta_m ) & \text{for } n > 0 ,
    \end{cases}
\end{equation}
and
\begin{equation}
    \pm\int_{\pm\teps_{m+1}}^{\pm\teps_m} \frac{\teps \, T_n(\teps)}{\sqrt{1 - \teps^2}} \, d\teps =
    \begin{cases}
        \pm (\sin\theta_{m+1} - \sin\theta_m) & \text{for } n = 0, \\[1ex]
        {\displaystyle\frac{1}{4}} ( \sin 2\theta_{m+1} - \sin 2\theta_m ) + {\displaystyle\frac{1}{2}} (\theta_{m+1} - \theta_m) & \text{for } n = 1, \\[2ex]
        {\displaystyle\frac{(\pm 1)^{n-1}}{2(n-1)}} \Bigl[ \sin(n-1)\theta_{m+1} - \sin(n-1)\theta_m \Bigr] & \\[2ex]
        \quad + \; {\displaystyle\frac{(\pm 1)^{n-1}}{2(n+1)}} \Bigl[ \sin(n+1)\theta_{m+1} - \sin(n+1)\theta_m \Bigr] & \text{for } n > 1 .
    \end{cases}
\end{equation}
Combining these results with Eqs.\ \eqref{eq:alpha,beta} and \eqref{eq:Delta-expansion} yields
\begin{equation}
\label{eq:alpha_m-KPM}
    \alpha_m^{\pm} = \frac{1}{\pi} \biggl[ g_0 \mu_0 (\theta_{m+1}-\theta_m) + 2 \sum_{n=1}^{N_C-1} \frac{(\pm 1)^n}{n} g_n \mu_n ( \sin n\theta_{m+1}-\sin n\theta_m ) \bigg]
\end{equation}
and
\begin{equation}
\label{eq:beta_m-KPM}
    \begin{split}      
    \beta_m^{\pm} & = \frac{1}{\pi} \biggl[ g_1 \mu_1 ( \theta_{m+1} - \theta_m ) + \sum_{n=1}^{N_C-2} \frac{(\pm 1)^n}{n} ( g_{n-1} \mu_{n-1} + g_{n+1} \mu_{n+1} ) \, ( \sin n\theta_{m+1} - \sin n\theta_m ) \\
    & \qquad \quad + \; \sum_{n=N_C-1}^{N_C} \frac{(\pm 1)^n}{n} g_{n-1} \mu_{n-1} ( \sin n\theta_{m+1} - \sin n\theta_m ) \biggr].
    \end{split}
\end{equation}      
\end{widetext}
Equations \eqref{eq:alpha_m-KPM} and \eqref{eq:beta_m-KPM} can be inserted into Eqs.\ \eqref{eq:recursion}--\eqref{eq:veps_0} to yield the Wilson-chain coefficients $\veps_n$ and $t_n$. We note in particular that $\veps_0 = g_1 \mu_1$ and $A = g_0 \mu_0$.

The KPM+NRG method has two distinct advantages over other approaches. First, the KPM allows one to access much larger system sizes than can be treated using ED and related techniques. Second, hybridization function moments that fully determine the Wilson-chain parameters can be expressed as sums over KPM coefficients weighted by trigonometric functions, without the need to perform numerical integration. However, truncating the KPM expansion at $N_C$ terms broadens spectral features located near the Fermi energy $\teps = 0$ over a width $\delta\teps = \pi/N_C$ (with a reduced broadening $\delta\teps = \pi/N_C^{3/2}$ near the band edges) \cite{RevModPhys.78.275}. Although this width is generally much smaller than the one that must be applied to ED calculations to allow application of the NRG, it nonetheless limits the length $N$ of the Wilson chain for which the coefficients $\veps_n$ and $t_n$ are faithfully reproduced, and thereby prevents access to the physics on energy and temperature scales smaller than of order $D\bLam^{-N/2}$.

\subsection{Comparison with Wilson-chain coefficients from direct integration}
\label{subsec:KPM+NRG-compare-coeffs}

One way to benchmark the KPM+NRG approach is to compare the Wilson-chain parameters it produces with ones for the same hybridization function obtained via other means. This section focuses on two examples, both of which involve particle-hole-symmetric hybridization functions and therefore have vanishing on-site coefficients $\veps_n$.

First we consider a flat-top hybridization function $\Delta(\omega)=\Delta_0 \Theta(D-|\eps|)$, where $\Theta(x)$ is the Heaviside function.
The reduced hybridization function, corresponding to Eq.\ \eqref{eq:Delta-power} with $r=0$, has KPM moments 
\begin{equation}
    \mu_n
    = \frac{1}{2} \int_{-1}^1 \!\! \cos(n\arccos x) \, dx
    = \begin{cases}
        (1 - n^2)^{-1} & \text{$n$ even} , \\
        0 & \text{$n$ odd}.
    \end{cases}
\end{equation}
Substituting these values into Eqs.\ \eqref{eq:alpha_m-KPM} and \eqref{eq:beta_m-KPM} yields values of $t_n$ that differ from their exact counterparts \cite{RevModPhys.47.773,RevModPhys.80.395}
\begin{equation}
\label{eq:xi_n-flat-DOS}
   t_n = \frac{(1+\Lambda^{-1}) (1-\Lambda^{-n-1}) \Lambda^{-n/2}}{2 \sqrt{1-\Lambda^{-2n-1}} \sqrt{1-\Lambda^{-2n-3}}}
\end{equation}
by a fractional error of less than $10^{-5}$ for expansion order $N_C=10^2$ and less than $10^{-9}$ for $N_C=10^5$.

\begin{table*}
\centering
\renewcommand{\arraystretch}{1.1}
\begin{tabular}{ r@{\hspace{2em}}p{2cm}@{\hspace{2em}}p{2cm}@{\hspace{2em}}p{2cm}@{\hspace{2em}}p{2cm}} 
\hline\hline
& \multicolumn{4}{c}{Scaled hopping coefficients \rule[-1.2ex]{0pt}{4ex}}  \\ 
\cline{2-5}
 $n$ & \multicolumn{1}{l}{$\;\;N_C=10^3$} & \multicolumn{1}{l}{\;\;$N_C=10^4$} & \multicolumn{1}{l}{\;\;$N_C=10^5$} & \multicolumn{1}{c}{direct \rule[-1.2ex]{0pt}{4ex}} \\
 \hline
0 & 1.0277400667 & 1.0277402378 & 1.0277402396 & 1.0277402396 \rule[-1.2ex]{0pt}{4ex}  \\
1 & 0.5598430497 & 0.5598153019 & 0.5598150205 & 0.5598150205  \\
2 & 1.0707215789 & 1.0707745240 & 1.0707750620 & 1.0707750620 \\
3 & 0.6180681897 & 0.6177961120 & 0.6177933421 & 0.6177933421  \\
4 & 1.0813785997 & 1.0818525992 & 1.0818574579 & 1.0818574579  \\
5 & 0.6270738676 & 0.6246306973 & 0.6246054709 & 0.6246054709  \\
6 & 1.0790024085 & 1.0831246862 & 1.0831683420 & 1.0831683420  \\
7 & 0.6462189195 & 0.6255939031 & 0.6253674691 & 0.6253674693 \\
8 & 1.0514730811 & 1.0829253734 & 1.0833149881 & 1.0833149885  \\
9 & 0.7591242845 & 0.6274630575 & 0.6254521939 & 0.6254521995  \\
10 & 0.9658350236 & 1.0799308710 & 1.0833312837 & 1.0833312949  \\
11 & 0.9359363668 & 0.6425740552 & 0.6254616893 & 0.6254616147  \\
12 & 0.9679116124 & 1.0567121630 & 1.0833331639 & 1.0833331068  \\
13 & 0.9834174258 & 0.7411020836 & 0.6254629740 & 0.6254626609  \\
14 & 0.9921226323 & 0.9725570448 & 1.0833303929 & 1.0833333082 \\
15 & 0.9965841359 & 0.9258040504 & 0.6253929505 & 0.6254627771  \\
16 & 0.9986021169 & 0.9645272603 & 1.0831458118 & 1.0833333305  \\
17 & 0.9994498243 & 0.9811130281 & 0.6244860070 & 0.6254627900  \\
18 & 0.9997890276 & 0.9908420551 & 1.0851426469 & 1.0833333330  \\
19 & 0.9999205683 & 0.9959756211 & 0.6791323218 & 0.6254627914 \\
20 & 0.9999705003 & 0.9983374119 & 0.9782897997 & 1.0833333333 \\
21 & 0.9999891612 & 0.9993415615 & 0.9055070894 & 0.6254627916  \\
22 & 0.9999960521 & 0.9997464420 & 0.9574932824 & 1.0833333333  \\
23 & 0.9999985724 & 0.9999042423 & 0.9767188899 & 0.6254627916  \\
24 & 0.9999994869 & 0.9999643537 & 0.9885773919 & 1.0833333333  \\
25 & 0.9999998166 & 0.9999868784 & 0.9949382199 & 0.6254627916  \\
26 & 0.9999999347 & 0.9999952133 & 0.9978956972 & 1.0833333333  \\
27 & 0.9999999769 & 0.9999982669 & 0.9991632833 & 0.6254627916  \\
\hline\hline
\end{tabular}
\caption{\label{tab:t_n-linear-DOS}%
Scaled Wilson-chain hopping coefficients $2 \Lambda^{n/2} t_n / (1 + \Lambda^{-1})$ for a linear-pseudogapped hybridization function [Eq.\ \eqref{eq:Delta-power} with $r=1$] discretized using $\Lambda = 3$ and $z=1$. Columns labeled with values of $N_C$ contain hoppings obtained using the KPM+NRG approach described in this appendix. The last column contains directly obtained hoppings, adapted from Ref.\ \onlinecite{PhysRevB.57.14254} (where the quantity denoted $t_n$ in this paper is instead written $\tau_{n+1}$).}
\end{table*}

A second example is the linear-pseudogapped hybridization function $\Delta(\omega)=\Delta_0|\omega/D| \, \Theta(D-|\eps|)$ that serves as a simplified model for two-dimensional Dirac semimetals such as graphene. This case corresponds to Eq.\ \eqref{eq:Delta-power} with $r=1$. Its KPM moments are
\begin{align}
    \mu_n & = \int_{-1}^1 \!\! |x| \cos(n\arccos x) \, dx \notag \\
    & = \begin{cases}
        (4 - n^2) ^{-1} & \text{if \: $n\:\text{mod}\:4 = 0$}, \\
        0 & \text{otherwise}.
    \end{cases}
\end{align}
Table \ref{tab:t_n-linear-DOS} lists the scaled hopping coefficients $2 \Lambda^{n/2} t_n / (1 + \Lambda^{-1})$ for $\Lambda = 3$ and $z = 1$ obtained by substituting the KPM moments into Eqs.\ \eqref{eq:alpha_m-KPM} and \eqref{eq:beta_m-KPM} and truncating the sums after $N_C = 10^3$, $10^4$, and $10^5$ terms. Also listed are the coefficients determined via direct computation of the integrals in Eqs.\ \eqref{eq:alpha,beta}, which can be considered exact to the precision shown. With increasing $n$, the KPM+NRG coefficients initially follow the alternating pattern of their exact counterparts [see Eq.\ \eqref{eq:tstar}] before crossing over for larger $n$ to approach $2 \Lambda^{n/2} t_n / (1 + \Lambda^{-1}) = 1$, the value characteristic of the flat-top hybridization function discussed earlier in this section. The crossover, which arises from the finite KPM energy resolution, takes place around the $n$ for which $\Lambda^{-n/2} \simeq \pi/N_C$ or $n\simeq 2\log(\pi/N_C) / \log\Lambda$. The KPM scaled hoppings for $N_C = 10^5$ differ from the exact values by less than $10^{-10}$ for all $n \le 6$; for $n\ge 7$, errors gradually grow, but the overall structure of the exact coefficients is preserved until $n\simeq 18$.

This second example provides evidence that the KPM+NRG accurately reproduces the Wilson-chain description of a hybridization function having nontrivial energy dependence down to the broadening energy scale associated with truncation of the KPM expansion at finite order $N_C$.

\subsection{Comparison with density-matrix renormalization-group results}
\label{subsec:KPM+NRG-compare-DMRG}

\begin{figure}[t!]
\begin{center}
\includegraphics[width = 0.49\textwidth]{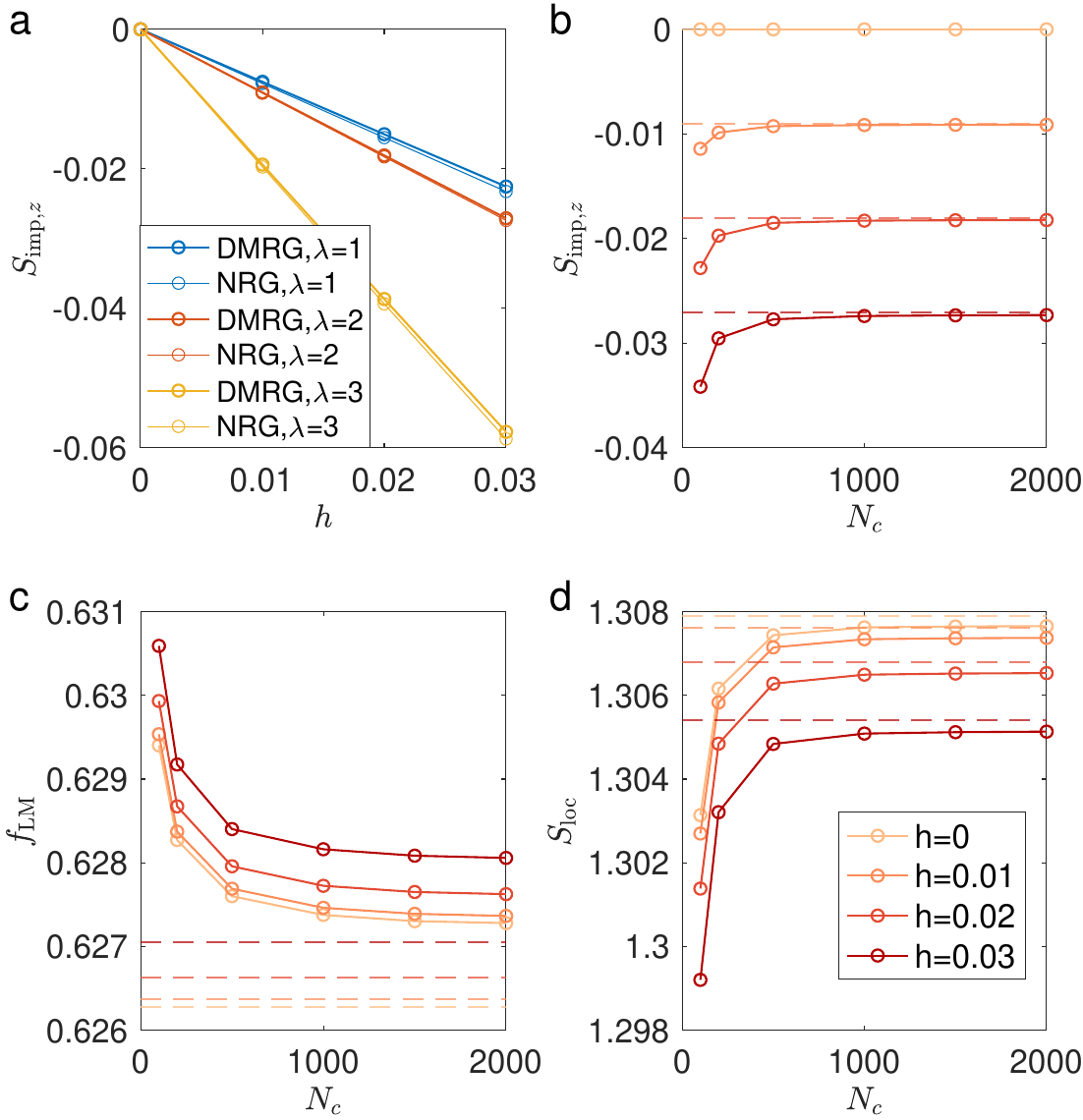}
\caption{\label{fig:compare-DMRG-Lambda=1.5}
Comparison between ground-state expectation values for the AAA model calculated using the KPM+NRG and DMRG methods. Data for an Anderson impurity having $U = -2\eps_d = D$ hybridizing with the local DOS at the middle site of a single realization $\phi=0$ of an \AubAnd\ chain consisting of $L = 500$ sites at half filling. NRG results are for discretization $\Lambda = 1.5$. (a) Impurity spin $z$ component $S_{\imp,z}$ as a function of applied field $h$ for $\lambda/t = 1$ (delocalized), $2$ (critical), and $3$ (localized), with KPM+NRG results being for $N_C=10^3$. (b)--(d) KPM+NRG expectation values vs KPM expansion order $N_C$ at $\lambda=2t$ and different local magnetic fields listed in the legend of the lower-right panel. At $N_c = 2000$, fractional differences from DMRG values (dashed horizontal lines) are smaller than 0.8\% for $S_{\imp,z}$, 0.16\% for $f_{\text{LM}}$, and $0.02\%$ for $S_{\loc}$.}
\end{center}
\end{figure}

In this section, physical observables computed using the KPM+NRG approach are compared with ones obtained via the DMRG method. These numerical methods have different strengths and weaknesses. KPM+NRG allows treatment of large systems (up to $L = 10^6$ in this work) but it is limited to temperatures exceeding a scale set by the finite KPM energy resolution [Eq.\ \eqref{eq:KPM-res}] and is affected by the discretization and truncation errors that are inherent to the NRG. The DMRG is subject to none of the aforementioned limitations, but its application is restricted to much smaller systems where the finite level spacing imposes a lower bound on the temperatures that can be reliably accessed. The comparisons reported in this section were all made for \AubAnd\ chains of length $L = 500$.

The observables that we compare are ground-state expectation values of operators involving impurity degrees of freedom: the impurity spin $z$ component \cite{Note2}
\begin{equation}
    S_{\imp,z} = \frac{1}{2} ( \hat{n}_{d\up} - \hat{n}_{d\dn} ),
\end{equation}
the local-moment fraction
\begin{equation}
    f_\text{LM} = \hat{p}_\up + \hat{p}_\dn
\end{equation}
where $p_\sigma = \hat{n}_{d\sigma} - \hat{p}_{\up\dn}$ with $\hat{p}_{\up\dn} = \hat{n}_{d\up} \hat{n}_{d\dn}$, and the local entanglement entropy
\begin{equation}
    S_\loc = -\sum_{i=0,\up,\dn,\up\dn} \hat{p}_i \ln \hat{p}_i
\end{equation}
with $\hat{p}_0=1-\hat{p}_\up-\hat{p}_\dn-\hat{p}_{\up\dn}$. 

Figure \ref{fig:compare-DMRG-Lambda=1.5} compares ground-state expectation values of the above operators computed using the KPM+NRG with $\Lambda=1.5$ with ones obtained using the DMRG. The top-left panel shows the variation of $\langle S_{\imp,z}\rangle$ with the local field $h$ for values of the AA potential strength $\lambda = t$ (delocalized phase), $\lambda = 2t$ (critical point), and $\lambda = 3t$ (localized phase), while the remaining panels focus on the evolution of KPM+NRG values with increasing KPM expansion order $N_C$. As detailed in the legend, the NRG results for $N_C = 2000$ all lie within 1\% of the DMRG values. This agreement provides evidence for the efficacy of the KPM+NRG approach.

\section{Self-similar hybridization functions}
\label{app:self-similar}

This appendix presents some details of our analysis of the uniform $1/5$ Cantor-set hybridization function $C(5)$ that are referenced in Sec.\ \ref{subsec:Cantor-chains} and also briefly discusses two other classes of self-similar hybridization functions, one fractal and the other nonfractal.

\subsection{Wilson-chain mapping of the $C(5)$ hybridization function}
\label{subsec:5-3-chain-mapping}

Section \ref{subsubsec:Cantor-Lambda=4M+1} demonstrates an exact equivalence for Fermi energy $\eps_F = 0$ between the $\Lambda=4M+1$, $z=1$ NRG treatments of a uniform $1/(4M+1)$ Cantor-set hybridization function $\Delta_{C(4M+1)}$ having foneractal dimension $D_{C(4M+1)}$ and a continuous hybridization function $\Delta_{P(r)}$ that diverges at $\teps=0$ according to a power $r = D_{C(4M+1)}-1$. As noted in Sec.\ \ref{subsubsec:Cantor-Lambda-to-1}, the equivalence arises due to a perfect alignment of the NRG energy bins with the self-similarity of the fractal hybridization function about the Fermi energy. This alignment is broken when the NRG discretization $\Lambda$ or the offset $z$ is changed.

In this section, we focus on the effect of reducing $\Lambda$ toward its continuum limit of $1$ at fixed $z=1$.
For the purposes of analysis, it is convenient to consider the sequence of discretizations $\Lambda_k$ defined in Eq.\ \eqref{eq:Lambda-seq}, where bin $m$ for $\Lambda=\Lambda_k$ is subdivided to form bins $2m$ and $2m+1$ for $\Lambda=\Lambda_{k+1}$, making it relatively simple to determine the $l\to\infty$ limits of $\alpha_m$ and $\beta_m$.

We illustrate the first few steps in the $\Lambda_k$ sequence for $M=1$ and $z=1$. For $\Lambda=\Lambda_1=5^{1/2}$, all bins numbered $m=2m'+1$ (with $m'$ a non-negative integer) fall entirely within a gap of $\tDelta_{C(5)}(\teps)$. This is reflected in
\begin{equation}
\label{eq:alpha,beta-Lambda=5r2}
\begin{aligned}
    \alpha_m &= \frac{1}{3^{\,m'+1}}, \;\;
    \beta_m = \frac{4}{15^{\,m'+1}} && m=2m', \\[0.5ex]
    \alpha_m &= \beta_m = 0 && m=2m'+ 1 .
\end{aligned}
\end{equation}

Next consider $\Lambda=\Lambda_2=5^{1/4}$. The integrated weight of $\tDelta_{C(5)}$ over each even-numbered $\Lambda_1$ bin is divided in the ratio 20:7 between $\Lambda_2$ bins $m=4m'$ and $4m'+1$. Since odd-numbered $\Lambda_1$ bins fall entirely in gaps, so too do $\Lambda_2$ bins numbered $m=4m'+2$ and $m=4m'+3$. One finds
\begin{equation}
\label{eq:alpha,beta-Lambda=5r4}
\begin{aligned}
    \alpha_m &= \frac{20}{3^{\,m'+4}}, \;\; 
    \beta_m = \frac{10744}{15^{\,m'+4}} && m = 4m', \\[0.5ex]
    \alpha_m &= \frac{7}{3^{\,m'+4}}, \;\;
    \beta_m = \frac{2756}{15^{\,m'+4}} && m=4m'+1, \\[0.5ex]
    \alpha_m &= \beta_m=0, && \text{otherwise}.
\end{aligned}
\end{equation}

The last case we consider explicitly is $\Lambda=\Lambda_3=5^{1/8}$, where 5 bins out of every consecutive sequence of 8 lie entirely within gaps. The integrated weight of $\tDelta_{C(5)}$ over $\Lambda_2$ bin $m=4m'$ is now split 3:2 between $\Lambda_3$ bins $m=8m'$ and $8m'+1$, while the entire weight that falls in $\Lambda_2$ bin $m=4m'+1$ is inherited by $\Lambda_3$ bin $m=8m'+2$. Then
\begin{equation}
\label{eq:alpha,beta-Lambda=5r8}
\begin{aligned}
    \alpha_m &= \frac{12}{3^{\,m'+4}}, \;\;
    \beta_m = \frac{6960}{15^{\,m'+4}} && m = 8m', \\[0.5ex]
    \alpha_m &= \frac{8}{3^{\,m'+4}}, \;\;
    \beta_m = \frac{3784}{15^{\,m'+4}} && m = 8m'+1, \\[0.5ex]
    \alpha_m &= \frac{7}{3^{\,m'+4}}, \;\;
    \beta_m = \frac{2756}{15^{\,m'+4}}, && m = 8m'+2, \\[0.5ex]
    \alpha_m &= \beta_m = 0 && \text{otherwise}.
\end{aligned}
\end{equation}

In the preceding examples and for still larger values of $k$, for any $m\ge 0$ such that $\alpha_m>0$, the ratios $\alpha_{m+2^k}/\alpha_m$ and $\beta_{m+2^k}/\beta_m$ are identical to their $k=0$ counterparts $\alpha_{m+1}/\alpha_m$ and $\beta_{m+1}/\beta_m$ deduced from Eqs.\ \eqref{eq:alpha,beta-Lambda=4M+1}. In this respect, $k>0$ members of the sequence $\Lambda_k$ still reflect the power-law divergence of the coarse-grained hybridization function that is so readily apparent for $k=0$. With increasing $k$, however, the self-similar gap hierarchy of the fractal hybridization function becomes increasingly apparent through $\alpha_m$ and $\beta_m$ moments that (a) for certain values of $m\mod 2^k$, vanish for every value of $m'=\lfloor m/2^k \rfloor$, (b) where nonvanishing, have scaled values $3^{m'}\alpha_m$ and $15^{m'}\beta_m$ that vary with $m \mod 2^k$.

\begin{figure}[t!]
\centering
\includegraphics[width = 0.49\textwidth]{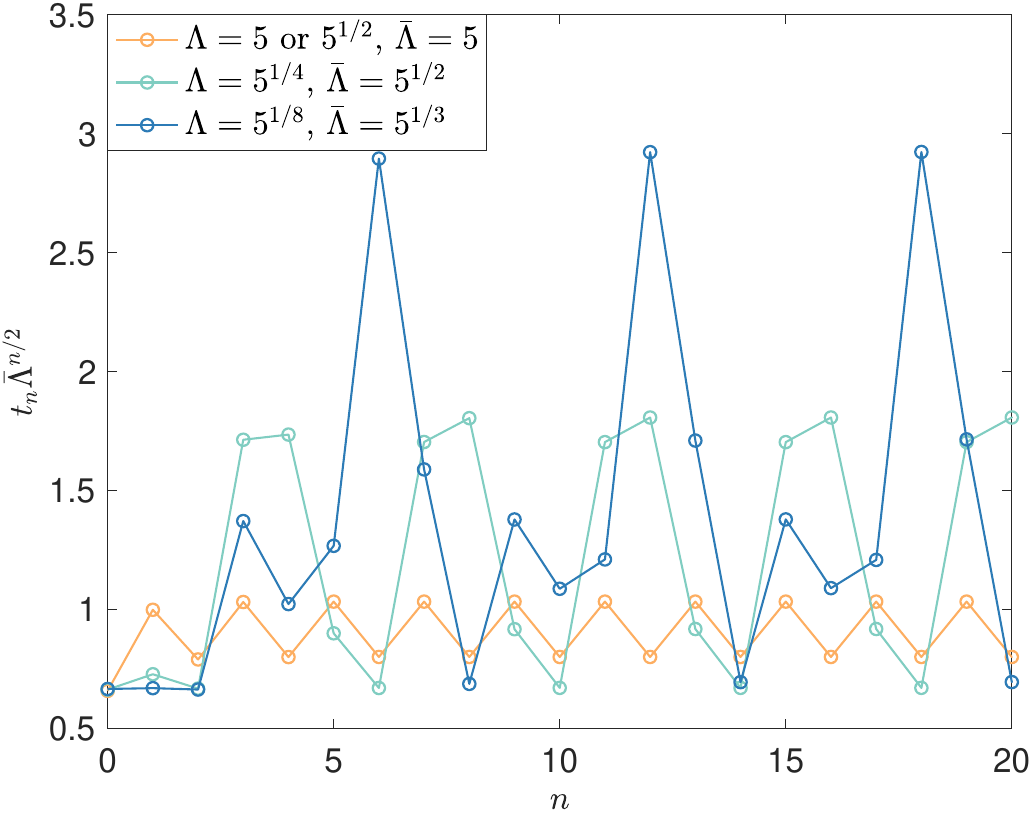}
\caption{\label{fig:5-3-Cantor-hopping}%
Scaled Wilson-chain hopping parameters $\xi_n=\bLam^{n/2} t_n$ for different NRG discretizations $\Lambda=5^{1/2^k}$, $z=1$ of a uniform $1/5$ Cantor-set hybridization function. The legend lists the value of $\bLam$ corresponding to each $\Lambda$.}
\end{figure}

One can apply Eqs.\ \eqref{eq:recursion}--\eqref{eq:A} to convert the set of bin moments $\alpha_m$ and $\beta_m$ into a set of Wilson-chain coefficients $t_n$ and $\veps_n$. The on-site energies $\veps_n$ necessarily vanish due to the particle-hole symmetry of $\tDelta_{C(4M+1)}$.
Figure \ref{fig:5-3-Cantor-hopping} plots the scaled hopping coefficients $\xi_n$, defined through Eq.\ \eqref{eq:xi_n}, for the first four discretizations $\Lambda=5^{1/2^k}$, $z=1$ of the $C(5)$ hybridization function. The figure legend specifies the value of $\bLam$ corresponding to each $\Lambda$. For $\Lambda = 5$, Eq.\ \eqref{eq:tstar} implies that
\begin{equation}
\label{eq:xi_n-periodicity}
    \lim_{n\to\infty} \xi_{n+P} = \xi_n
\end{equation}
with $\bLam=\Lambda$ and period $P=2$. For $\Lambda=5^{1/2}$, the hopping parameters $t_n$ (as well as their scaled counterparts $\xi_n$) are identical to those for $\Lambda=5$. Formally, this conclusion follows from (i) the structure of Eqs.\ \eqref{eq:recursion}--\eqref{eq:A}, where $\alpha_m^+ = 0$ leads to $u_{mn}=0$ and $\alpha_m^- = 0$ leads to $v_{mn}=0$, so bins with $\alpha_m^\pm$ can be completely disregarded in the computation of $t_n$ and $\veps_n$, and (ii) the fact that the nonzero moments in Eqs.\ \eqref{eq:alpha,beta-Lambda=5r2} can be made identical to those in Eqs.\ \eqref{eq:alpha,beta-Lambda=4M+1} under a simple re-indexing $m\to m/2$. By contrast, support of $\tDelta_{C(5)}(\teps)$ within the energy range $\Lambda^{-(m'+1)} < \teps < \Lambda^{-m'}$ spanned by one $\Lambda=5$ energy bin is split over $q=2$ bins for $\Lambda=5^{1/4}$. This imparts additional structure to the scaled hopping coefficients, which now obey Eq.\ \eqref{eq:xi_n-periodicity} with $\bLam=5^{1/2}$ and $P=4$. The same pattern holds for $\Lambda=5^{1/8}$, with $q=3$, $\bLam=5^{1/3}$, and $P=6$.

\subsection{$C(4M+3)$ fractal hybridization functions}
\label{subsec:Cantor-4M+3}

Section \ref{sec:Cantor} focuses on a reduced hybridization functions $\Delta_{C(4M+1)}(\eps)$ ($M = 1$, $2$, $\ldots$) describing uniform $1/(4M+1)$ Cantor sets. These hybridization functions can be constructed by iteration of a finite subdivision rule prescribed in Sec.\ \ref{subsec:fractal-host}.

\begin{figure}[b!]
\centering
\includegraphics[width = 0.48\textwidth]{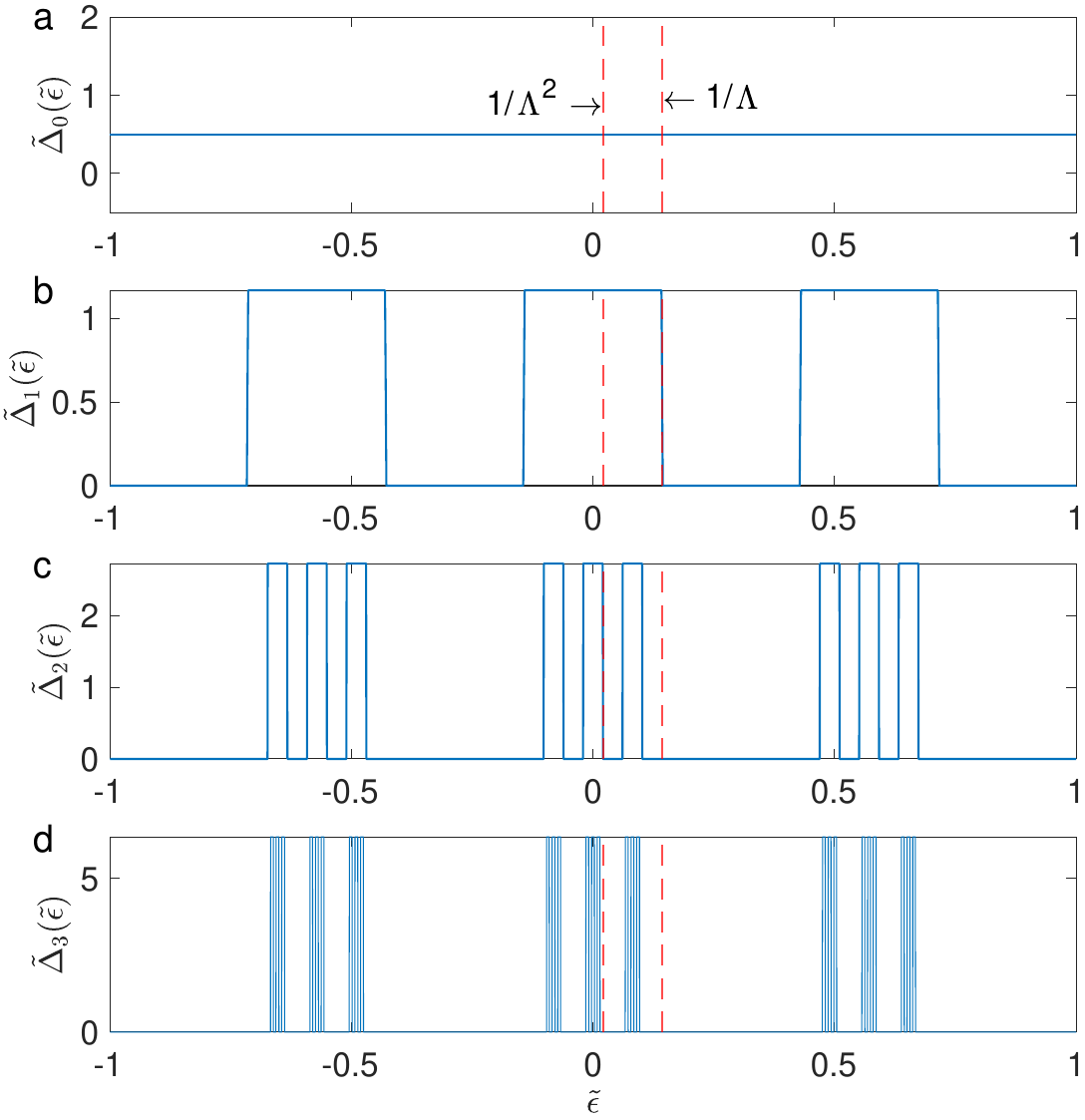}
\caption{\label{fig:7-3-Cantor-DOS}%
Reduced hybridization functions $\tDelta_l(\teps)$ approximating a fractal $1/7$ Cantor set: (a) Uniform initial hybridization function $\tDelta_0(\teps)$. (b)-(d) First three approximants $\Delta_l(\teps)$ formed by iteratively dividing each interval into seven equal parts labeled 1 through 7 and removing the four odd-numbered parts. The vertical red dashed lines mark the lower bounds $\Lambda^{-m}, \, m = 1, \, 2$ of the first two logarithmic bins in the NRG discretization of the hybridization function for $\Lambda=7$ and $z=1$.}
\end{figure}

This section addresses properties of a class of uniform $1/(4M+3)$ Cantor-set hybridization functions $\Delta_{C(4M+3)}(\eps)$, where $M$ is a positive integer. These functions may be constructed by iteration of a finite subdivision rule in which each energy range over which $\Delta_{l-1}(\eps)>0$ is divided into $4M+3$ equal parts, labeled $1$ through $4M+3$ in order of ascending energy. In order that $\Delta_{C(4M+3)}(\eps)$ has nonvanishing weight arbitrarily close to $\eps=0$, one sets $\Delta_l(\eps)=0$ throughout each of the $2M+2$ odd-numbered intervals and sets $\Delta_l(\eps) = (4M+3)(2M+1)^{-1} \Delta_{l-1}(\eps)$ throughout the $2M+1$ even-numbered intervals so that $\int_{-D}^D \Delta_l(\eps) \, d\eps = \pi V^2$ for all $l$. Figure \ref{fig:7-3-Cantor-DOS} shows the first three iterations of this process for the case $M=1$. Whereas $\Delta_{C(4M+1)}(\eps)$ illustrated in Fig.\ \ref{fig:5-3-Cantor-DOS} retains the half-bandwidth $D$ of its zeroth-order approximant, $\Delta_{C(4M+3)}(\eps)$ has a smaller half-bandwidth $[1-2\sum_{l=1}^{\infty} (4M+3)^{-l}] D = [2M / (2M+1)]D$. Nonetheless, there are close parallels between the two classes of hybridization function.

Let us consider the NRG mapping of $\tDelta_{C(4M+3)}$ for $\Lambda=4M+1$ and $z=1$. Figure \ref{fig:7-3-Cantor-DOS} shows that the bin boundary $\teps_m$ defined in Eq.\ \eqref{eq:teps_m} lies at the edge of a retained interval in the $l=m$ approximant but lies in a gap for all $l>m$. As a consequence, the integrals $\alpha_m=\alpha_m^\pm$ and $\beta_m=\beta_m^\pm$ [Eq.\ \eqref{eq:alpha,beta}] cease to change once $l > m$. Due to the self-similarity of $\tDelta_{C(4M+3)}$ under $\teps\to\teps/(4M+3)$, one can readily show that for $l\to\infty$ and for $m\ge 0$,
\begin{equation}
\label{eq:alpha,beta-Lambda=4M+3}
\begin{aligned}
    \alpha_m[C(4M+3)]&
      = \frac{M}{(2M+1)^{m+1}} , \\
    \beta_m[C(4M+1)]&
      = \frac{2M(M+1)}{[(2M+1)(4M+3)]^{m+1}}.
\end{aligned}
\end{equation}
We note that $\alpha_m[C(4M+3)]$ is identical to $\alpha_m[P(r)]$ in Eqs.\ \eqref{eq:alpha,beta-power-law} for $\Lambda=4M+3$ provided that the power entering Eq.\ \eqref{eq:Delta-power} is chosen to be
\begin{equation}
    r = \frac{\log(2M+1)}{\log(4M+3)} - 1 = D_{C(4M+3)} - 1 ,
\end{equation}
where $D_{C(4M+3)}$ is the fractal dimension of $C(4M+3)$ given in Eq.\ \eqref{eq:D_Cbar}.
This choice also yields
\begin{align}
\label{eq:a_4M+3}
\frac{\beta_m[C(4M+3)]}{\beta_m[P(r)]}
    &= \frac{2M(M+1)}{4M^2+5M+1} \biggl[ 1 + \frac{\log(4M+3)}{\log(2M+1)} \biggr] \notag \\
    &\equiv a_{4M+3}.
\end{align}
Following arguments presented in Sec.\ \ref{sec:Cantor}, we conclude that the $\Lambda=4M+3$, $z=1$ treatment of the Anderson impurity model with a $C(4M+3)$ Cantor-set hybridization function yields the same properties (up to a suitable rescaling of all energy and temperature scales by the factor $a_{4M+3}$) as the corresponding treatment of a power-law hybridization function $\Delta_{P(D_{C(4M+3)}-1)}$.

Now we consider the next few steps in the sequence $\Lambda=(4M+3)^{1/k}$ for $M=1$, $z=1$, and $k=1$, 2, 3, $\ldots$. For $\Lambda=7^{1/2}$, all bins numbered $m=2m'+1$ (with $m'$ a non-negative integer) fall entirely within a gap of $\tDelta_{C(7)}(\teps)$, resulting in
\begin{equation}
\label{eq:alpha,beta-7-3-Lambda=5r2}
\begin{aligned}
    \alpha_m &= \frac{1}{3^{\,m'+1}}, \;\;
    \beta_m = \frac{4}{21^{\,m'+1}} && m=2m', \\[0.5ex]
    \alpha_m &= \beta_m = 0 && m=2m'+ 1 .
\end{aligned}
\end{equation}
Equations \eqref{eq:alpha,beta-7-3-Lambda=5r2} imply that the Wilson-chain coefficients $t_n$ and $\veps_n$ are identical for $\Lambda=7$ and $\Lambda=7^{1/2}$. This is analogous to the equivalence of $\Lambda=5$ and $5^{1/2}$ in the NRG treatment of the $C(5)$ Cantor set considered in Sec.\ \ref{subsec:5-3-chain-mapping}.

For $\Lambda=7^{1/3}$, bin boundary $\eps_1=\Lambda^{-1}$ falls in a gap of the $l=2$ approximant to $\tDelta_{C(7)}(\teps)$. The integrated weight of $\tDelta_{C(7)}$ over each $\Lambda=7$ bin is divided for $\Lambda=7^{1/3}$ in the ratio 2:1 between bins $m=3m'$ and $3m'+1$, while bins $m=3m'+2$ inherit zero weight. In this case
\begin{equation}
\label{eq:alpha,beta-7-3-Lambda=5r3}
\begin{aligned}
    \alpha_m &= \frac{2}{3^{\,m'+2}}, \;\; 
    \beta_m = \frac{60}{21^{\,m'+2}} && m = 3m', \\[0.5ex]
    \alpha_m &= \frac{1}{3^{\,m'+2}}, \;\;
    \beta_m = \frac{24}{21^{\,m'+2}} && m=3m'+1, \\[0.5ex]
    \alpha_m &= \beta_m=0, && m=3m'+2.
\end{aligned}
\end{equation}

The last case that we consider is $\Lambda=7^{1/4}$.
The weight that falls in each even-numbered $\Lambda=7^{1/2}$ bin is divided for $\Lambda=7^{1/4}$ in the ratio 1:2 between bins $m=4m'$ and $4m'+1$. Since odd-numbered $\Lambda=7^{1/2}$ bins fall entirely in gaps, so too do $\Lambda=7^{1/4}$ bins numbered $m=4m'+2$ and $m=4m'+3$. One finds
\begin{equation}
\label{eq:alpha,beta-7-3-Lambda=5r4}
\begin{aligned}
    \alpha_m &= \frac{1}{3^{\,m'+2}}, \;\; 
    \beta_m = \frac{32}{21^{\,m'+2}} && m = 4m', \\[0.5ex]
    \alpha_m &= \frac{2}{3^{\,m'+2}}, \;\;
    \beta_m = \frac{52}{21^{\,m'+2}} && m=4m'+1, \\[0.5ex]
    \alpha_m &= \beta_m=0, && \text{otherwise}.
\end{aligned}
\end{equation}

\begin{figure}[t!]
\centering
\includegraphics[width = 0.49\textwidth]{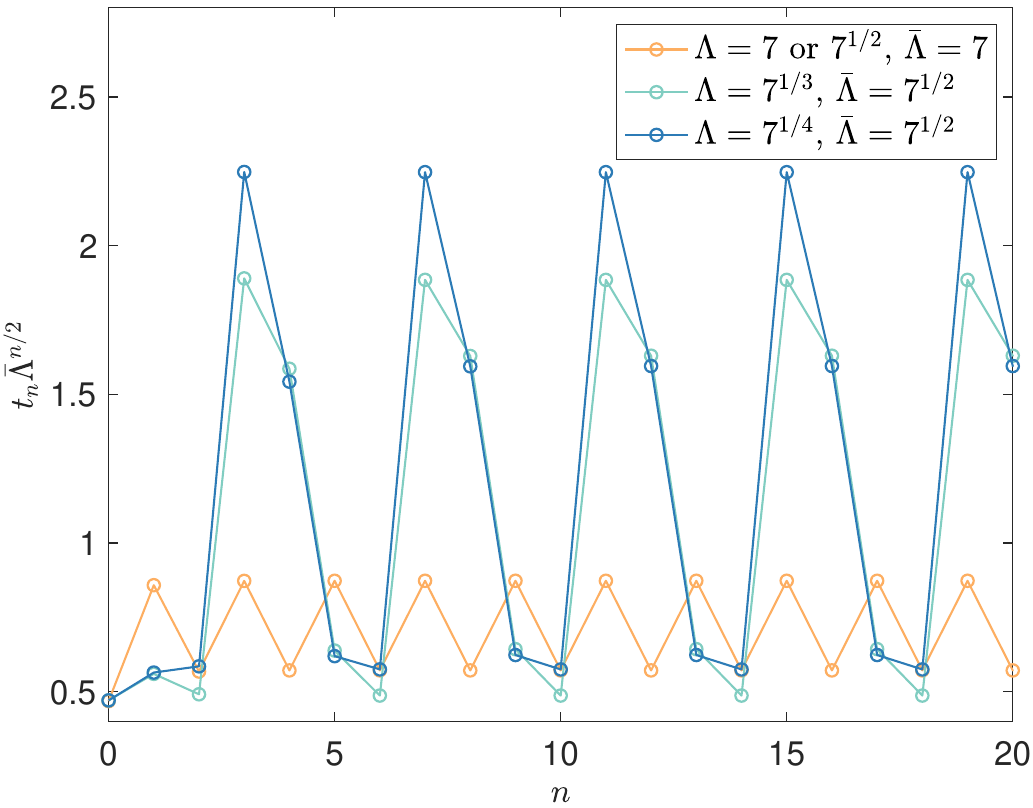}
\caption{\label{fig:7-3-Cantor-hopping}%
Scaled Wilson-chain hopping parameters $\xi_n=\bLam^{n/2} t_n$ for different NRG discretizations $\Lambda=7^{1/k}$, $z=1$ of a uniform $1/7$ Cantor-set hybridization function. The legend lists the value of $\bLam$ corresponding to each $\Lambda$.}
\end{figure}

Figure \ref{fig:7-3-Cantor-hopping} plots the scaled hopping coefficients $\xi_n$, defined through Eq.\ \eqref{eq:xi_n}, for the first four discretizations $\Lambda=7^{1/k}$, $z=1$ of the $C(7)$ hybridization function. The figure legend specifies the value of $\bLam$ corresponding to each $\Lambda$. The values of $\xi_n$ satisfy Eq.\ \eqref{eq:xi_n-periodicity} with $P=2$, $4$, and $4$ for $\Lambda = 7$ (equivalent to $7^{1/2}$), $7^{1/3}$, and $7^{1/4}$, respectively.

\subsection{Nonfractal self-similar hybridization functions}
\label{subsec:nonfractal}

In Section \ref{sec:Cantor}, we separate properties that can be attributed to fractality of the hybridization function from ones that arise purely due to self-similarity of $\tDelta(\teps)$ under rescaling of energies about the Fermi energy: $\teps\to\teps/b$ with $b>1$.

To this end, it is instructive to study the family of hybridization functions $\tDelta_{S(b)}(\teps)$ defined in Eq.\ \eqref{eq:Delta-log-bins}. Given the simple form of the hybridization function when viewed on a logarithmic energy axis, it is natural to start by considering the $\Lambda = b$, $z=1$ NRG mapping, for which
\begin{equation}
\label{eq:alpha,beta-S(b)-Lambda=b}
\begin{aligned}
    \alpha_m &= \frac{1}{2} (1-b^{-1}) b^{-m}, \\
    \beta_m & = \frac{1}{2} (1+b^{-1/2}) \, b^{-m} \, \alpha_m
\end{aligned}
\end{equation}
for $m = 0$, 1, 2, $\ldots$. The expression for $\alpha_m$ is identical to that for a flat-top hybridization function [i.e., $\tDelta_{P(0)}(\teps)$ defined through Eq.\ \eqref{eq:Delta-power} with $r = 0$]  when discretized using $\Lambda = b$, $z=1$, while $\beta_m(S(b))/\beta_m(P(0)) = (1 + b^{-1/2}) / (1 + b^{-1})$. We thus see that the $\Lambda=b$, $z=1$ treatment of the Anderson impurity model with an energy-dependent but nonfractal hybridization function that contains a discrete self-similarity scale $b>1$ yields the same properties (up to a suitable rescaling of all energy and temperature scales) as the corresponding treatment of an energy-independent (and therefore self-similar under any rescaling) hybridization.

The continuum limit can be approached via a sequence of NRG discretizations $\Lambda=b^{1/2k}$, $z=1$ for $k = 1$, 2, 3, $\ldots$. The support of $\tDelta_{S(b)}(\teps)$ is distributed among $k$ out of a total of $2k$ NRG bins that cover each energy interval $b^{-(m'+1)} < \teps \le b^{-m'}$ (with $m'$ being a non-negative integer). It is straightforward to show that
\begin{equation}
\label{eq:alpha,beta-S(b)}
\begin{aligned}
    \alpha_m &= \frac{1}{2} (1+b^{-1/2}) \, (1-b^{-1/2k}) \, b^{-m/2k}, \\
    \beta_m & = \frac{1}{2} (1+b^{-1/2k}) \, b^{-m/2k} \, \alpha_m
\end{aligned}
\end{equation}
for $m$ such that $2m'k \le m < (2m'+1)k$ with $m' = 0$, 1, 2, $\ldots$, while $\alpha_m=\beta_m = 0$ for $(2m'+1)k \le m < 2(m'+1)k$.

These bin moments $\alpha_m$ and $\beta_m$ can be processed through Eqs.\ \eqref{eq:recursion}--\eqref{eq:A} to obtain Wilson-chain coefficients. Due to the particle-hole symmetry of $\tDelta_{S(b)}(\teps)$, $\veps_n = 0$ for all $n$. The hopping coefficients, when scaled using Eq.\ \eqref{eq:xi_n} with $\bLam=\Lambda^2=b^{1/k}$, satisfy Eq.\ \eqref{eq:xi_n-periodicity} with $P=2k$. As is the case for the Cantor sets considered in Appendices \ref{subsec:5-3-chain-mapping} and \ref{subsec:Cantor-4M+3}, the coarse-grained equivalence between a self-similar hybridization function and a continuous power-law hybridization function breaks down upon approach to the continuum limit $\Lambda = 1$.

\section{Wilson-chain mapping of the critical \AubAnd\ Anderson model}
\label{app:AAA-NRG}

This appendix contains details about the Wilson-chain description of the full \AubAnd\ Anderson model, where the impurity hybridizes with the LDOS at a particular site in the host. We focus on the case of greatest interest, where the AA model is at its critical point, $\lambda=2t$.

\begin{figure}[t!]
\centering
\includegraphics[width=0.49\textwidth]{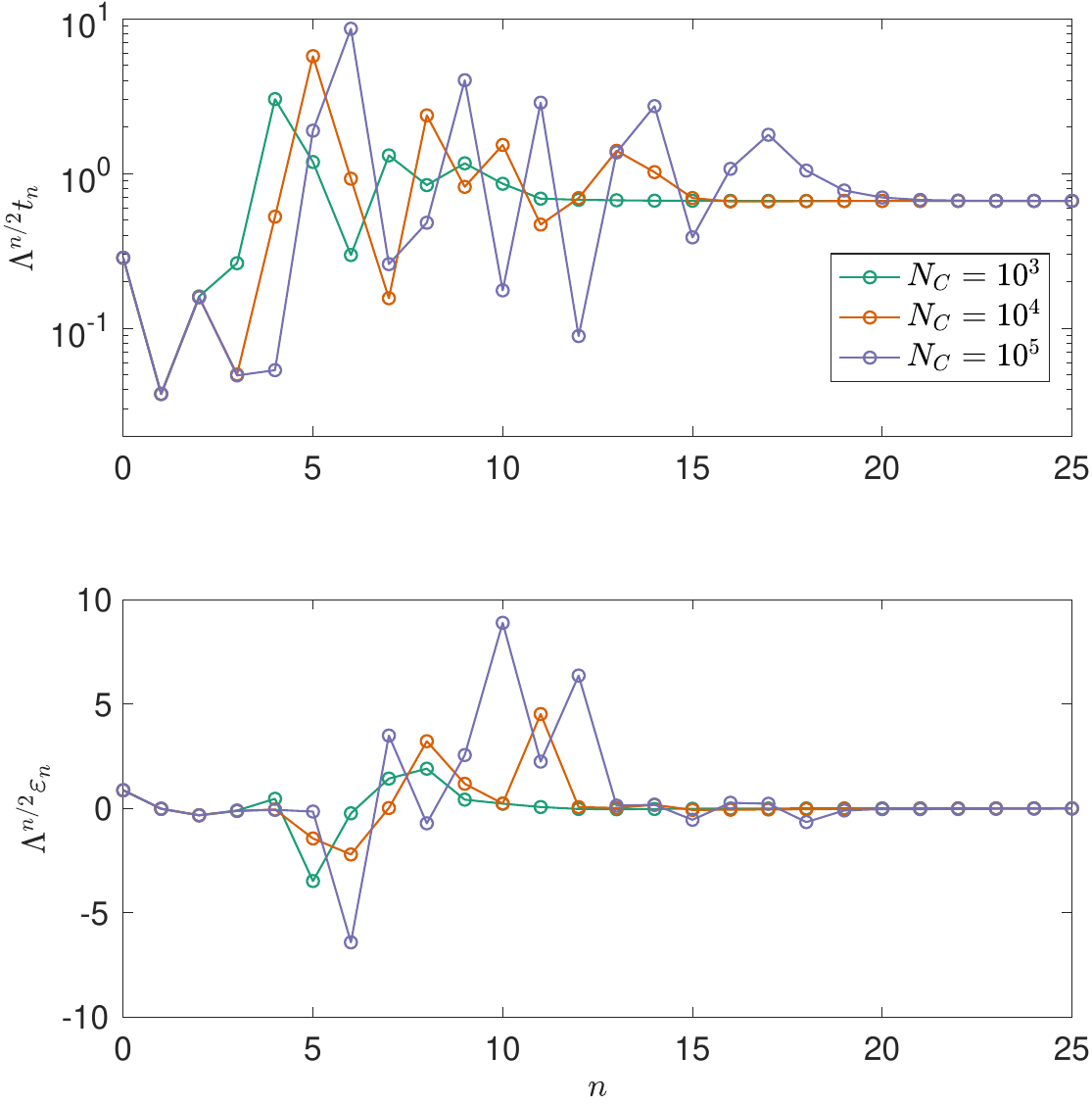}
\caption{\label{fig:AAA-hopping}%
Scaled Wilson-chain hopping coefficients (upper panel) and scaled onsite energies (lower panel) vs site index $n$ for an Anderson impurity hybridizing with the middle site of a critical \AubAnd\ chain. Data for $\lambda=2t$, $\phi=0$, $L=10^6$, $n_c = 0.309$, $\Lambda=3$, and different KPM truncation orders $N_C$ specified in the legend.}
\end{figure}

Figure \ref{fig:AAA-hopping} shows the scaled hopping $\Lambda^{n/2} t_n$ and scaled onsite energy $\Lambda^{n/2} \veps_n$ obtained in the KPM+NRG treatment of an impurity hybridizing with the middle site of the $\phi=0$ realization of the AA model at potential strength $\lambda=2t$ and band filling $n_c = 0.309$. These scaled tight-binding coefficients are plotted for three different KPM expansion orders: $N_C = 10^3$, $10^4$, and $10^5$. For a given $N_C$, the values of $\Lambda^{n/2} t_n$ initially fluctuate significantly as $n$ increases from $0$, rather reminiscent of the behavior of the scaled hopping coefficients for the fractal $1/5$ Cantor-set hybridization function shown in Fig.\ \ref{fig:5-3-Cantor-hopping}. For larger chain site indices $n$ where $\Lambda^{-n/2} \lesssim \pi/N_C$, $\Lambda^{n/2} t_n$ approaches an constant value arising from the KPM broadening of the hybridization function.
Focusing on a filling $n_c\ne 0.5$ in Fig.\ \ref{fig:AAA-hopping} allows $\veps_n$ to take nonzero values. The scaled onsite energy $\Lambda^{n/2} \veps_n$ also fluctuates with increasing $n$, but unlike the scaled hopping coefficient, its value is appreciable only for $5 < n < 14$, identifying the corresponding energy range $\Lambda^{-7} < |\teps| < \Lambda^{-5/2}$ as the one in which particle-hole asymmetry of the hybridization function plays the greatest role. The fluctuations of both $\Lambda^{n/2} t_n$ and $\Lambda^{n/2} \veps_n$ grow with increasing $N_C$ as structure in the hybridization function becomes better resolved.

The presence of scaled tight-binding coefficients that fluctuate widely in magnitude from one Wilson-chain site to the next imposes an additional challenge for the NRG iterative diagonalization of the discretized Hamiltonian, which is based on the fundamental assumption that the addition of site $N+1$ creates a modest perturbation of the low-lying eigensolution of a chain consisting of sites $0 \le n \le N$. This challenge can be overcome by retaining a larger number of many-body eigenstates at the end of each iteration, but this comes at additional the cost of additional computational time. Retaining up to $N_s = 4^7$ many-body eigenstates, we find little change in $T\chi_\imp$ and $S_\imp$ on reducing $\Lambda$ from $5$ to $3$. We have used the latter value when computing $S_\imp$ vs $T$ or $T\chi_\imp$ vs $T$ curves, but have employed $\Lambda=5$even , or even $\Lambda=8$, and retained fewer than $4^7$ eigenstates when calculating distributions of the Kondo temperature over large numbers of samples.

\bibliography{ref}

\end{document}